\documentclass[preprint,amsmath,amssymb,aps]{revtex4-1}
\usepackage{graphicx}
\usepackage{lineno}
\usepackage{dcolumn}
\usepackage{booktabs}
\usepackage{bm}
\usepackage{subfig}
\usepackage{amsmath}
\usepackage{nomencl}
\makenomenclature 
\usepackage{ifthen}

\renewcommand\nomgroup[1]{%
 \item[\bfseries
 \ifthenelse{\equal{#1}{A}}{}{%
 \ifthenelse{\equal{#1}{B}}{Subscripts/Superscript}{%
 \ifthenelse{\equal{#1}{C}}{Abbreviations}{} } }%
]}

\usepackage{color}

\graphicspath{{FIG/}}

\begin{document}

\preprint{APS/123-QED}

\title{
On the formulations of interfacial force in the phase-field-based lattice Boltzmann method }

\author{Chunhua Zhang}
\affiliation{
State Key Laboratory of Coal Combustion, Huazhong University of Science and Technology, Wuhan 430074, China}
\author{Hong Liang}
\affiliation{
Department of Physics, Hangzhou Dianzi University, Hangzhou 310018, China}
\author{Zhaoli Guo}
\email{zlguo@hust.edu.cn}
\affiliation{
State Key Laboratory of Coal Combustion, Huazhong University of Science and Technology, Wuhan 430074, China}
\date{\today}

\begin{abstract}
Different formulations of interfacial force have been adopted in phase-field-based lattice Boltzmann method for two-phase flows. Although they  are identical mathematically, their numerical performances may be different due to truncation errors in the discretization. In this paper, four-type formulations of interfacial force available  in the literature, namely stress tensor form (STF), chemical potential form (CPF), pressure form (PF) and continuum surface force (CSF) form, are compared and discussed. A series of benchmark problems, including stationary droplet, two  merging droplets, Capillary wave, rising bubble and drop deformation in shear flow, are simulated. Numerical results show that  CPF is a good choice for small surface deformation problems while STF is preferred for dynamical problems, both STF and CSF demonstrate good numerical stability.
\end{abstract}
\maketitle

\mbox{}

\printnomenclature

\nomenclature[A]{$\phi$}{order parameter}
\nomenclature[A]{$M$}{mobility, $m N^{-1} s^{-1}$}
\nomenclature[A]{$\sigma$}{coefficient of surface tension, $N m^{-1}$}
\nomenclature[A]{$W$}{interface thickness,$m$}%
\nomenclature[A]{$\kappa$}{related to the gradient of order parameter}%
\nomenclature[A]{$\beta$}{related to the bulk free energy density}%
\nomenclature[A]{$\psi$}{total free energy}%
\nomenclature[A]{$f_0$}{bulk energy density}%
\nomenclature[A]{$\mu_{\phi}$}{chemical potential, $N m$}%
\nomenclature[A]{$\rho$}{density, $kg m^{-3}$}%
\nomenclature[A]{$\bm u$}{velocity, $m s^{-1}$}%
\nomenclature[A]{$ P$}{pressure, $Pa$}%
\nomenclature[A]{$\nu$}{kinematic viscosity, $m^2 s^{-1}$}%
\nomenclature[A]{$\mu$}{dynamic viscosity, $Pa s$}%
\nomenclature[A]{$\bm F_{sf}$}{surface tension force, $N m^{-3}$}%
\nomenclature[A]{$\bm g$}{gravitational acceleration, $m s^{-2}$}%

\nomenclature[A]{$f_i$}{distribution function for fluid flow}%
\nomenclature[A]{$h_i$}{distribution function for order parameter}%
\nomenclature[A]{$c$}{Lattice speed}%
\nomenclature[A]{$\bm c_i$}{ discrete velocity vectors}%
\nomenclature[A]{$c_s$}{speed of sound}%
\nomenclature[A]{$\tau_f$}{relaxation time}%
\nomenclature[A]{$\tau_h$}{relaxation time}%
\nomenclature[A]{$\omega_i$}{weight factors }%
\nomenclature[A]{$\delta x$}{space step}%
\nomenclature[A]{$\delta t$}{time step}%

\nomenclature[A]{$\bm n$}{unit normal vector of the interface}%
\nomenclature[A]{T}{nondimensional time}
\nomenclature[A]{$\text{Oh}$}{Ohnesorge number}
\nomenclature[A]{\text{Ca}}{Capillary number}
\nomenclature[A]{\text{Pe}}{Peclet number}
\nomenclature[A]{\text{Cn}}{Cahn number}
\nomenclature[A]{\text{Re}}{Reynolds number}

\nomenclature[B]{sf}{surface tension force}%
\nomenclature[B]{eq}{equilibrium}%
\nomenclature[B]{1,2}{fluid 1, fluid 2}%
\nomenclature[B]{c}{characteristic}%

\nomenclature[C]{CE}{Chapman-Enskog}%
\nomenclature[C]{NSAC}{ Navier Stokes Allen Cahn}%
\nomenclature[C]{NSCH}{ Navier Stokes Cahn Hilliard}%
\nomenclature[C]{NSK}{ Navier Stokes Kortweg}%
\nomenclature[C]{VOF}{volume of fluid}%
\nomenclature[C]{LS}{ level set}
\nomenclature[C]{LBM}{Lattice Boltzmann method}%
\nomenclature[C]{STF}{stress tensor form}%
\nomenclature[C]{PF}{pressure form}%
\nomenclature[C]{CPF}{chemical potential form}%
\nomenclature[C]{CSF}{continuum surface force}%

\section{Introduction}
Multiphase flows are ubiquitous in both natural processes and industrial applications, such as droplet dynamics~\cite{cristini2004theory}, lab-on-chip devices~\cite{clime2009numerical},
surfactant behavior~\cite{liu2018hybrid}, underground water flows~\cite{shad2010multiphase} and enhanced oil recovery~\cite{jafari2019application}. A number of numerical methods have been developed for simulating such flows, which  can be divided into two categories, i.e, interface tracking approach and interface capturing approach. In the former, interfaces are explicitly tracked,  such as the marker and cell method~\cite{mckee2008mac} and front-tracking method~\cite{tryggvason2011direct}. In the latter, interfaces are implicitly tracked and an interface function that marks the location of the interface is governed by the advection (diffusion) equations, such as volume of fluid (VOF) method~\cite{hirt1981volume}, level set (LS) method~\cite{chang1996level} and phase field method~\cite{jacqmin1999calculation}.

Among these methods, the phase field method is an increasingly popular choice for multiphase fluids simulations. The basic idea is to introduce a so-called order parameter that
has distinct values in the bulk phases but varies smoothly over the interfacial region. The order parameter defined as the volume fraction or mass fraction  is usually governed by the phase field equations, such as  the Cahn-Hilliard equation or the Allen-Cahn equation, which leads to the Navier-Stokes-Cahn-Hilliard (NSCH) system or the Navier-Stokes-Allen-Cahn (NSAC) system. If the fluid density is taken as an order parameter, the flow can be described by  the Navier-Stokes-Kortweg (NSK) system~\cite{korteweg1901forme,anderson1998diffuse}. Although  the momentum equations with interfacial force in  NSCH, NSAC and NSK  are very similar,
the properties of these equations are different. In the NSCH and NSAC equations, the pressure serves as an auxiliary variable associated with the incompressibility (or quasi-incompressibility) condition. In the NSK equations, the pressure is connected to the density via an equation of state. In the traditional computational fluid dynamics (CFD),  many discretization methods have been developed to numerically solve the above governing equations. Recently, the lattice Boltzmann method (LBM)  has grown as an alternative tool for multiphase flow simulations~\cite{liang2014phase,liu2014lattice,fakhari2018phase}.
The LBM is a mesoscopic method based on certain kinetic models. In  LBM, the fluid is represented by a discrete set of particle distribution functions which  only perform propagation and collision processes on a fixed lattice. The macroscopic  quantities of the flow are calculated by taking the moments of the particle distribution functions. LBM is simple and easy to be implemented compared with the traditional CFD to discretize the macroscopic governing equations. However, it can be shown that the corresponding phase field equation and hydrodynamic equations can be recovered from the lattice Boltzmann equations through the Chapman-Enskog (CE) analysis.

In  computational methods for multiphase flows, approximating the surface tension force accurately is critical to capture correct flow behaviors.
A number of mathematica models for the interfacial force are available in phase-field-based lattice Boltzmann methods so far. In fact, the interfacial force  can be strictly  derived based on the entropy principle of rational thermodynamics~\cite{jacqmin1999calculation,lowengrub1998quasi,abels2012thermodynamically,lam2018thermodynamically}. The resulting interfacial force appears as a gradient of the stress tensor  of the order parameter in the modified momentum equation.  These formulations can be  called stress tensor  form (STF). The stress form can be further simplified by redefining the pressure. Then, the interfacial force can be expressed as the forms dependent on the gradients of the order parameter~\cite{lee2006eliminating,he1999lattice,fakhari2010investigation,shah2018numerical,khan2019simulation}.  These formulations can be called  pressure form (PF). If the chemical potential related to the order parameter is employed, the interfacial force can also be expressed as the forms dependent on the chemical potential~\cite{zu2013phase,jacqmin1996energy,ding2007diffuse,fakhari2010phase}, which can be named as chemical potential form (CPF). Mathematically, the STF, PF and CPF are equivalent. In addition, in the continuum surface force (CSF) model of Brackbill \emph{et al.}~\cite{brackbill1992continuum}, the interfacial force is treated as a volumetric force proportional to the normal vector and curvature of the interface and a surface Dirac function localizing the interfacial force to the interface, which has been widely used in the VOF and LS methods.  Based on the CSF model, Kim~\emph{et al.}~\cite{kim2005continuous} proposed a CSF type interfacial force for phase field methods. The basic idea is to replace the level set by the order parameter and take the square of gradient of the order parameter as the surface Dirac function. An advantage of the CSF formulation is that the  pressure field can be calculated explicitly while the calculated pressure field with the previous interfacial forces includes some gradient terms of the order parameter except the true pressure.  The surface Dirac function in CSF model can also be defined in other ways. For instance, Lee and Kim~\emph{et al.}~\cite{lee2012regularized}  compared various types of surface Dirac  functions in the CSF model. They argued that the absolute value of the gradient of the order parameter has the best performances in their considered numerical experiments. These formulations  are called CSF form of the interfacial force in the present work. It's worth noting that the calculation of the normal vectors and the curvature at the interface is critical in the CSF models.

Although most of the above interfacial force formulations are mathematically equivalent, the performance of  each formulation may be different in practical computations. For example, Lee and Fischer ~\emph{et al.}~\cite{lee2006eliminating} compared the parasitic currents between the pressure form and  potential form in LBM, and the results showed that potential form yielded much smaller parasitic currents. Chao and Mei~\emph{et al.}~\cite{chao2011filter} compared the interface force distribution between the pressure form and the CSF form, and the results showed that the pressure form  could generate wiggles over the interface region while  the  CSF form produced no such unphysical results. However, there is a lack of systematic study of the performance of these four  interfacial force formulatiions widely used in LBM, and this paper will focus on this topic.

The paper is organized as follows. In section 2,  the governing equations of the phase field model for binary fluids are presented   and the formulas of surface tension force are summarized. The phase-field-based lattice Boltzmann method is briefly introduced in section 3. In section 4, several benchmark problems are investigated and the results are compared. Finally, conclusions are drawn in Section 5.

\section{Mathematical formulation}
\label{sec2}
\subsection{Governing equations}
In this study, we consider the NSCH equations for multiphase flows. The Cahn-Hillard equation  is expressed as~\cite{cahn1958free,jacqmin1999calculation}
\begin{equation}\label{eq:CH}
\frac{\partial \phi}{\partial t}+\nabla\cdot(\phi \bm u)=\nabla\cdot M\nabla\mu_{\phi} ,
\end{equation}
where $\phi$ is the order parameter to identify different phases, $M$  is the mobility, $\mu_{\phi}$  is  the chemical potential that is defined as
\begin{equation}\label{eq:chemical_potential}
\mu_{\phi}=\frac{\delta \psi}{\delta \phi}=\frac{\partial f_0}{\partial \phi}-\kappa \nabla^2 \phi,
\end{equation}
where $\psi$ is the system free energy,
\begin{equation}\label{eq:free_energy}
\psi=\int_V \left[ f_0(\phi) + \frac{\kappa}{2} |\nabla\phi|^2 \right]dV,
\end{equation}
where  $f_0=\beta(1-\phi^2)^2$  is the bulk energy density, the second term is the interface energy density,
 $\beta$ and $\kappa$ are determined by the surface tension $\sigma$ and the interface width $W$.

For a plane interface at equilibrium,  the equilibrium profile for the order parameter can be obtained  by solving $\mu_{\phi}=0$,
\begin{equation}\label{eq:equilibrium_profile}
  \phi(r)=\tanh\left(\sqrt{\frac{2 \beta}{\kappa}}r\right),
\end{equation}
where $r$ is the signed distance function which is  the coordinate normal to the interface. $\sqrt{\kappa/2\beta}$  has a length scale of interface thickness.
As the surface tension  is interpreted as  energy per unit surface area,
the surface tension  for a flat interface with equilibrium profile can be calculated by
\begin{equation}\label{eq}
\begin{aligned}
\sigma=& \int_{-\infty}^{+\infty}\left(f_0(\phi)+\frac{\kappa }{2}|\nabla\phi|^2 \right)dr \\
     =& \kappa \int_{-\infty}^{+\infty}|\nabla\phi|^2 dx =\frac{4}{3}\sqrt{2\beta\kappa},
\end{aligned}
\end{equation}
In Ref.\cite{jacqmin1999calculation}, $\sqrt{\kappa/2\beta}$ is defined as $W/2$, which leads to
 \begin{equation}\label{eq}
\beta=\frac{3}{4}\frac{\sigma}{W}, \quad  \kappa=\frac{3}{8}W\sigma.
\end{equation}

The dynamics of a fluid mixture of two incompressible viscous fluids can be described by the Navier-Stokes equations  with interfacial force~\cite{jacqmin1999calculation,ding2007diffuse}
\begin{equation}\label{eq:continue}
  \nabla\cdot \bm u=0,
\end{equation}
\begin{equation}\label{eq:NS}
\frac{\partial(\rho \bm u)}{\partial t} +\nabla\cdot(\rho \bm u\bm u)
=-\nabla P_{sf}+\nabla\cdot\mu(\nabla\bm u+\nabla \bm u^T)+ \bm F_g+ \bm {F}_{sf},
\end{equation}
where $\rho$ is the fluid density, $\bm u$ is the flow velocity, $P_{sf}$ is the generalized pressure dependent on the definition of the interfacial  force, $\mu$ is the dynamic viscosity, $\bm F_g=(\rho-\rho_{0}) \bm g$ is the gravitational force with $\bm g$ being the gravitational acceleration and $\rho_{0}$ being the background density, $\bm F_{sf}$ is the interfacial force. The subscript $sf(=stf,cpf,pf,csf) $ denotes different formulations of interfacial force.

The mixture density $\rho$ and viscosity $\mu$  can be given by
\begin{equation}\label{eq}
  \rho=\rho_1\frac{1+\phi}{2}+\rho_2\frac{1-\phi}{2}
\end{equation}
\begin{equation}\label{eq}
  \mu=\mu_1\frac{1+\phi}{2}+\mu_2\frac{1-\phi}{2}
\end{equation}
where the subscripts $1$ and $2$ indicate fluid 1 and fluid 2.

To non-dimensionalize the equations in NSCH system,  the following  dimensionless variables are used,
\begin{equation}\label{eq}
\bm u'=\frac{\bm u}{U_c}, \quad  \bm x'=\frac{\bm x}{L_c},\quad   t'=\frac{t}{T_c}, \quad  p'=\frac{P_{sf}}{p_c}, \quad  \mu'_\phi=\frac{\mu_\phi }{\mu_{\phi,c}}, \quad
\bm F_{sf}'=\frac{\bm F_{sf} L_c^2 }{\sigma},
\end{equation}
where $U_c, L_c, T_c(=L_c/U_c), p_c(=\rho_c U_c^2), \mu_{\phi,c}(=4\beta)$ are respectively the reference velocity,  length,  time,  pressure and  chemical potential.  In this paper, the density and dynamical viscosity of fluid $1$ are chosen as the reference quantities, i.e, $\rho_c=\rho_1, \mu_c=\mu_1$. With the above variables and dropping the primes, the dimensionless governing equations can be written as
\begin{equation}\label{eq:nondimensionalization1}
\partial_t \phi+ \nabla \cdot (\phi \bm u)=\frac{1}{\text{Pe}}\nabla\cdot (M\nabla \mu_{\phi}),
\end{equation}
\begin{equation}\label{eq:nondimensionalization2}
\partial_t (\rho\bm u) + \nabla\cdot(\rho \bm u\bm u)=-\nabla P_{sf}+ \frac{1}{\text{Re}}\nabla\cdot \mu(\nabla \bm u+\nabla \bm u^T)+ \frac{1}{\text{We}}\bm F_{sf}+\frac{1}{\text{Fr}^2} \bm F_g,
\end{equation}
\begin{equation}\label{eq:nondimensionalization3}
\nabla\cdot \bm u=0,
\end{equation}
with
\begin{equation}\label{eq}
\begin{aligned}
\mu_\phi &=\phi(\phi^2-1)-\frac{\text{Cn}^2}{8}\nabla^2\phi, \\
\rho &=\frac{1+\phi}{2}+\frac{1-\phi}{2}\frac{\rho_2}{\rho_1}, \\
\mu &=\frac{1+\phi}{2}+\frac{1-\phi}{2}\frac{\mu_2}{\mu_1}.
\end{aligned}
\end{equation}
The dimensionless groups used above are the Reynolds number $\text{Re}$, Peclet number $\text{Pe}$, Weber number $\text{We}$, Frounde number $\text{Fr}$  and  Cahn number $\text{Cn}$, which are respectively defined by
\begin{equation}\label{eq}
\text{Re}=\frac{\rho_c U_c L_c}{\mu_c},\quad \text{Pe}=\frac{U_c L_c}{4 M\beta},\quad  \text{\text{We}}=\frac{\rho_c L_c U_c^2}{\sigma},
\quad \text{Fr}=\frac{U_c}{\sqrt{gL_c}},\quad
\text{Cn}=\frac{W}{L_c},
\end{equation}

\subsection{Interfacial force formulations}
\label{subsec:SF}
Based on the energetic variational approach or the free energy inequality, the surface tension force  in the  momentum equation  can be defined as~\cite{yue2004diffuse,abels2012thermodynamically,abels2017diffuse}
\begin{equation}\label{eq:SF1}
\bm F_{stf-1}=- \nabla \cdot\kappa ( \nabla\phi\otimes\nabla\phi),
\end{equation}
where $\nabla\phi\otimes \nabla\phi$ is the usual tensor product and denotes the induced  elastic stress due to the mixing of the different species. In this case,  the  generalized pressure $P_{sf}$ in Eq.~(\ref{eq:NS}) includes both the hydrostatic pressure $p_h$ due to the incompressibility and the contributions from the induced stress, $P_{stf-1}=p_h+\kappa|\nabla\phi|^2$. In Ref.~\cite{starovoitov1994model,jacqmin2000contact}, the surface tension force term is defined as
\begin{equation}\label{eq:SF2}
\bm F_{stf-2}= \nabla\cdot \kappa(|\nabla\phi|^2 \bm I -\nabla\phi\otimes\nabla\phi),
\end{equation}
which implies that the principle axes of the tensor are perpendicular to the tangent plane of the interface. The normal stress perpendicular to the tangent plane of the interface is zero and the two tangent normal stresses are equal. In this case, the generalized  pressure in Eq.~(\ref{eq:NS}) becomes the true pressure, namely, $P_{stf-2}=p_h$ ~\cite{jacqmin1999calculation,jacqmin2000contact}.

For simplicity, we assume that the surface tension $\sigma$ is constant. By using the following identity
\begin{equation}\label{eq:identity}
\begin{aligned}
\kappa\nabla\cdot(\nabla\phi\otimes \nabla\phi)
&=\frac{\kappa}{2}  \nabla|\nabla\phi|^2+\kappa\nabla\phi\Delta\phi\\
&=\nabla\left( \frac{\kappa}{2} |\nabla\phi|^2 +\kappa\phi\Delta\phi \right)-\kappa\phi\nabla\Delta\phi \\
&=\nabla\left( \frac{\kappa}{2} |\nabla\phi|^2 +f_0 \right)-\mu_{\phi}\nabla\phi \\
&=\nabla\left( \frac{\kappa}{2} |\nabla\phi|^2 +f_0-\phi\mu_{\phi} \right)+\phi\nabla\mu_{\phi},
\end{aligned}
\end{equation}
and absorbing the gradient terms into pressure $p_h$, the surface tension force can be expressed as
\begin{equation}\label{eq:SF3456}
\begin{aligned}
\bm F_{cpf-1} &=-\phi\nabla\mu_{\phi}, &  \quad  \bm F_{cpf-2}  &=\mu_{\phi}\nabla\phi, \\
\bm F_{pf-1}  &=-\kappa\nabla\phi\Delta\phi, &\quad \bm F_{pf-2} &=\kappa\phi\nabla\Delta \phi.
\end{aligned}
\end{equation}
The corresponding  generalized pressure is redefined as
\begin{equation}\label{eq:P3456}
\begin{aligned}
P_{cpf-1} &=p_h+f_0-\phi\mu_{\phi} -\frac{\kappa}{2}|\nabla\phi|^2, \\
P_{cpf-2} &=p_h+f_0-\frac{\kappa}{2}|\nabla\phi|^2,\\
P_{pf-1} &=p_h-\frac{\kappa}{2}|\nabla\phi|^2,\\
P_{pf-2} &=p_h+\kappa\phi\Delta\phi-\frac{\kappa}{2}|\nabla\phi|^2.
\end{aligned}
\end{equation}
$ \bm F_{cpf-1}$ and $\bm F_{cpf-2}$ are termed as chemical potential form. $\bm F_{pf-1}$ and $\bm F_{pf-2}$ are the pressure form.
It is noted that
 $\bm F_{stf-1}$ is used in~\cite{yang2006numerical,shen2009efficient} and  $\bm F_{stf-2}$ is used in~\cite{starovoitov1994model,lee2002modeling,kim2005diffuse,zhang2019fractional},
 $\bm F_{cpf-1}$ is used in~\cite{zu2013phase,wang2015multiphase,chen2018simplified} and $\bm F_{cpf-2}$ is used  in~\cite{jacqmin1996energy,liang2014phase,ding2007diffuse,fakhari2010phase},
$\bm F_{pf-1}$ is used  in~\cite{shah2018numerical,khan2019simulation} and $\bm F_{pf-2} $ is used in~\cite{he1999lattice,fakhari2010investigation}.

Based on the CSF model, the surface tension force  can be   given by~\cite{brackbill1992continuum,popinet2018numerical}
\begin{equation}\label{LS_CSF}
\bm F_{csf}=  \sigma \widetilde{\kappa}\delta_s \bm n,
\end{equation}
where $\bm n$ is the unit normal vector, $\widetilde{\kappa}=-\nabla\cdot\bm n$ is the local mean curvature, $\delta_s $ is the surface Dirac function used to ensure the force acting on the interfacial region.   To match the surface tension of the sharp interface model, the Dirac function should satisfy
\begin{equation}\label{eq:diracfunction}
  \int_{-\infty}^{\infty}\delta_s d r=1.
\end{equation}
There are many possible choices for $\delta_s$. Kim~\cite{kim2005continuous} proposed to use $\alpha |\nabla \phi|^2$ as  the Dirac function with $\alpha=3W/8$,
\begin{equation}\label{eq:SF7}
\bm F_{csf-1}=-\kappa\nabla\phi|\nabla\phi|\nabla\cdot\bm n.
\end{equation}
Lee and Kim~\emph{et.al}~\cite{lee2012regularized} proposed  $\alpha |\nabla\phi|$ as the Dirac function  with $\alpha=0.5$,
\begin{equation}\label{eq:SF8}
\bm F_{csf-2}=-\frac{\sigma}{2}\nabla\phi (\nabla\cdot\bm n).
\end{equation}
The derivation of $\alpha$ is referred to Appendix.\ref{ap3:dirac}.
In above interfacial force formulations, Eqs.~(\ref{eq:SF1}),~(\ref{eq:SF2}) and (\ref{eq:SF3456}) are identical mathematically. In fact, these formulations can be rewritten as
\begin{equation}\label{eq:relationSF7}
\begin{aligned}
\bm F_{stf-1}  &=\bm F_{csf-1}-\left[\nabla\frac{\kappa |\nabla\phi|^2}{2}
+\frac{\kappa\nabla\phi(\nabla\phi\cdot\nabla|\nabla\phi|)}{|\nabla\phi|}\right], \\
\bm F_{stf-2}  &=\bm F_{csf-1} -\left[-\nabla\frac{\kappa |\nabla\phi|^2}{2}
+\frac{\kappa\nabla\phi(\nabla\phi\cdot\nabla|\nabla\phi|)}{|\nabla\phi|}\right],\\
\bm F_{cpf-1} &=\bm F_{csf-1} -\left[\nabla(\phi\mu_{\phi})-\nabla f_0
+\frac{\kappa\nabla\phi(\nabla\phi\cdot\nabla|\nabla\phi|)}{|\nabla\phi|}\right],\\
\bm F_{cpf-2} &=\bm F_{csf-1} -\left[-\nabla f_0
+\frac{\kappa\nabla\phi(\nabla\phi\cdot\nabla|\nabla\phi|)}{|\nabla\phi|}\right],\\
\bm F_{pf-1}  &=\bm F_{csf-1} -\frac{\kappa\nabla\phi(\nabla\phi\cdot\nabla|\nabla\phi|)}{|\nabla\phi|},\\
\bm F_{pf-2}  &=\bm F_{csf-1} -\left[-\nabla(\kappa\phi\Delta\phi)
+\frac{\kappa\nabla\phi(\nabla\phi\cdot\nabla|\nabla\phi|)}{|\nabla\phi|}\right].
\end{aligned}
\end{equation}
It is clear that there are some gradient terms in $\bm F_{stf},\bm F_{cpf}$ and $\bm F_{pf}$. This is why the previous formulations cannot be used to calculate the pressure field explicitly~\cite{kim2005continuous}.

By using Eq.~(\ref{eq:equilibrium_profile}), the following relations can be obtained
\begin{equation}\label{eq:D_eqphi}
\begin{aligned}
|\nabla\phi| &=\frac{2}{W}(1-\phi^2),\\
\frac{\nabla\phi(\nabla\phi\cdot\nabla|\nabla\phi|)}{|\nabla\phi|} &= \frac{1}{2}\nabla|\nabla\phi|^2.
\end{aligned}
\end{equation}

Inserting Eq.~(\ref{eq:D_eqphi}) into Eq.~(\ref{eq:relationSF7}) leads to
\begin{equation}\label{eq:relationSF7-simplify}
\begin{aligned}
\bm F_{stf-1}  &=\bm F_{csf-1}-\nabla\kappa |\nabla\phi|^2, \\
\bm F_{cpf-1}  &=\bm F_{csf-1} -\nabla(\phi\mu_{\phi} ),\\
\bm F_{pf-1}   &=\bm F_{csf-1} -\nabla \frac{\kappa}{2}|\nabla\phi|^2 ,\\
\bm F_{pf-2}   &=\bm F_{csf-1} -\nabla(-\kappa\phi\Delta\phi +\frac{\kappa}{2}|\nabla\phi|^2 ),\\
\bm F_{stf-2}  &=\bm F_{cpf-2}=\bm F_{csf-1}.\\
\end{aligned}
\end{equation}
Therefore,  $\bm F_{csf-1}$, $\bm F_{stf-2}$ and $\bm F_{cpf-2}$ are identical when the system is at equilibrium. The main difference between  $\bm F_{csf-1}$ and $\bm F_{csf-2}$ is the definition of the Dirac delta function. All above formulations have been used to mimic the interfacial force in the phase-field-based LBM. In Sec. V, the performance of the LBM models with the above eight formulations of surface tension force will be compared.

\section{Phase-field-based Lattice Boltzmann Method}
We adopted the multiphase LBM of He \emph{et al}~\cite{he1999lattice} for the hydrodynamic equations and the improved LBM of Zhang \emph{et al}~\cite{zhang2019high} for Cahn-Hilliard equation. The evolutions of the  distribution functions $f_i$ and $h_i$ are respectively expressed as
\begin{equation}\label{fh}
f_i(\bm x+\bm c_i\delta t,t+\delta t)-f_i(\bm x,t) =
-\frac{1}{\tau_f}[f_i(\bm x,t)-f_i^{eq}(\bm x,t)]+ \delta t \left(1-\frac{1}{2\tau_f}\right) F_i,
\end{equation}

\begin{equation}\label{fh}
h_i(\bm x+\bm c_i\delta t,t+\delta t)-h_i(\bm x,t) =
-\frac{1}{\tau_h}[h_i(\bm x,t)-h_i^{eq}(\bm x,t)]+ \delta t \left(1-\frac{1}{2\tau_h}\right) H_i,
\end{equation}
where $f_i(\bm x,t)$ and $h_i(\bm x,t)$ are the distribution functions for the hydrodynamics and order parameter fields respectively, $\bm c_i$ is the discrete velocity in the i-th direction, $\delta t$ is the time step, $\tau_f$ and $\tau_h$ are the dimensionless relaxation times related to the shear viscosity and mobility respectively, $F_i$ and $H_i$ are the discrete force terms. To recover the correct governing equations, the equilibrium distributions $f_i^{eq}$ and $h_i^{eq}$ are defined as
\begin{equation}\label{eq}
f_i^{eq}=\omega_i [P_{sf} +c_s^2\rho s_i(\bm u) ]
\end{equation}
\begin{equation}\label{eq:heq_CH}
 h_i^{eq}=\left\{
  \begin{aligned}
  &\phi+(\omega_0-1)\eta\mu_{\phi},& & \text {$i=0$}\\
  &\omega_i\eta\mu_{\phi} +\omega_i \phi \frac{\bm c_i \cdot \bm u}{c_s^2},&  & \text {$i \neq 0$}
  \end{aligned}
  \right.
\end{equation}
with
\begin{equation}\label{eq}
s_i(\bm u)=\frac{\bm c_i\cdot \bm u}{c_s^2}+ \frac{\bm u\bm u:(\bm c_i \bm c_i -c_s^2\bm I)}{2c_s^4},
\end{equation}
where $\omega_i$ is the weighting coefficient corresponding to the discrete velocity $\bm c_i$, $c_s=c/\sqrt{3}$ is the lattice sound speed, $c=\delta x/\delta t$ is  the lattice speed with $\delta x$ being the lattice length scale,
 and $\eta$ is an adjustable parameter for the mobility.
In this work, the two-dimensional nine-velocity (D2Q9) model is used in which the discrete velocity is
\begin{equation}\label{eq}
\begin{aligned}
&\left(\bm c_0, \bm c_1, \bm c_2, \bm c_3, \bm c_4, \bm c_5, \bm c_6,\bm c_7, \bm c_8\right)\\
&=c \left(
\begin{array}{ccccccccc}
0 & 1 & 0 & -1 & 0 & 1 & -1 & -1 & 1 \\
0 & 0 & 1 & 0 & -1 & 1 & 1 & -1 & -1\\
\end{array}
 \right)
\end{aligned}
\end{equation}
and the corresponding weighting coefficients are $\omega_0=4/9$, $\omega_{1-4}=1/9$ and $\omega_{5-8}=1/36$.

The force terms $F_i$ and $H_i$ are given by
\begin{equation}\label{eq:P_LBM}
F_i=(\bm c_i-\bm u)\cdot\left[\Gamma_i(\bm u) (\bm F_{sf}+\bm F_g)+s_i(\bm u) \nabla c_s^2\rho\right]
\end{equation}
\begin{equation}\label{eq:CH_LB_force}
H_i=\bar{\omega}_i\frac{3\tau_2}{\tau_h\tau_1\delta t}\nabla\cdot(\bm u\phi) + \omega_i \frac{\bm c_i \cdot \partial_t(\phi\bm u)}{c_s^2}.
\end{equation}
where $\Gamma(\bm u)=\omega_i + s_i(\bm u)$, $\tau_2=(\tau_h^2-\tau_h+\frac{1}{6})\delta t^2$, $\tau_1=(\tau_h-0.5)\delta t$ and $\bar{\omega}_i$ is a new weight coefficient and satisfies $\sum_i \bar{\omega}_i=\sum_i \bar{\omega}_i\bm c_i=0, \sum_i \bar{\omega}_i \bm c_i \bm c_i=c_s^2\bm I$.  In particular, if $\tau_g=0.5+\sqrt{3}/6$, the above improved LBM for CH can be simplified to the one of Liang \emph{et al}~\cite{liang2014phase}. However, the relationship between the Peclet number and Cahn number should be $\text{Pe}\sim \text{Cn}^{-1}$ to achieve the sharp-interface limit with continuous mesh refinement~\cite{magaletti2013sharp}. Then, the relaxation time  may have a value except the optimum one in some situations and the improved LBM should be considered.

The macroscopic quantities are calculated by
\begin{equation}\label{eq:mac}
\begin{aligned}
P_{sf} &=\sum_i f_i +\frac{\delta t}{2}c_s^2\bm u\cdot \nabla\rho, \\
\bm u &=\frac{1}{c_s^2\rho}\left[\sum_i\bm c_i f_i +\frac{\delta t}{2}c_s^2 (\bm F_{sf} +\bm F_g) \right] \\
\phi &=\sum_i h_i,
\end{aligned}
\end{equation}

Through the Chapman-Enskog expansion, the macroscopic governing equations recovered from the above LBM are
\begin{align}\label{eq:recoverd_vel}
\frac{1}{c_s^2\rho}\partial_t P_{sf}+\nabla\cdot\bm u &=0,\\
\label{eq:recoverd_mom}
\partial_t(\rho\bm u)+\nabla\cdot(\rho\bm u\bm u) &=
-\nabla P_{sf} + \nabla\cdot\mu(\nabla\bm u+\nabla\bm u^T)
+ \bm F_{sf} +\bm F_g,\\
\partial_t\phi+\nabla\cdot(\phi\bm u)&=\nabla\cdot M\nabla\mu_{\phi},
\end{align}
where the  viscosity $\mu$ and the mobility $M$ are defined as $\mu=\rho c_s^2(\tau_f-0.5)\delta t$ and $M=c_s^2\eta(\tau_h-0.5)\delta t$, respectively.

The gradient terms in each formulation of interfacial force  can be calculated with different schemes. In the present work, we will use the isotropic central scheme~\cite{guo2011force},
\begin{equation}\label{eq:ICS-1st}
\nabla\Psi =\frac{1}{c_s^2\delta t}\sum_{i=1}^{8}\omega_i\bm c_i\Psi(\bm x+\bm c_i\delta t),\\
\end{equation}
\begin{equation}\label{eq:ICS-sec}
\nabla^2\Psi =\frac{2}{c_s^2\delta t}\sum_{i=1}^{8}\omega_i\left[\Psi(\bm x+\bm c_i\delta t)-\Psi(\bm x)\right],
\end{equation}
where $\Psi$ denotes arbitrary quantity. For a node located at wall boundary, a second-order one-side finite difference is employed.

\section{Boundary conditions}
\begin{figure}[!htb]
  \centering
  \includegraphics[width=0.5\textwidth]{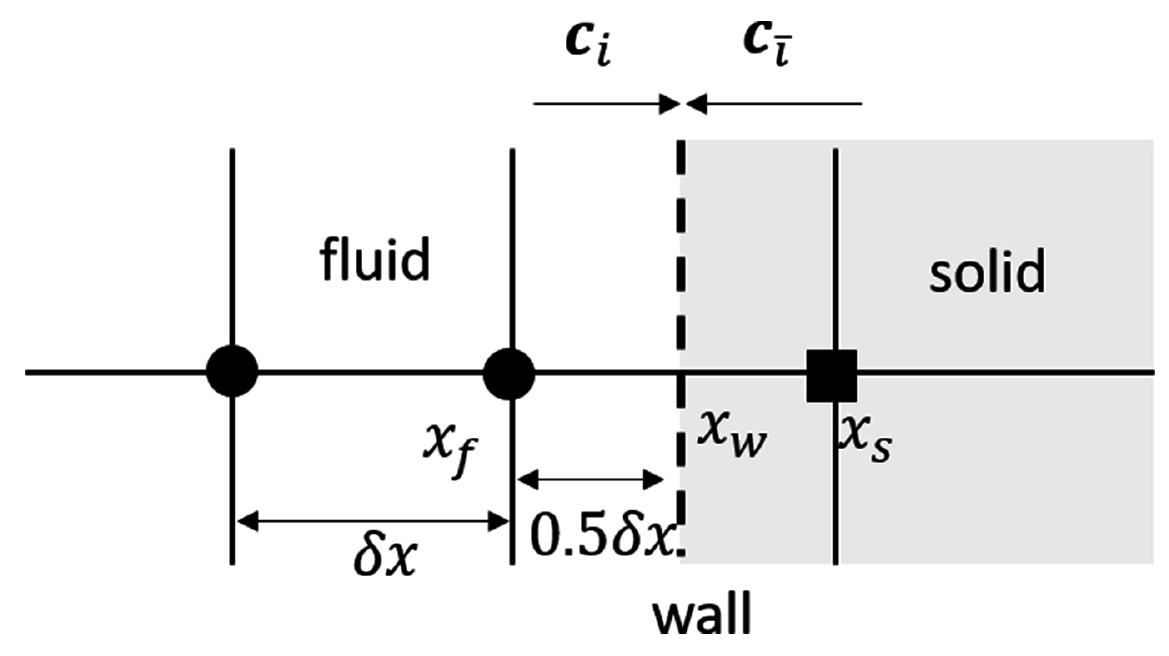}
  \caption{ Illustration for the half-way bounce-back. The thin solid straight line is the grid line and the dashed line corresponds to the computational boundary.
The black circles are the fluid nodes and the black square is the solid node. The arrow represents the particle's direction, the rightmost grey shaded domain is the solid region. }\label{half-bounce-back}
\end{figure}

Boundary treatment is one of the most important tasks in numerical methods.
In LBM, the classical boundary condition to model walls is the bounce-back method, which can be realised by both the full-way bounce-back and the half-way bounce-back~\cite{kruger2017lattice}. As the half-way bounce-back can be implemented without solid nodes and is more accurate for unsteady flows, we will only consider the half-way bounce-back in the practical calculation.
As shown in Fig~\ref{half-bounce-back}, following Ladd's half-way bounce-back scheme, the unknown distribution function  is determined by~\cite{ladd1994numerical,liu2014lattice}
\begin{equation}\label{eq}
\begin{aligned}
f_{\bar{i}}(\bm x_f, t+\delta t) &=f_i^+(\bm x_f,t)- 2\omega_i \rho(\bm x_w,t) \bm c_i \cdot \bm u_w,\\
g_{\bar{i}}(\bm x_f,t+\delta t) &=g_i^+(\bm x_f,t) - 2\omega_i \phi(\bm x_w,t) \frac{\bm c_i \cdot \bm u_w}{c_s^2},
\end{aligned}
\end{equation}
where $f_{\bar{i}}$ and $g_{\bar{i}}$ are the distribution function with the velocity $\bm c_{\bar{i}}=-\bm c_{i}$, the  superscript '+' denotes the post-collision value of the corresponding distribution function and $\bm u_w$ is the prescribed wall velocity. For a stationary boundary with $\bm u_w=0$, the above equations can be used for the non-slip boundary.

For the order parameter, the following boundary conditiions are employed,
\begin{equation}\label{eq:orderparameterboundary}
\bm n_w\cdot \nabla \phi=0, \qquad   \bm n_w \cdot \nabla \mu_\phi=0,
\end{equation}
where $\bm n_w$  is the unit outward normal defined at the solid boundary. Eq.(\ref{eq:orderparameterboundary}) means that the order parameter conserves mass over the entire domain. In addition, the density $\rho(\bm x_w,t)$ can be approximated by $\rho(\bm x_f,t)$. Here we use $\nabla \phi \cdot n_w=0$  to interpolate the density at the wall.

\section{numerical results and discussion}
In this section, the performance of each interfacial force formulation is validated by a series of benchmark tests, including stationary droplet, two merging droplets, capillary wave, rising bubble and the deformation droplet in a shear flow. For each test, the results obtained by the lattice Boltzmann equation (LBE) model with different interfacial force formulations are compared with the theoretical solutions or the available reference solutions in the literature. In Eq.~(\ref{eq:CH_LB_force}), the time derivative  is calculated by explicit Euler scheme, and $\bar{\omega}_0=\omega_0-1$, $\bar{\omega}_i=\omega_i$ for $i>0$.
The Peclet number is set to be $1.0/\text{Cn}$ and the interface width is set to be four grids unless otherwise stated.

\subsection{Stationary droplet}
We first make a comparison among different interfacial force formulations by simulating a stationary droplet. Theoretically, the exact solution is zero velocity for all time.
Initially, a circle droplet with radius $R$ is placed at the center of the domain $L\times L$. The order parameter is set to be
\begin{equation}\label{eq}
\phi(x,y)=\tanh\left(2\frac{R-\sqrt{(x-x_c)^2+(y-y_c)^2}}{W}\right),
\end{equation}
where $(x_c,y_c)$ is the center coordinate of the droplet. Periodic boundary conditions are applied to all the boundaries. The initial velocity field is set to be zero.
The physical parameters are set to be $L=1 \text{m}$, $R=0.25\text{m}$,  $\rho_1$=4\text{kg/$m^3$}, $\rho_2$=1 \text{kg/$m^3$} , $\nu_1=\nu_2=0.25$ \text{$m^2$/s} and $\sigma=0.357 \text{N/m}$. Three uniform grids  of $60\times 60, 120\times 120,240\times 240$ are used.
The characteristic velocity is $U_c= \sigma/\mu_1$.

We first examine the shape of the droplet at equilibrium.
The interface profile of the droplet obtained by all interfacial force formulations are similar and  agree well with the initial interface profile, and the results are not shown here.
It is also found that the deviation between the numerical results given by all formulations and the analytical interface profile becomes small as the value of mobility decreases, which is also consistent with the results in \cite{liang2014phase}. Since the definition of characteristic velocity is artificial to some extent, the relationship of $\text{Pe}\sim 1/\text{Cn}$ may be unable to produce the closest results to the exact one.

From the Laplace law,  the numerical surface tension can be calculated  by $\sigma_{\text{num}}=R_{\text{num}}\times(p_{\text{in}}-p_{\text{out}})$. The relative error, $\text{Err}=|\sigma_{\text{num}}-\sigma_{\text{exact}}|/\sigma_{\text{exact}}\times 100\%$, is listed in Table~\ref{tab:test1_laplace}.
It can be seen that
the error decreases as the grid resolution increases.
For all meshes, $\bm F_{csf-2}$ gives the smallest error while  $\bm F_{pf-2}$ gives the largest one .
\begin{table}[!htb]
\centering
\caption{Comparison of numerical surface tension based on Laplace law ($\sigma=0.357$)} \label{tab:test1_laplace}
\setlength{\tabcolsep}{0.92mm}{%
\begin{tabular}{ccccccccc}
\hline
$\bm F_{sf}$ & \multicolumn{2}{c}{$60 \times 60$} & & \multicolumn{2}{c}{$120 \times 120$} &  &\multicolumn{2}{c}{$240 \times 240$}\\
\cline{2-3} \cline{5-6} \cline{8-9}
 & $\sigma_{num}$     & $Err(\%)$  &  & $\sigma_{num}$     & $Err(\%)$        &     &$\sigma_{num}$ &  $Err(\%)$  \\
\hline
$\bm F_{stf-1}$&0.3401  &4.736&       &0.3404 &4.656&       &0.3410 &4.480   \\

$\bm F_{stf-2}$&0.3401  &4.740&       &0.3404 &4.660&       &0.3410 &4.484    \\

$\bm F_{cpf-1}$&0.3505  &1.833 &      &0.3511 &1.652&       &0.3519& 1.418    \\

$\bm F_{cpf-2}$&0.3507  &1.765 &      &0.3512 &1.619&       &0.3520& 1.404    \\

$\bm F_{pf-1}$& 0.3512  &1.629 &       &0.3517 &1.475&       &0.3524& 1.277    \\

$\bm F_{pf-2}$&0.3289  &7.877 &       &0.3290 &7.846&       &0.3296& 7.688    \\

$\bm F_{csf-1}$&0.3404  &4.657 &      &0.3404 &4.637&       &0.3410& 4.475    \\

$\bm F_{csf-2}$&0.3599  &0.825 &      &0.3558 &0.326&       &0.3566& 0.108    \\
\hline
\end{tabular}}
\end{table}

\begin{figure}[!htb]
\centering
\subfloat[]{\includegraphics[width=0.25\textwidth]{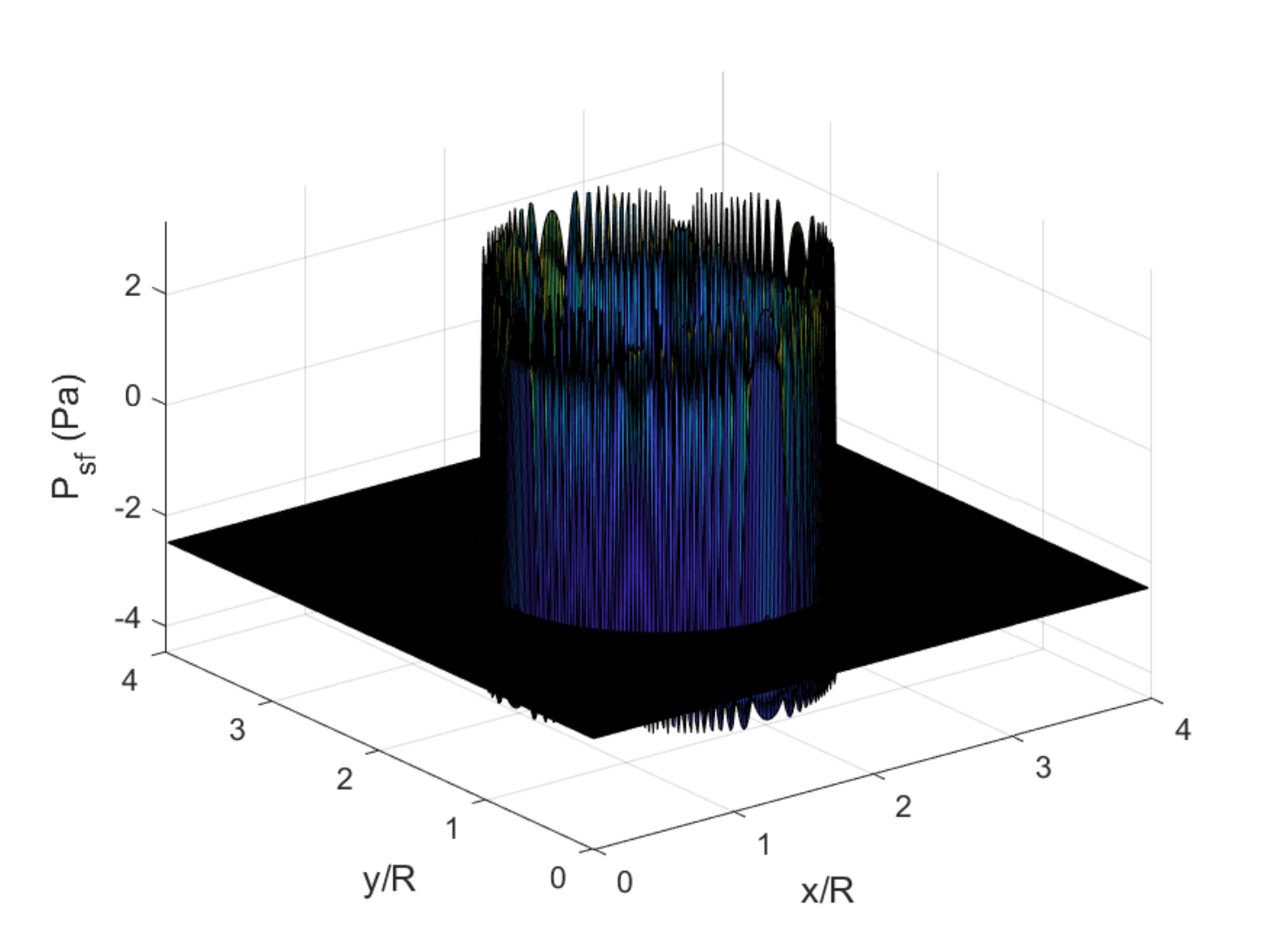}}~
\subfloat[]{\includegraphics[width=0.25\textwidth]{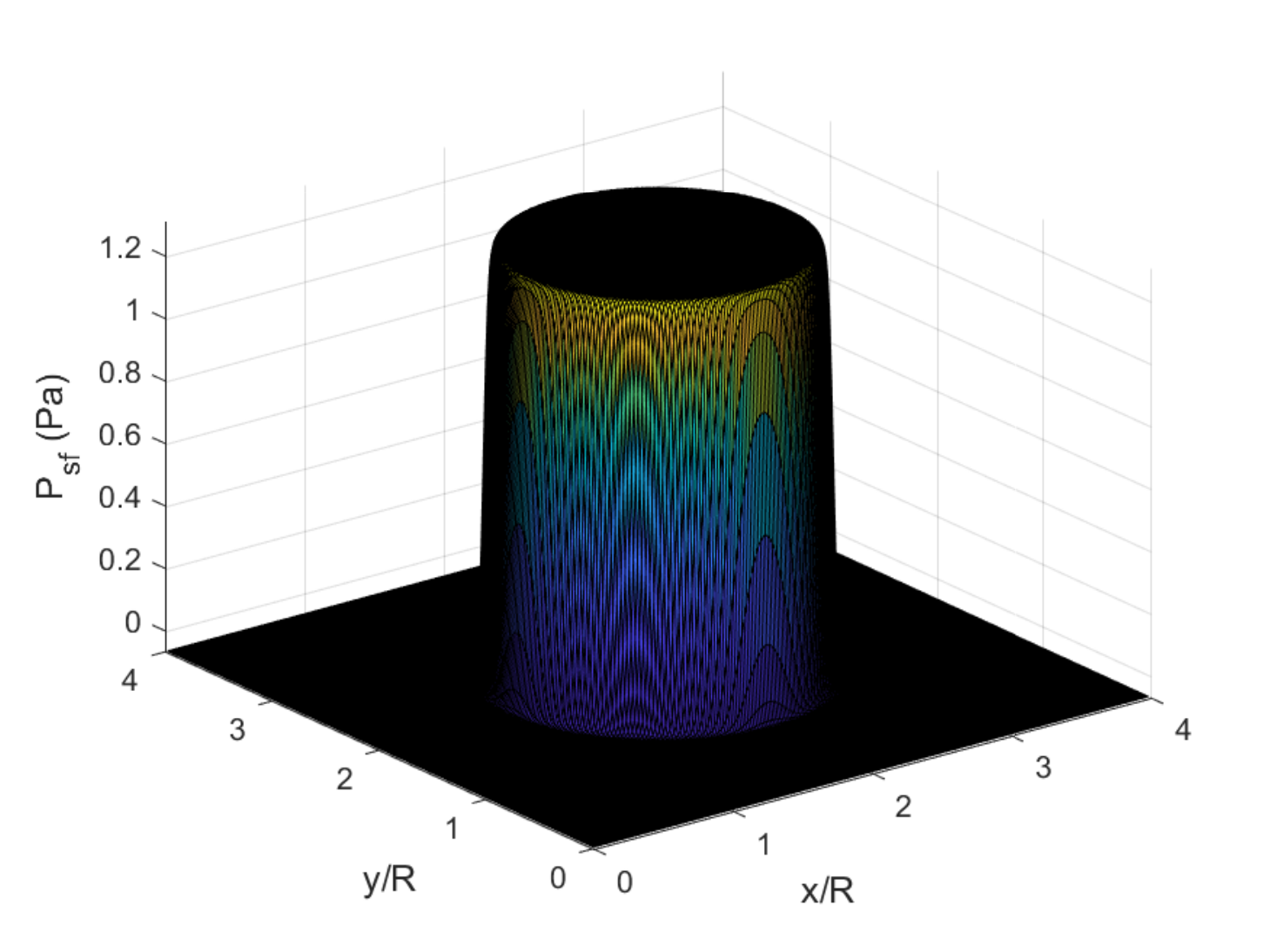}}~
\subfloat[]{\includegraphics[width=0.25\textwidth]{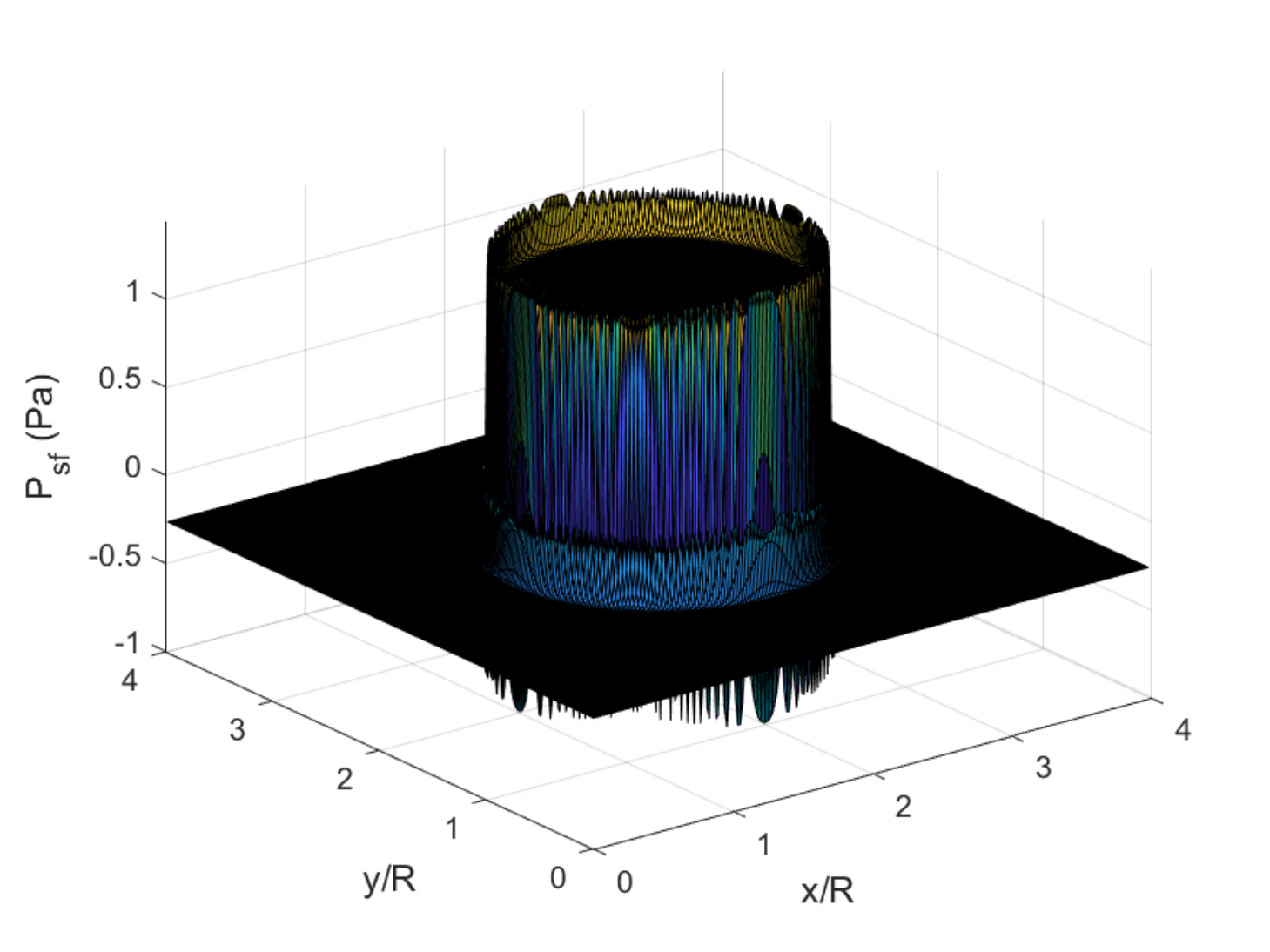}}~
\subfloat[]{\includegraphics[width=0.25\textwidth]{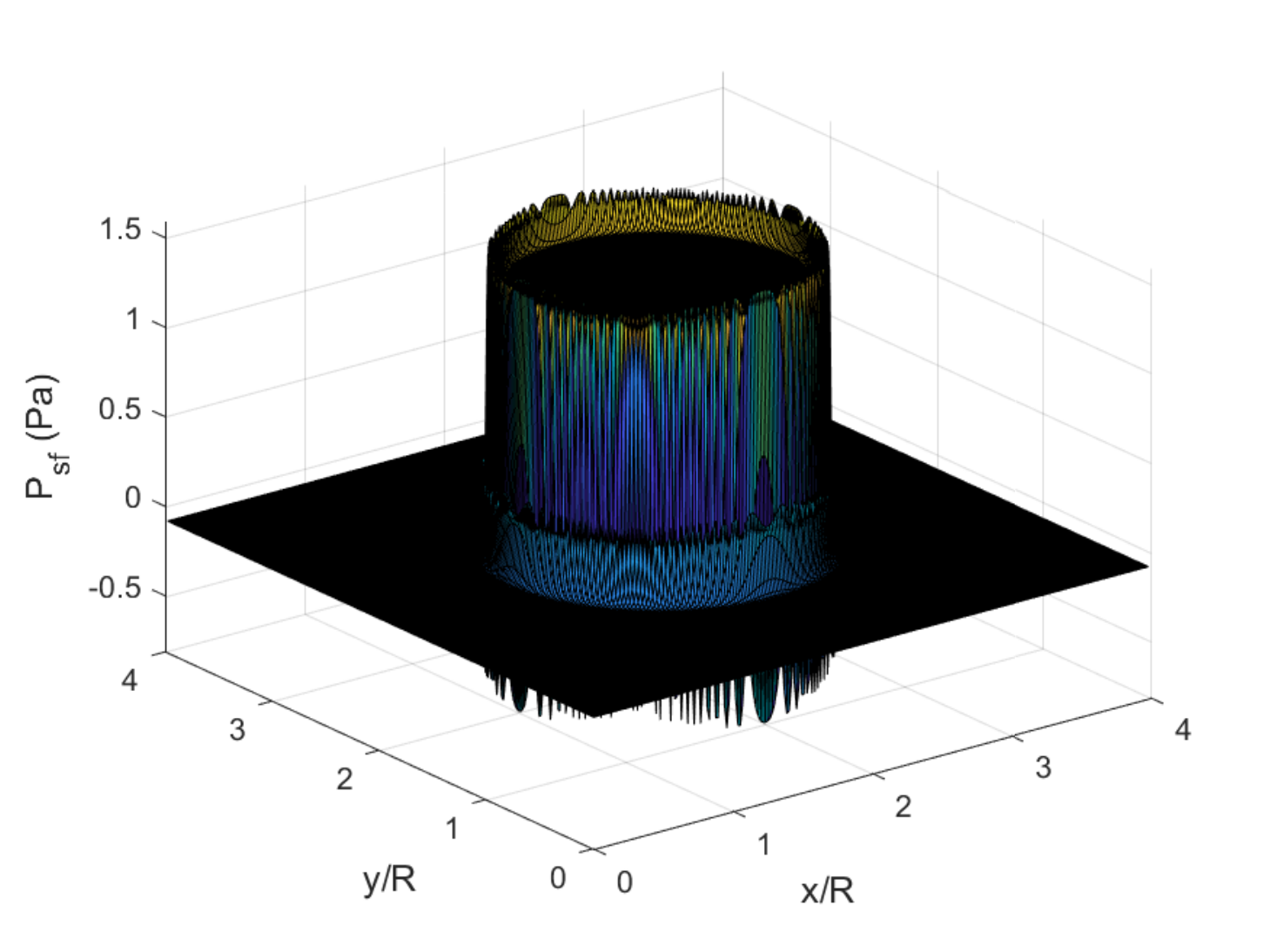}} \\
\subfloat[]{\includegraphics[width=0.25\textwidth]{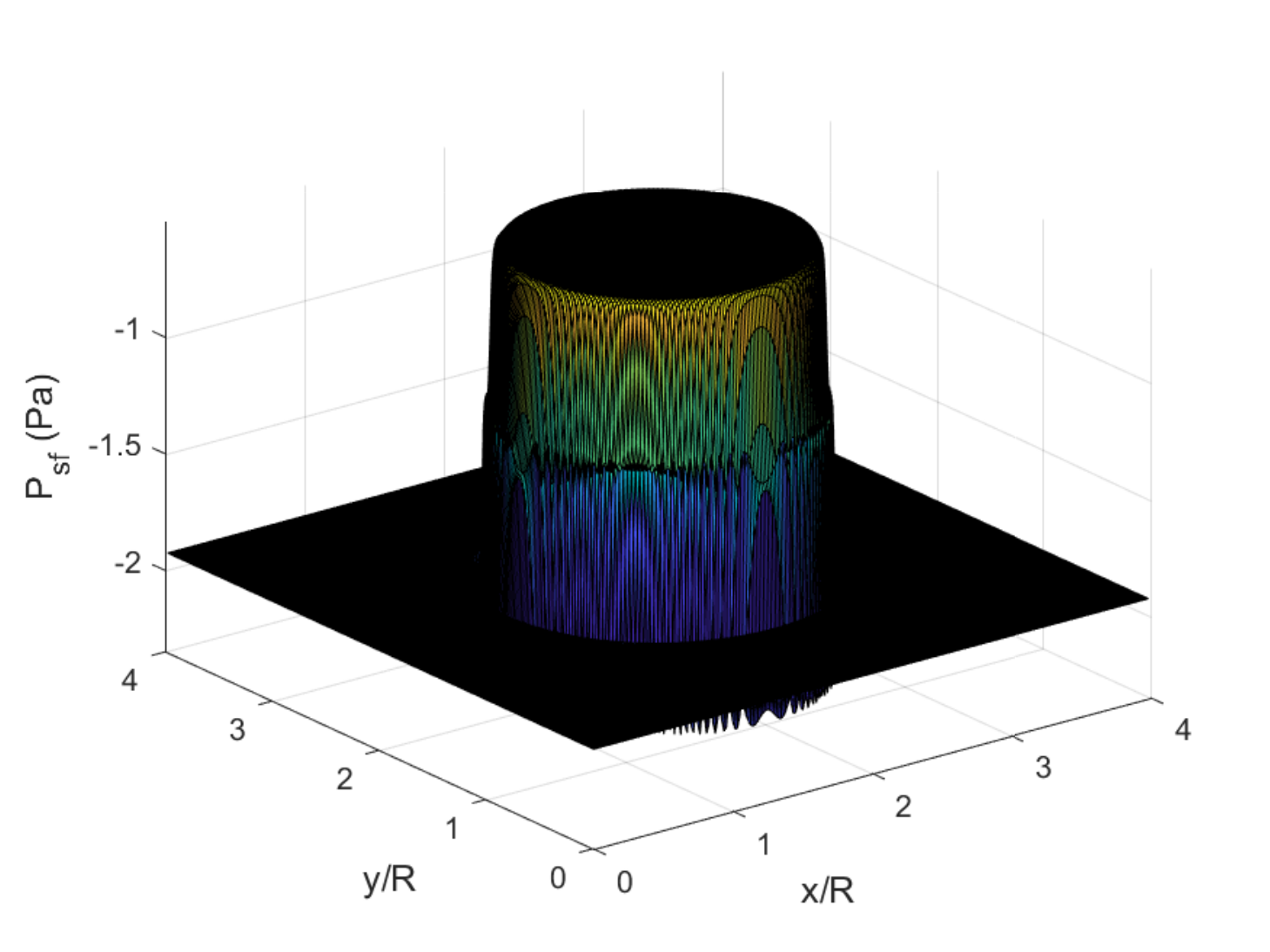}}~
\subfloat[]{\includegraphics[width=0.25\textwidth]{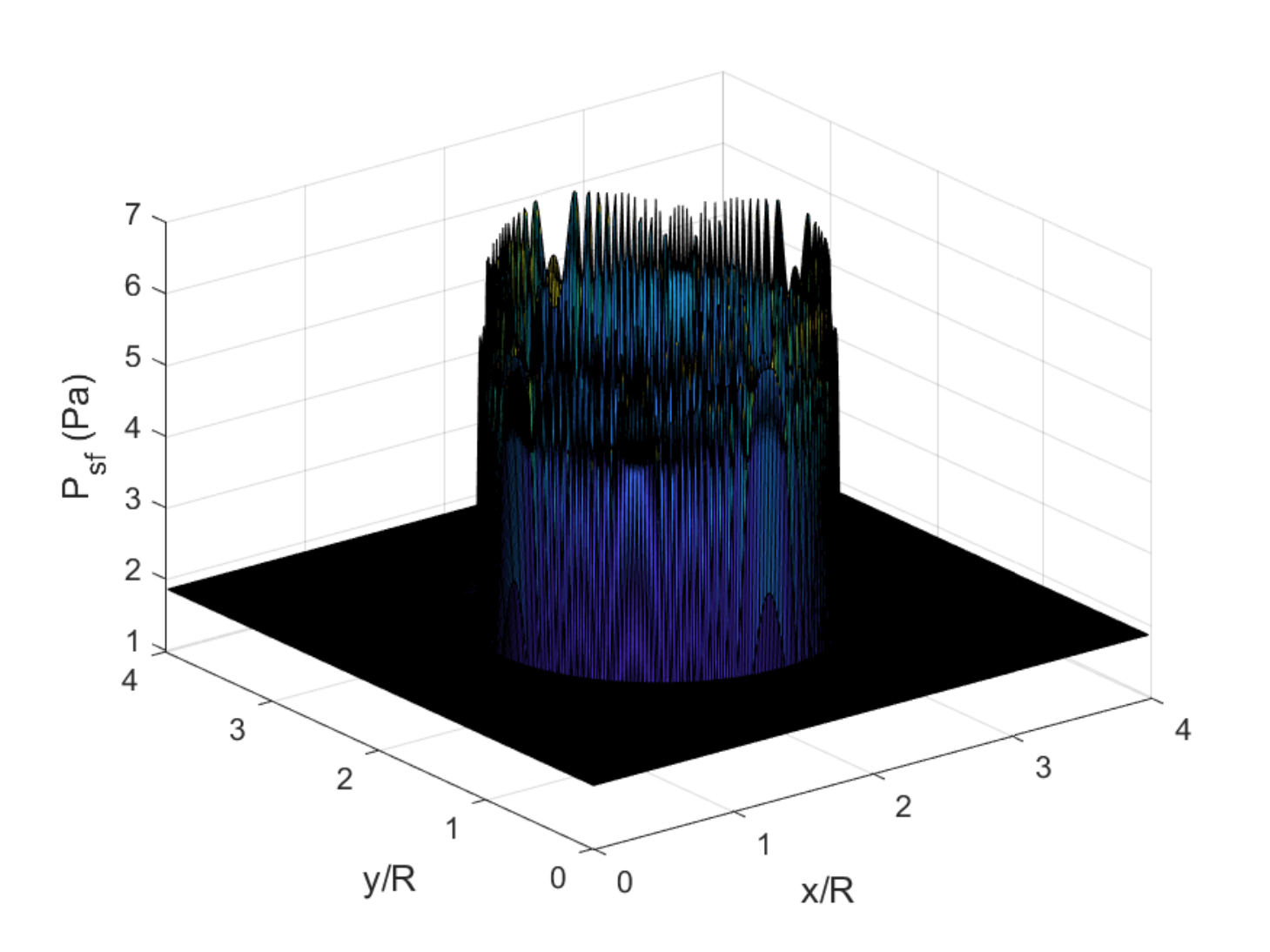}}~
\subfloat[]{\includegraphics[width=0.25\textwidth]{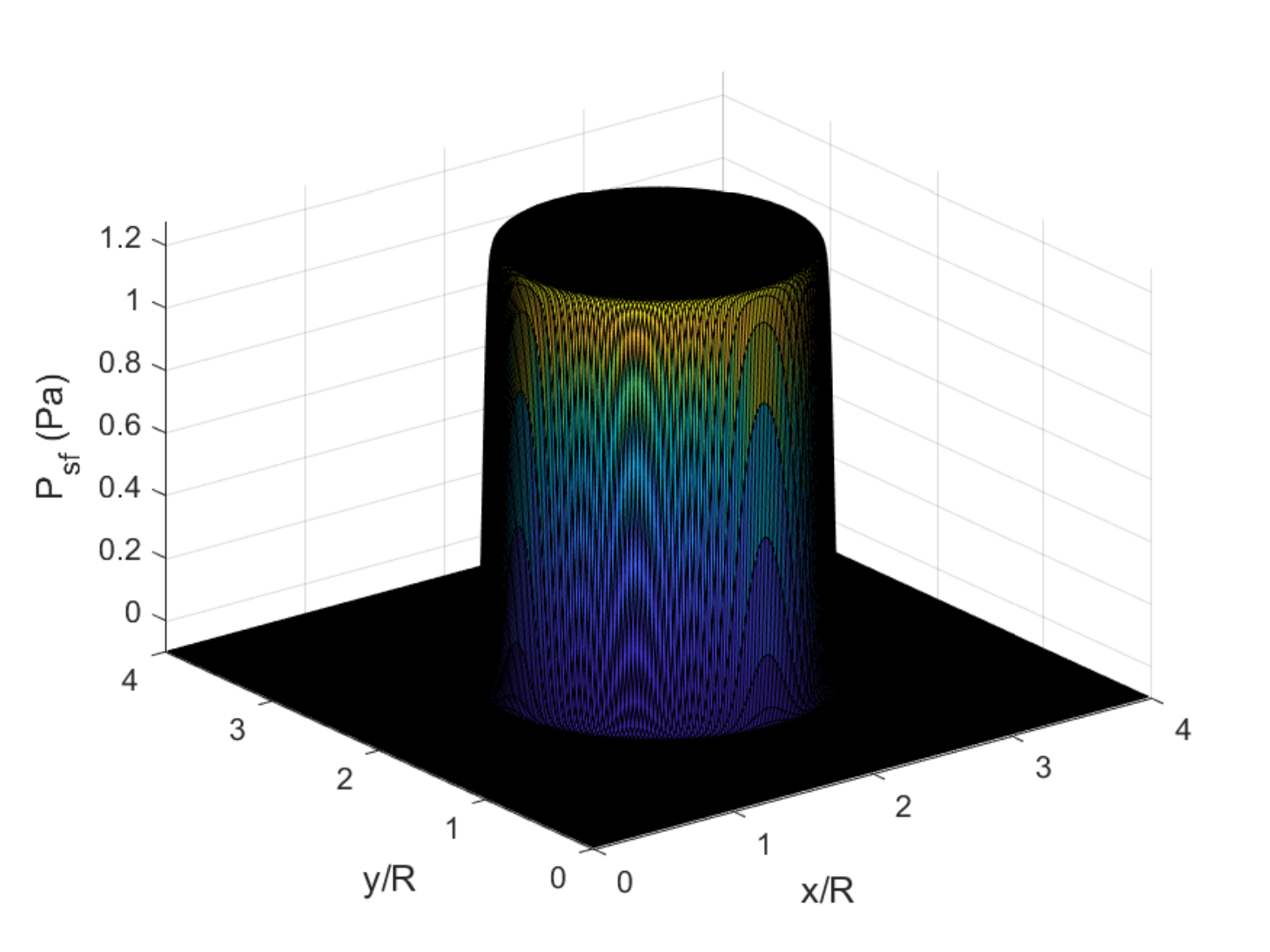}}~
\subfloat[]{\includegraphics[width=0.25\textwidth]{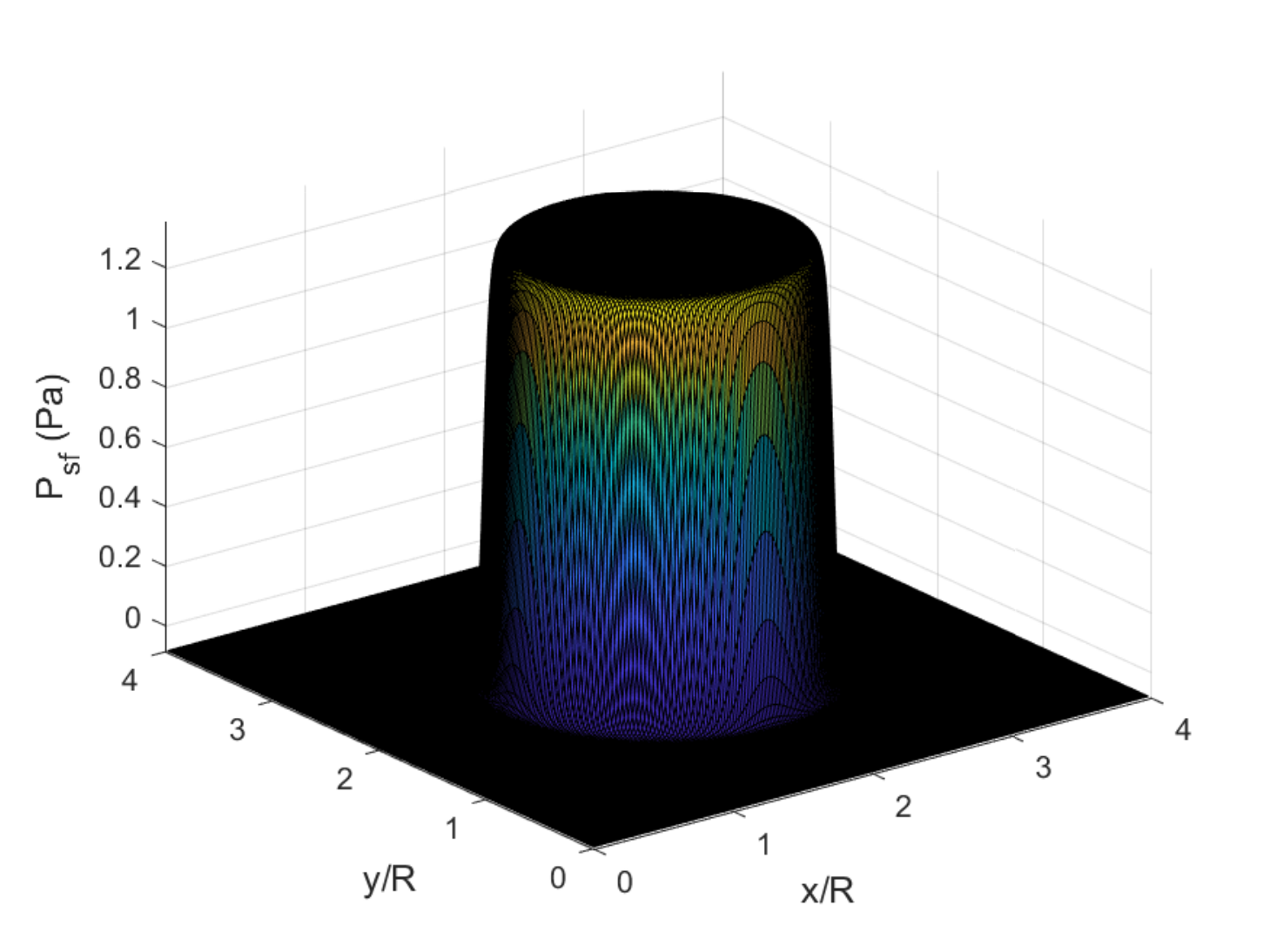}} ~
\caption{The pressure field on $240\times 240$ grid for (a) $\bm F_{stf-1}$,(b) $\bm F_{stf-2}$,(c)$\bm F_{cpf-1}$,(d)$\bm F_{cpf-2}$,(e)$\bm F_{pf-1}$,(f)$\bm F_{pf-2}$,(g)$\bm F_{csf-1}$ and (h) $\bm F_{csf-2}$.}
\label{test1-P}
\end{figure}

The pressure field  $p_h$  on $240\times 240$ meshes   is  presented in  Figure~\ref{test1-P}.
It can be seen that the pressure inside the droplet is higher than that in the surrounding fluid. However,  $\bm F_{stf-2}$, $\bm F_{csf-1}$ and $\bm F_{csf-2}$ give  smooth pressure field across the interface while the others give  obvious oscillation near the interface.

The spurious velocity for each formulation is also examined. The  magnitudes of spurious velocities denoted by  $\text{Ca}=\mu_1 |\bm {u}_{max}|/\sigma $ are presented in Table~\ref{tab:maximum_velocity}. It can be seen  that
both $\bm F_{cpf-1}$ and $\bm F_{cpf-2}$ give small spurious velocities while the others  give  larger ones.
\begin{table}[!htb]
\centering
\caption{The maximum spurious velocities of stationary droplet at equilibrium. }\label{tab:maximum_velocity}
\setlength{\tabcolsep}{0.92mm}{%
\begin{tabular}{ccccccccc}
\hline
$\bm F_{sf}$ & \multicolumn{2}{c}{$60 \times 60$} &    & \multicolumn{2}{c}{$120 \times 120$} &       &\multicolumn{2}{c}{$240 \times 240$ }    \\
\cline{2-3}            \cline{5-6}                \cline{8-9}
 & $|\bm u|_{max} $     & $\text{Ca}$ &  & $|\bm u|_{max} $      &$\text{Ca}$        &     &$|\bm u|_{max} $   &$\text{Ca}$  \\
\hline
$\bm F_{stf-1}$ & $2.327\times 10^{-4}$  & $6.518\times 10^{-4}$ &      &$2.520\times 10^{-4}$ & $7.058\times 10^{-4}$&
                 &$2.528\times 10^{-4}$  & $7.081\times 10^{-4}$       \\

$\bm F_{stf-2}$ &$1.716\times 10^{-4}$  &$4.808\times 10^{-4}$  &      &$1.969\times 10^{-4}$  &$5.515\times 10^{-4}$  &
                &$2.204\times 10^{-4}$  &$6.173\times 10^{-4}$    \\

$\bm F_{cpf-1} $&$ 4.180\times 10^{-5}$  & $1.171\times 10^{-4}$  &    & $9.065\times 10^{-6}$  & $2.539\times 10^{-5}$  &
                &$5.861\times 10^{-7}$  & $1.642\times 10^{-6}$   \\

$\bm F_{cpf-2} $&$4.459\times 10^{-5}$  &$ 1.249\times 10^{-4}$  &      & $1.242\times 10^{-5}$ & $3.480\times 10^{-5}$  &
                & $2.170\times 10^{-6}$  &$6.078\times 10^{-6}$     \\

$\bm F_{pf-1}$& $2.237\times 10^{-4}$  & $6.265\times 10^{-4}$ &    & $3.217\times 10^{-4}$ & $9.012\times 10^{-4}$ &
              & $3.611\times 10^{-4}$ & $1.011\times 10^{-3}$   \\

$\bm F_{pf-2}$& $6.940\times 10^{-4}$  & $1.944\times 10^{-3}$ &    & $8.562\times 10^{-4}$ & $2.398\times 10^{-3}$ &
              & $9.442\times 10^{-4}$ & $2.645\times 10^{-3}$   \\

$\bm F_{csf-1}$& $2.765\times 10^{-5}$  & $7.745\times 10^{-5}$ &   & $1.184\times 10^{-5}$ & $3.317\times 10^{-5}$ &
               & $9.756\times 10^{-6}$ & $2.733\times 10^{-5}$   \\

$\bm F_{csf-2}$& $1.528\times 10^{-4}$  & $4.279\times 10^{-4}$ &    & $1.273\times 10^{-4}$ & $3.566\times 10^{-4}$ &
               & $6.957\times 10^{-5}$ & $1.949\times 10^{-4}$   \\
\hline
\end{tabular}}
\end{table}

\begin{figure}[!htb]
\centering
\subfloat[]{\includegraphics[width=0.25\textwidth]{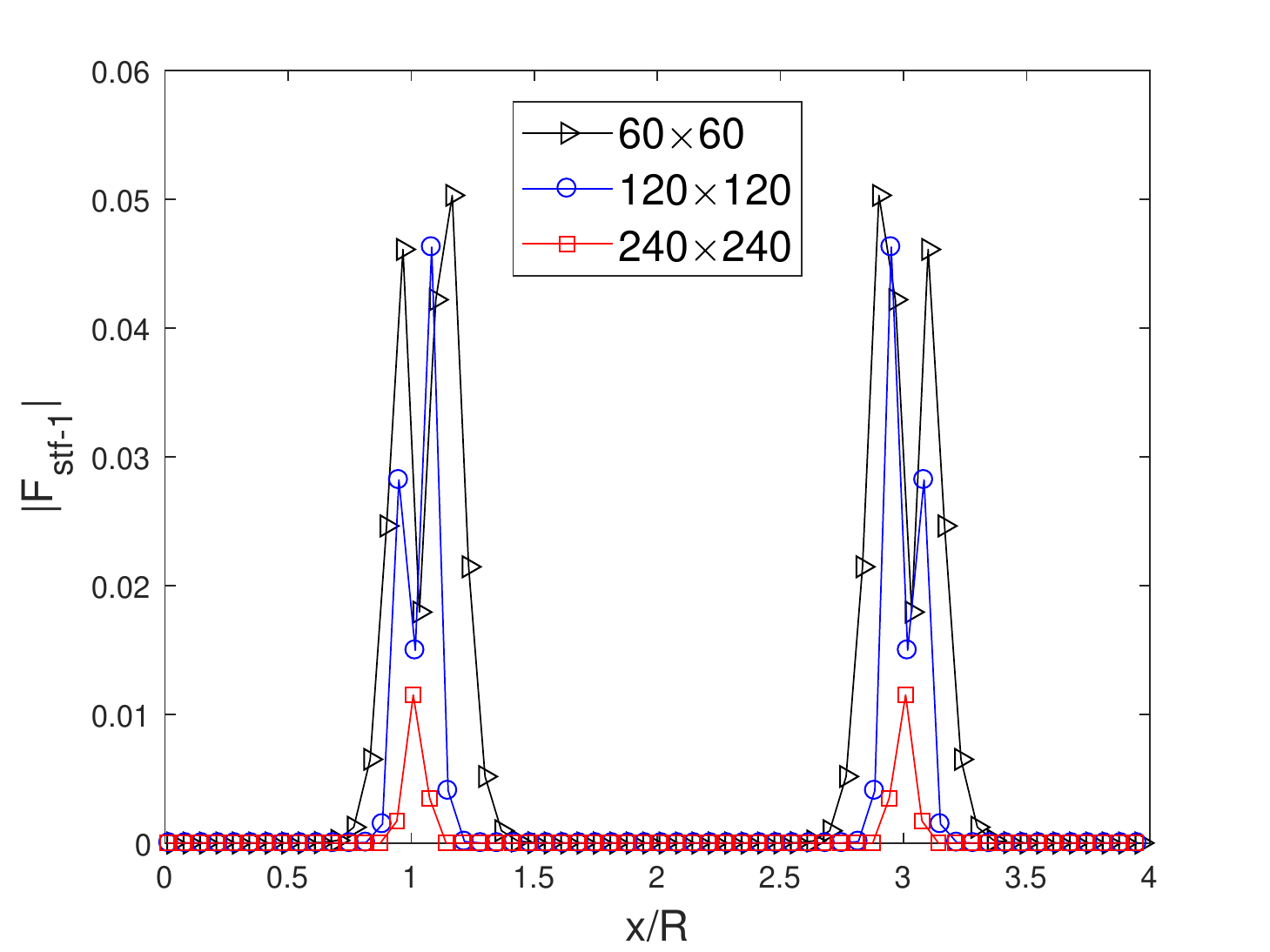}}~
\subfloat[]{\includegraphics[width=0.25\textwidth]{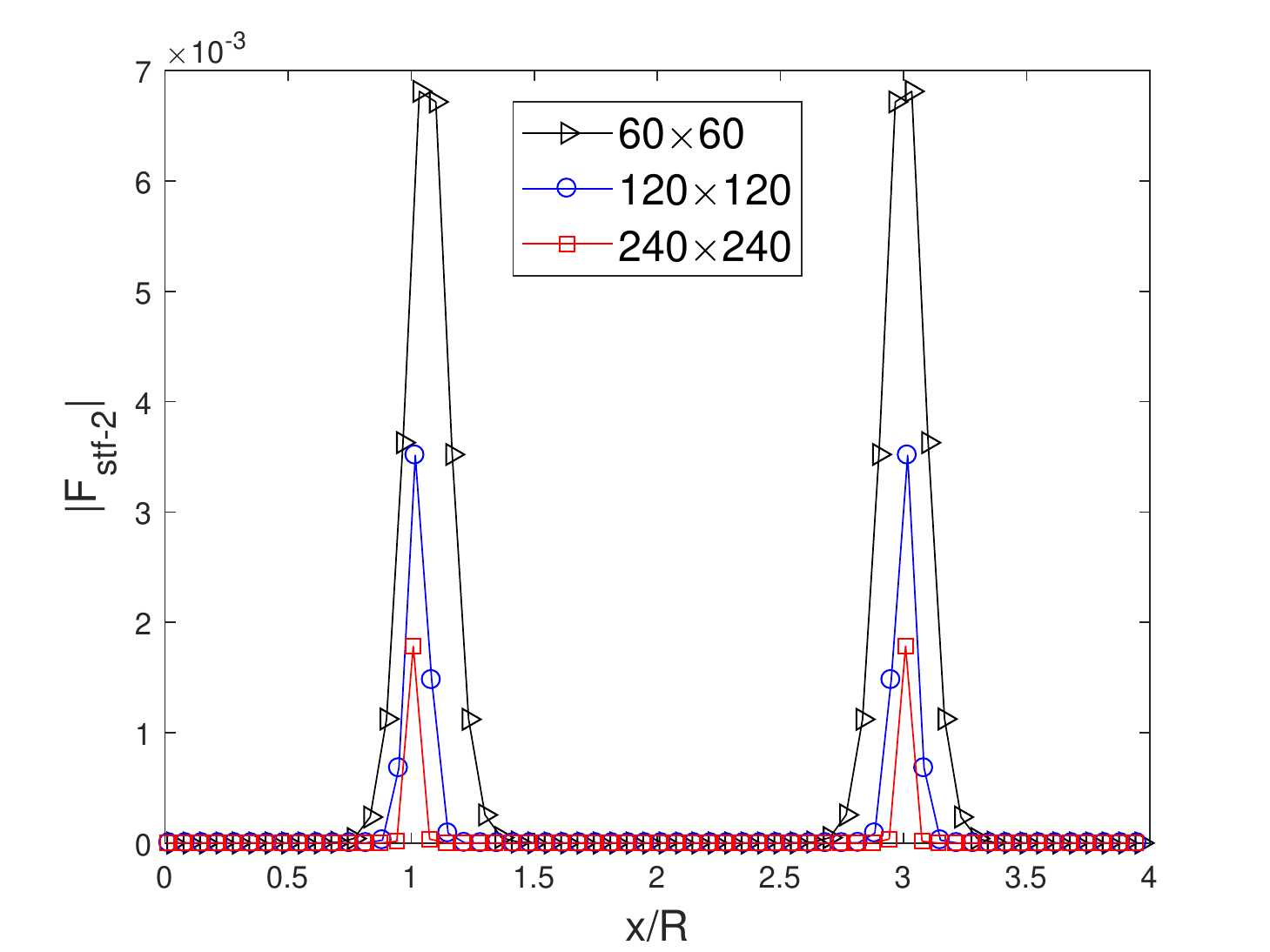}}~
\subfloat[]{\includegraphics[width=0.25\textwidth]{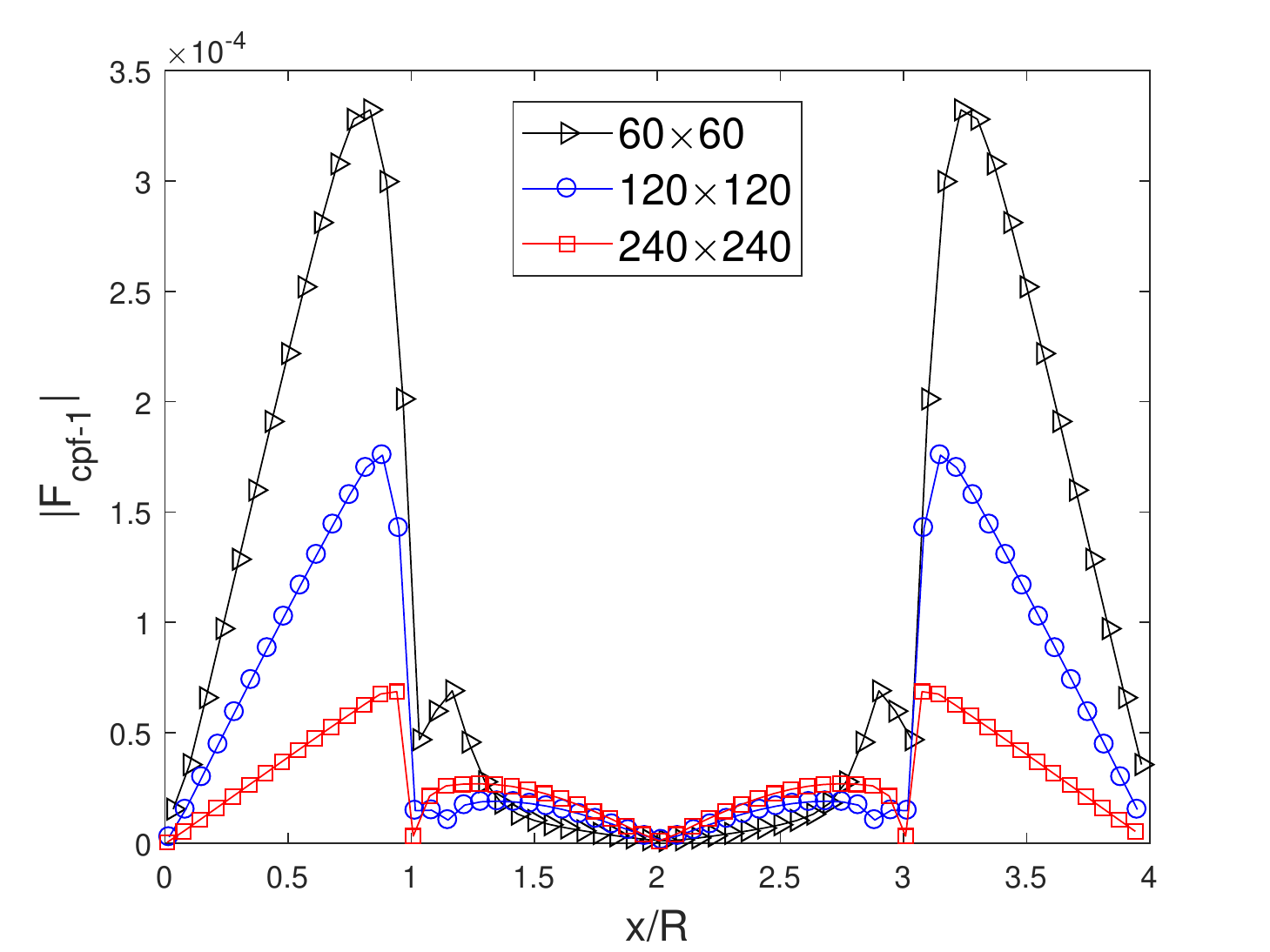}}~
\subfloat[]{\includegraphics[width=0.25\textwidth]{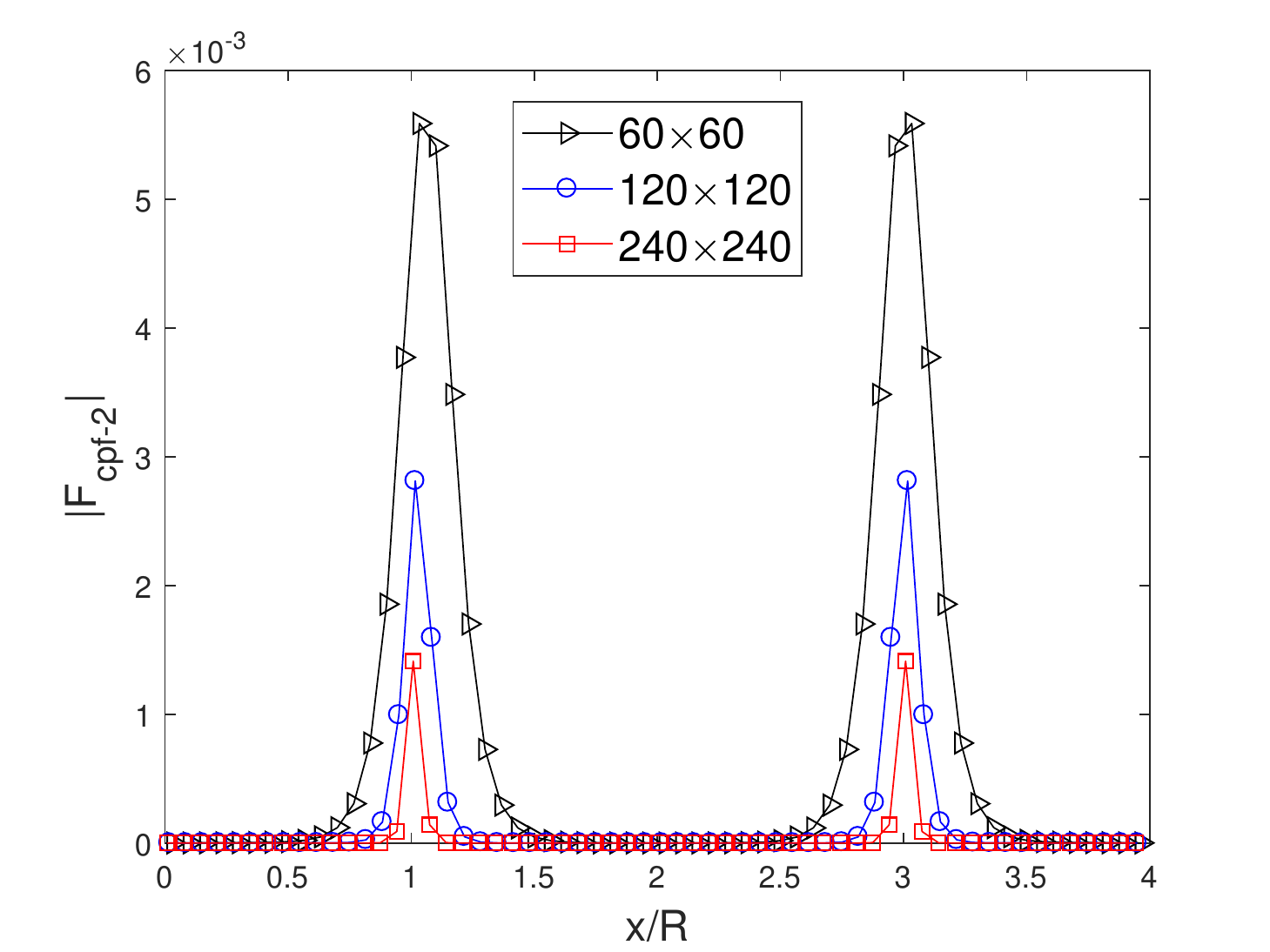}} \\
\subfloat[]{\includegraphics[width=0.25\textwidth]{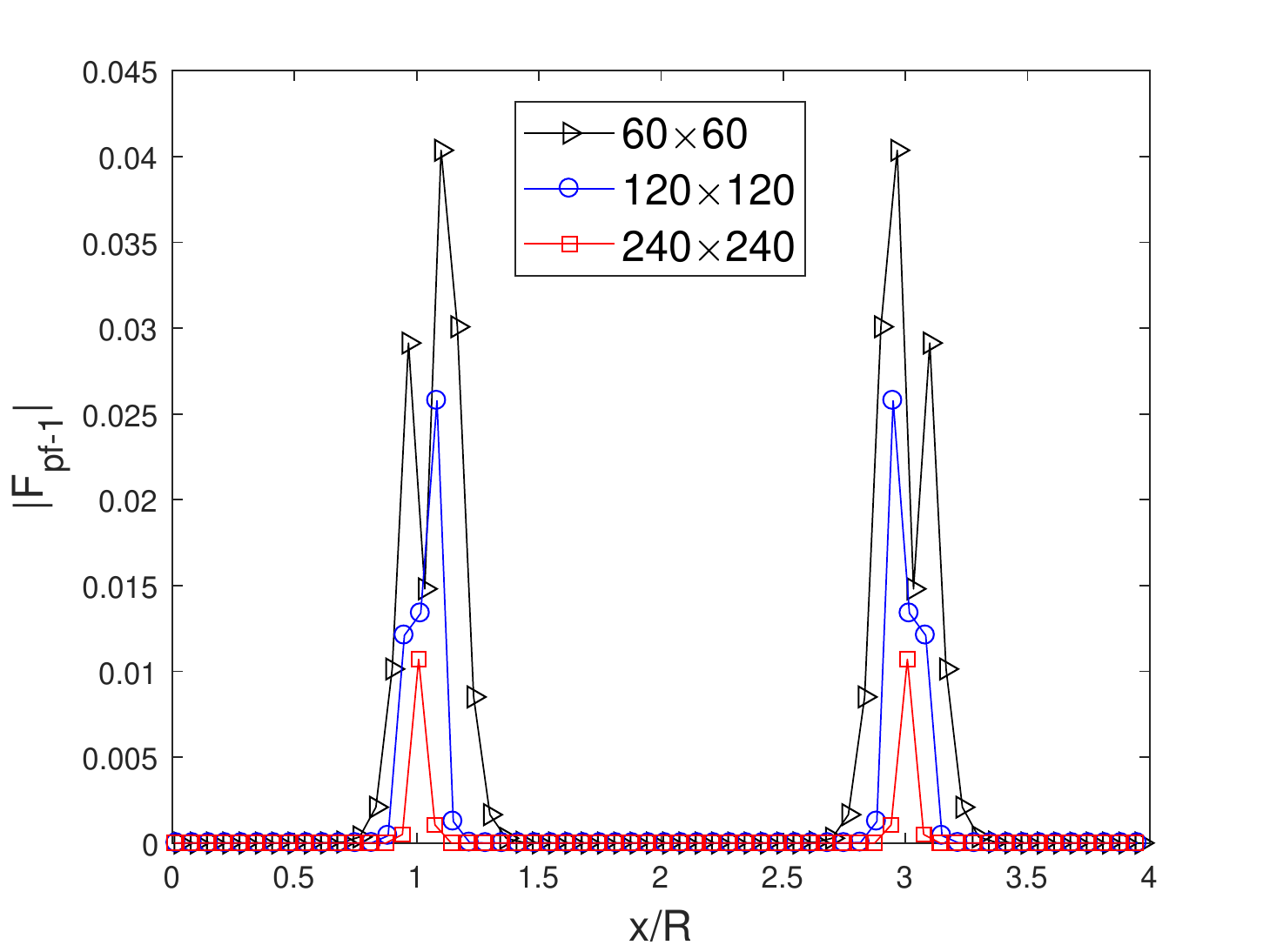}}~
\subfloat[]{\includegraphics[width=0.25\textwidth]{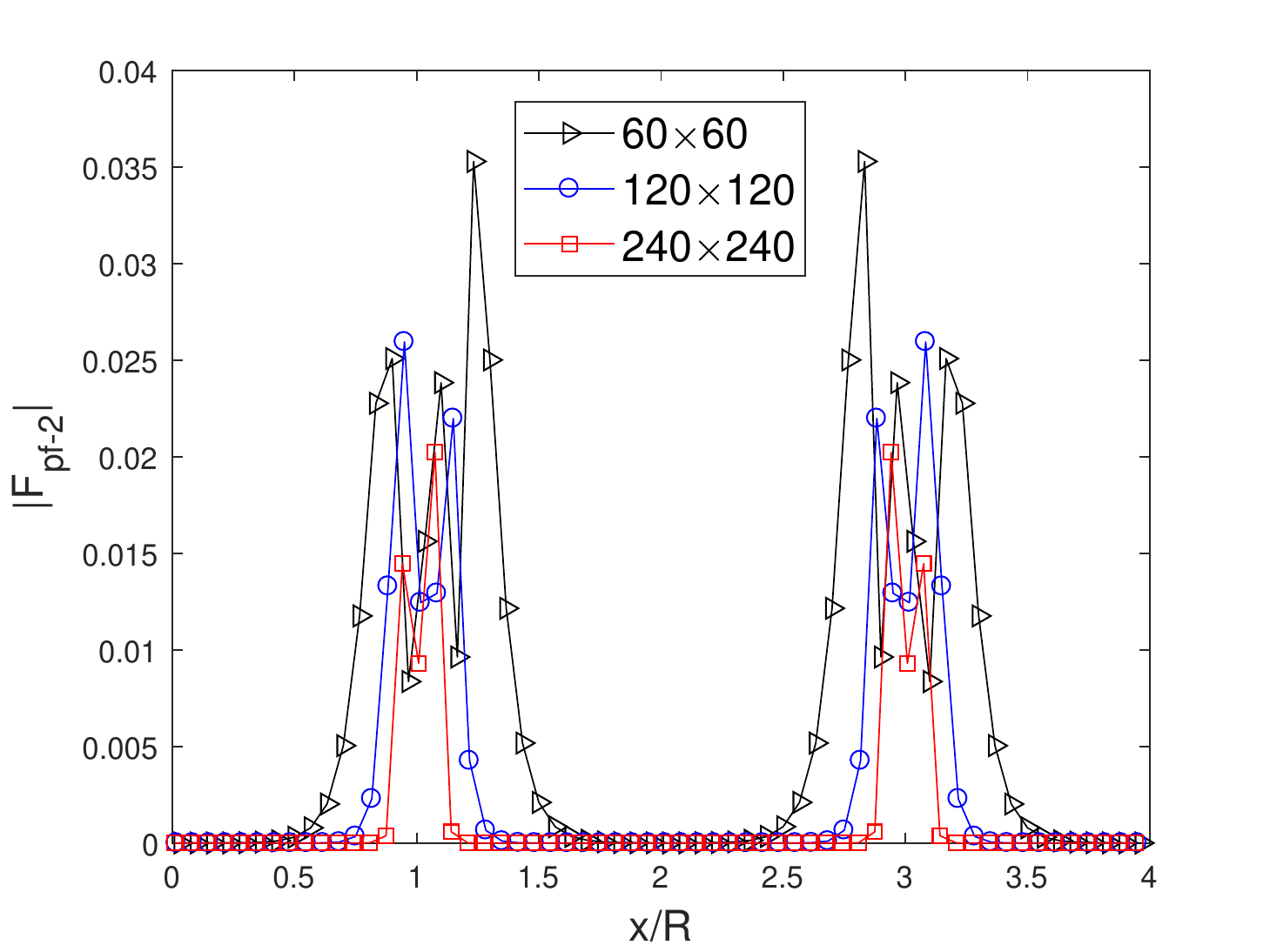}}~
\subfloat[]{\includegraphics[width=0.25\textwidth]{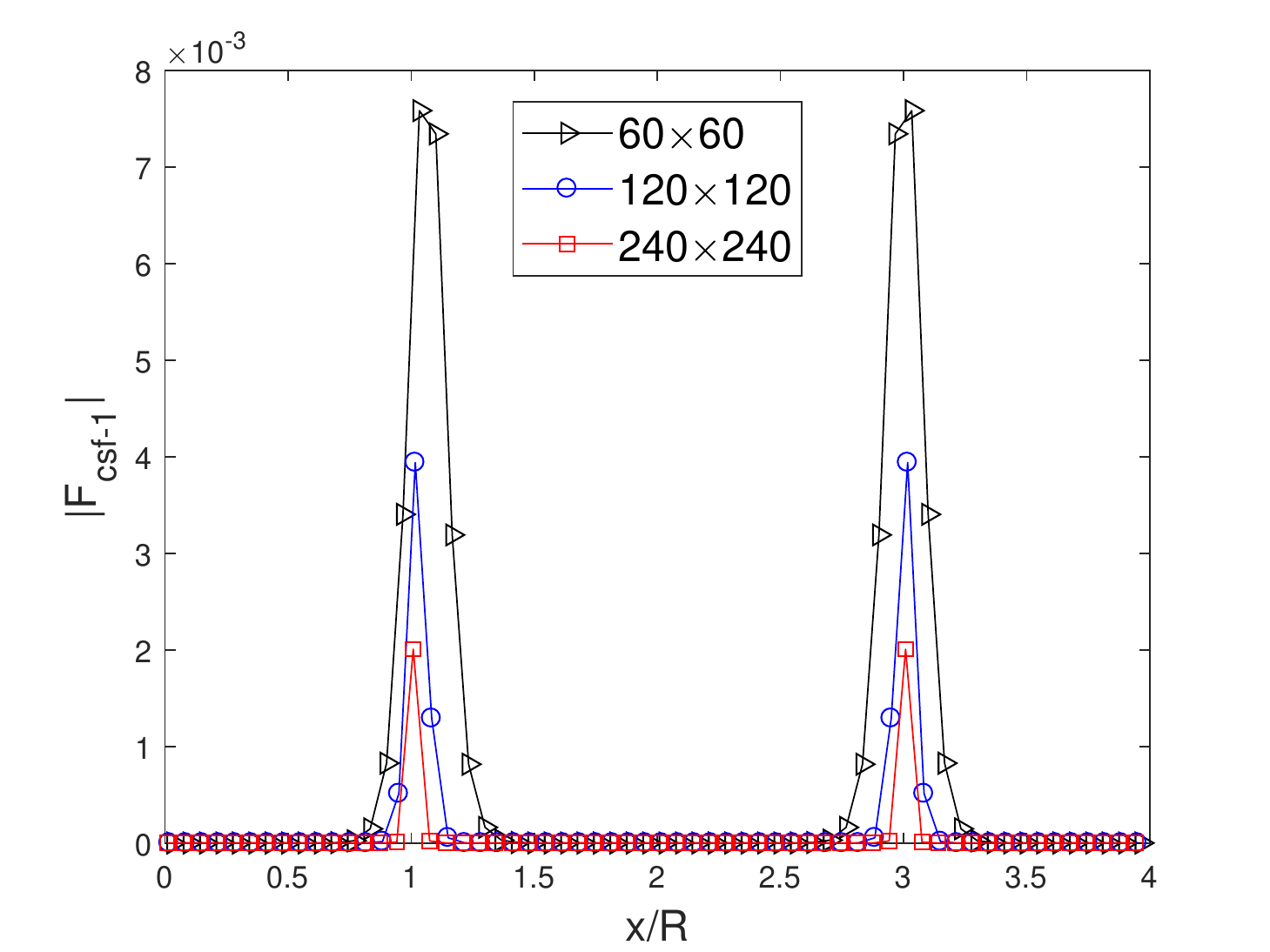}}~
\subfloat[]{\includegraphics[width=0.25\textwidth]{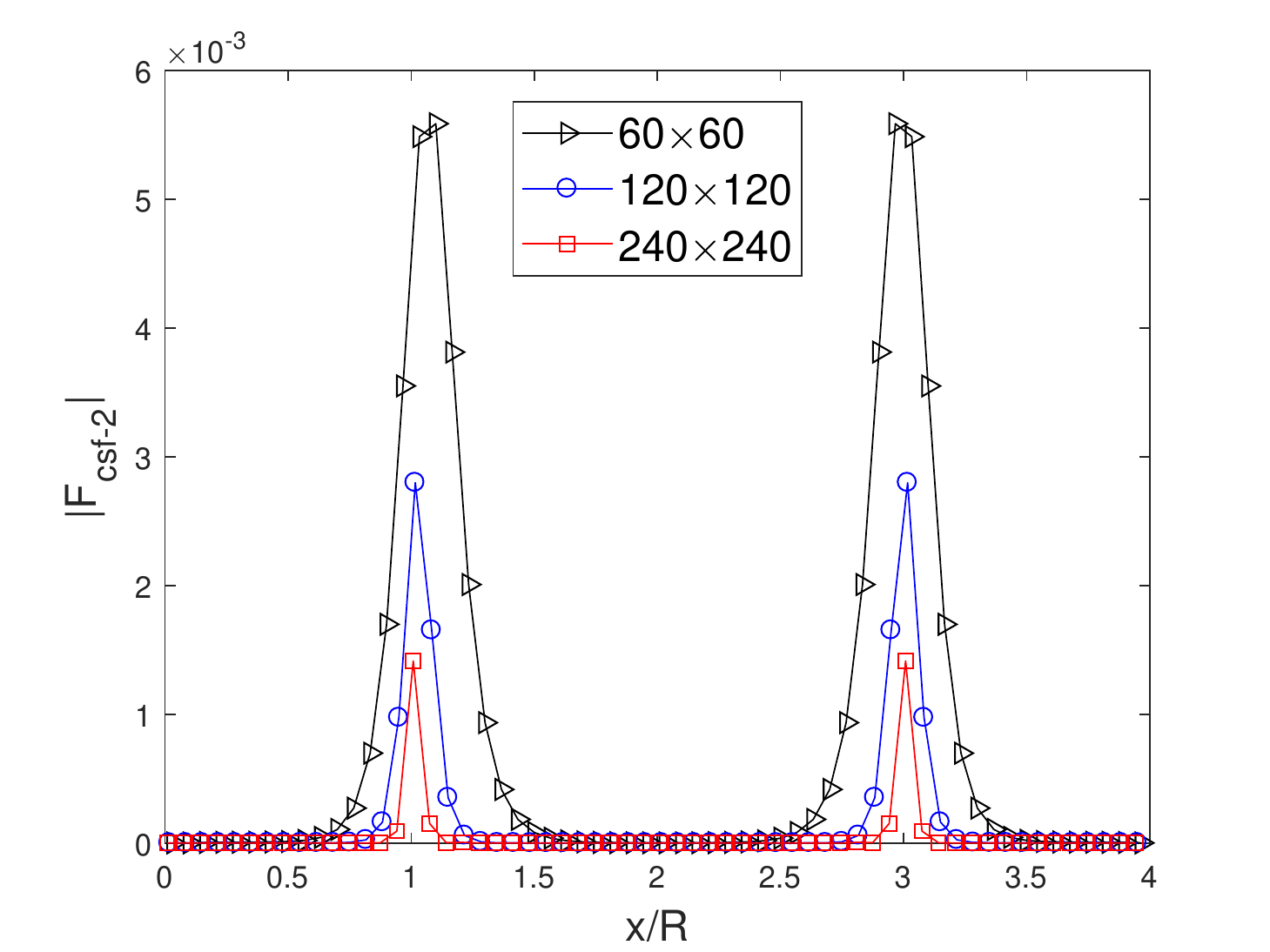}}~
\caption{ The  interfacial force profiles along the midline of the drop for (a) $\bm F_{stf-1}$,(b) $\bm F_{stf-2}$,(c)$\bm F_{cpf-1}$,(d)$\bm F_{cpf-2}$,(e)$\bm F_{pf-1}$,(f)$\bm F_{pf-2}$,(g)$\bm F_{csf-1}$ and (h) $\bm F_{csf-2}$.}
\label{test1-surfaceforce}
\end{figure}

Finally,  the absolute values of  interfacial force across the drop center with different formulations are compared. The results are shown in Figure~\ref{test1-surfaceforce}.
 Theoretically, the interfacial force should be  zero everywhere except in the vicinity of the interface.
However, the absolute values of $\bm F_{cpf-1}$ have non-zero values in the whole domain. This may cause earlier motion of the interface although the amplitude of interfacial force is small. Since the interface width is fixed, the  range of nonzero interfacial force  decreases with increasing grid resolution.
In addition, based on the definition of each formulation and the equilibrium state,  the profile of the interfacial force should be symmetric with respect to the phase interface ($\phi=0$). However,  the interfacial force profiles of  $\bm F_{stf-2},\bm F_{cpf-2}, \bm F_{csf-1}, \bm F_{csf-2}$  are symmetric while the profiles of the others are asymmetrical.

\subsection{Droplets merging}
\begin{figure}[htb]
\centering
\subfloat[]{\includegraphics[width=0.25\textwidth]{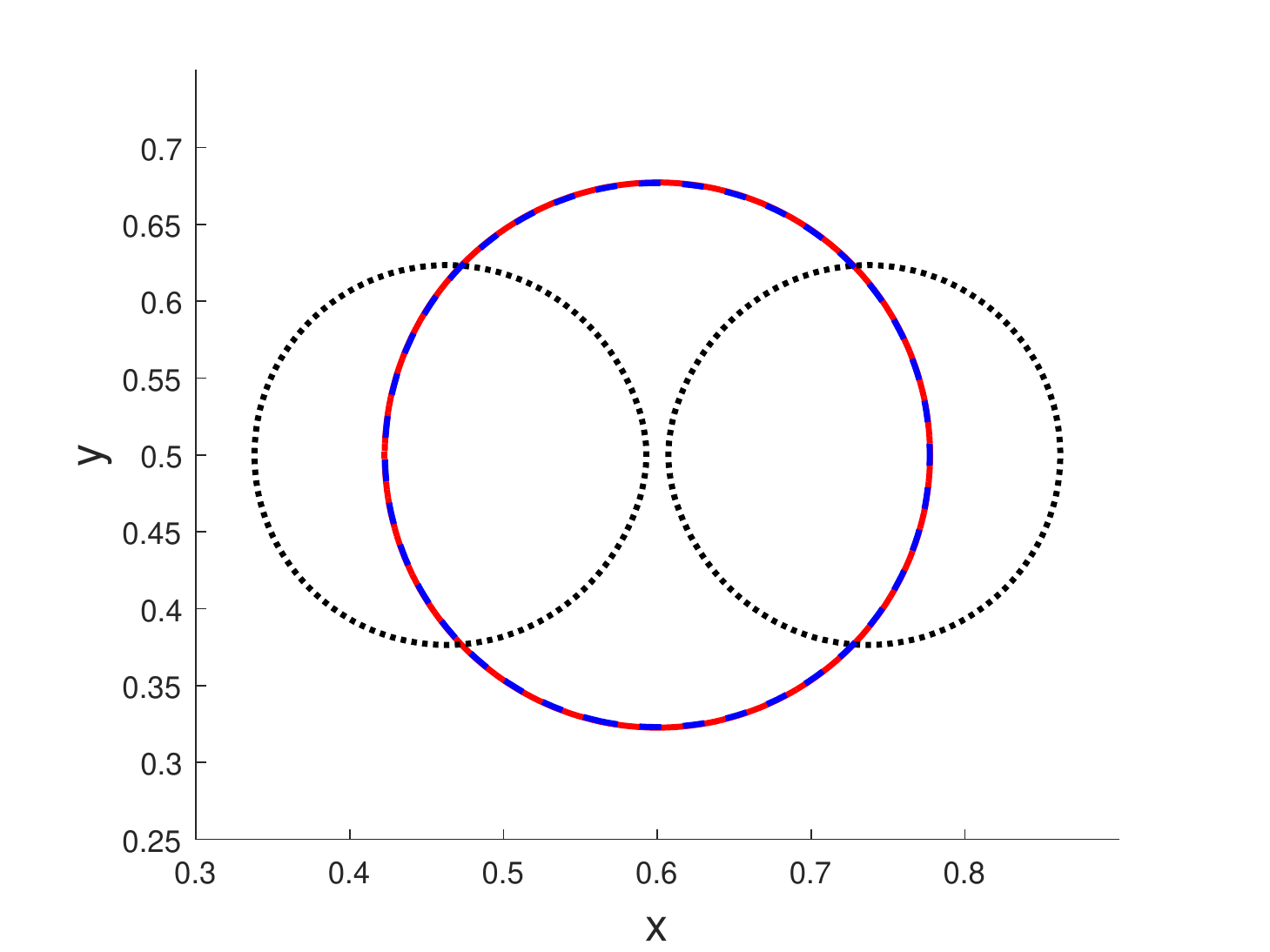}}~
\subfloat[]{\includegraphics[width=0.25\textwidth]{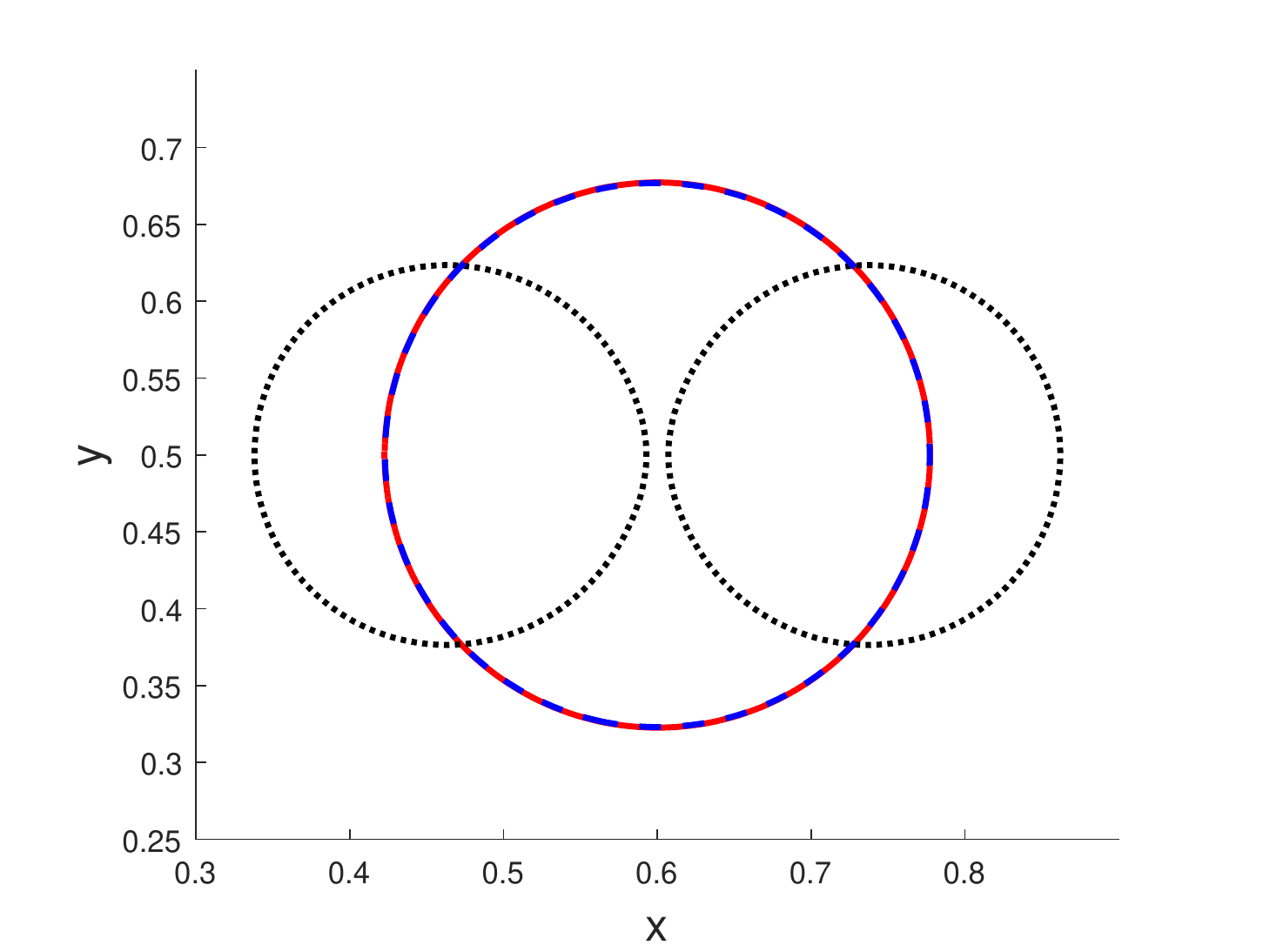}}~
\subfloat[]{\includegraphics[width=0.25\textwidth]{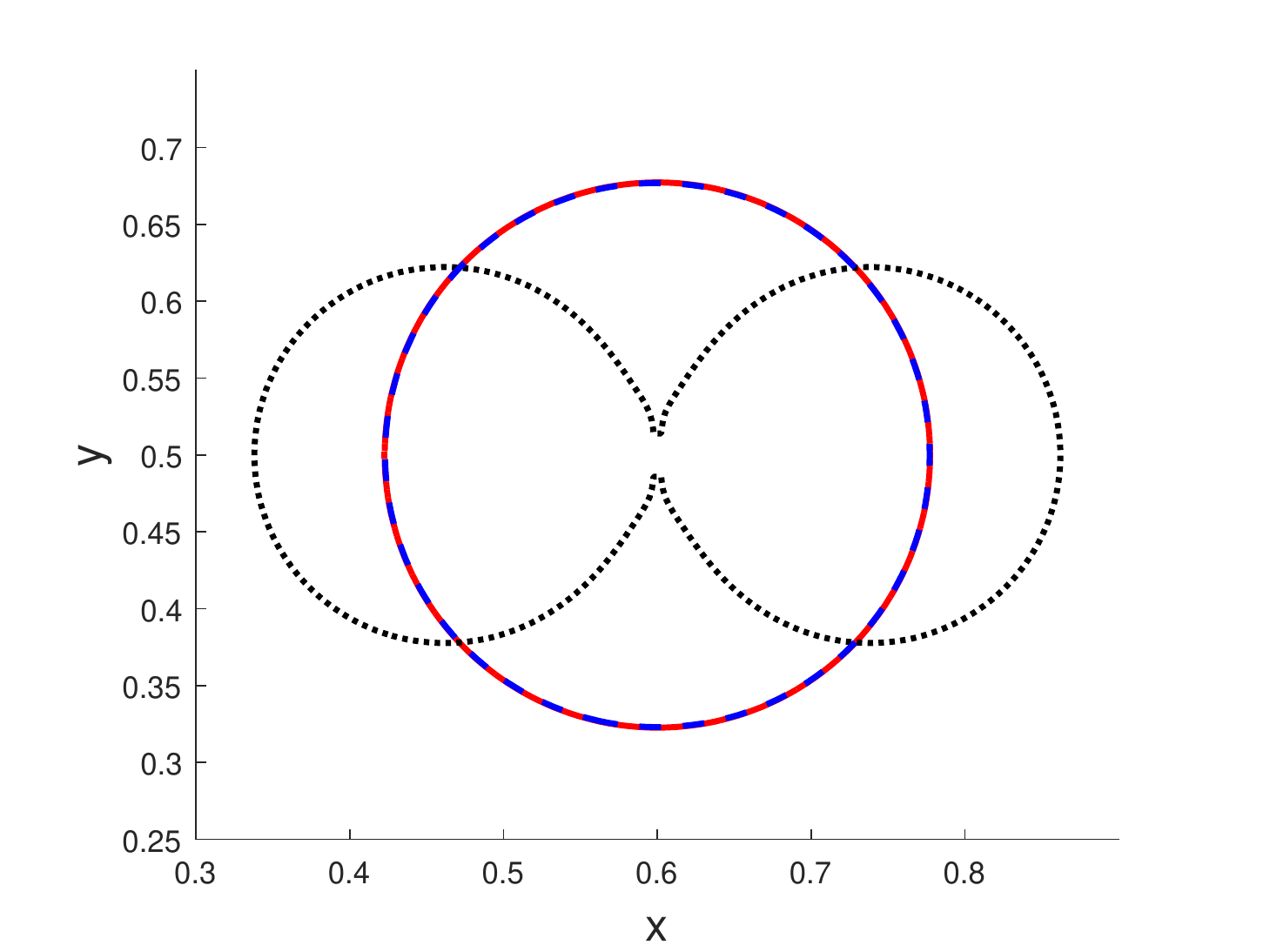}}~
\subfloat[]{\includegraphics[width=0.25\textwidth]{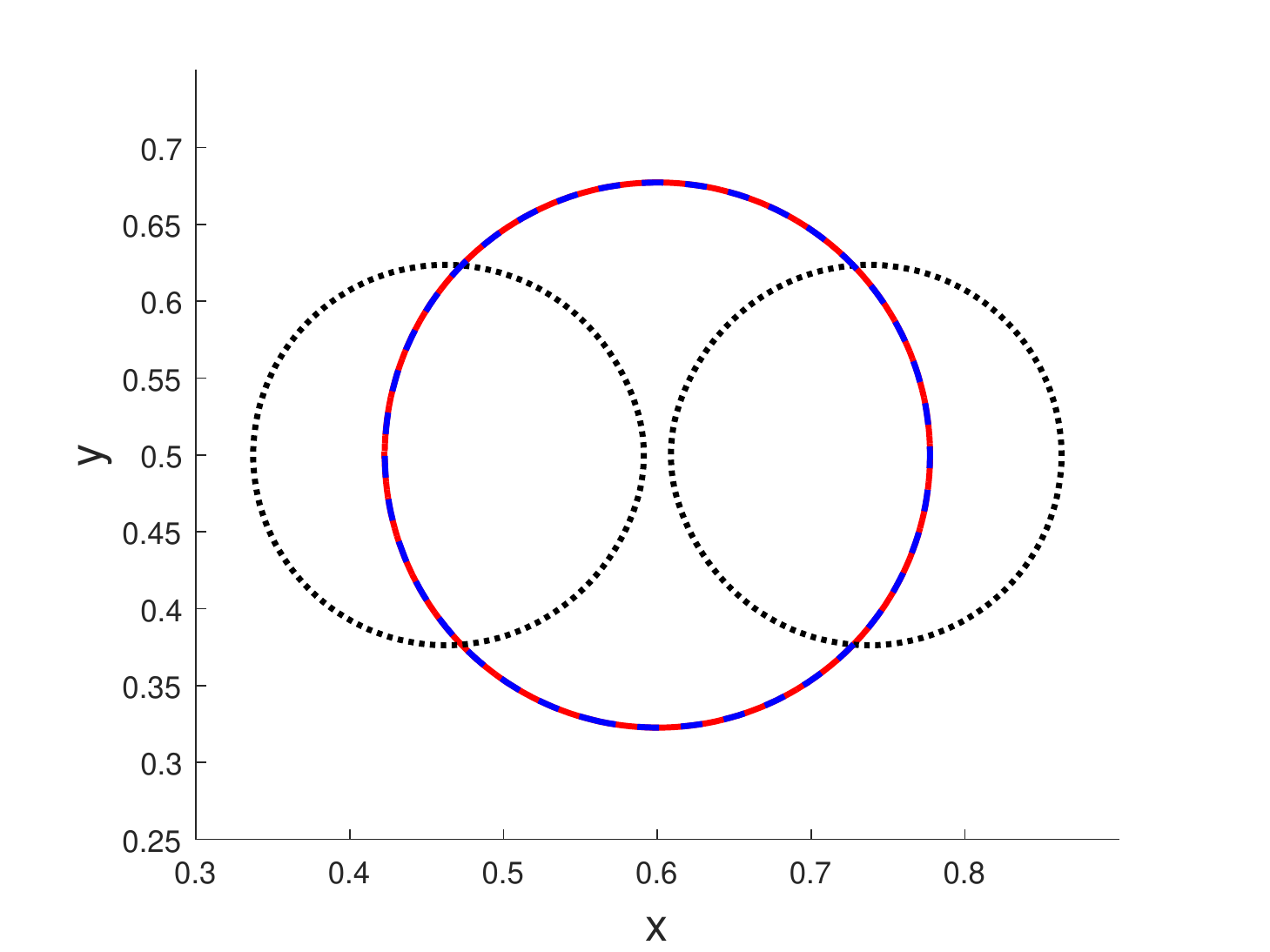}}\\
\subfloat[]{\includegraphics[width=0.25\textwidth]{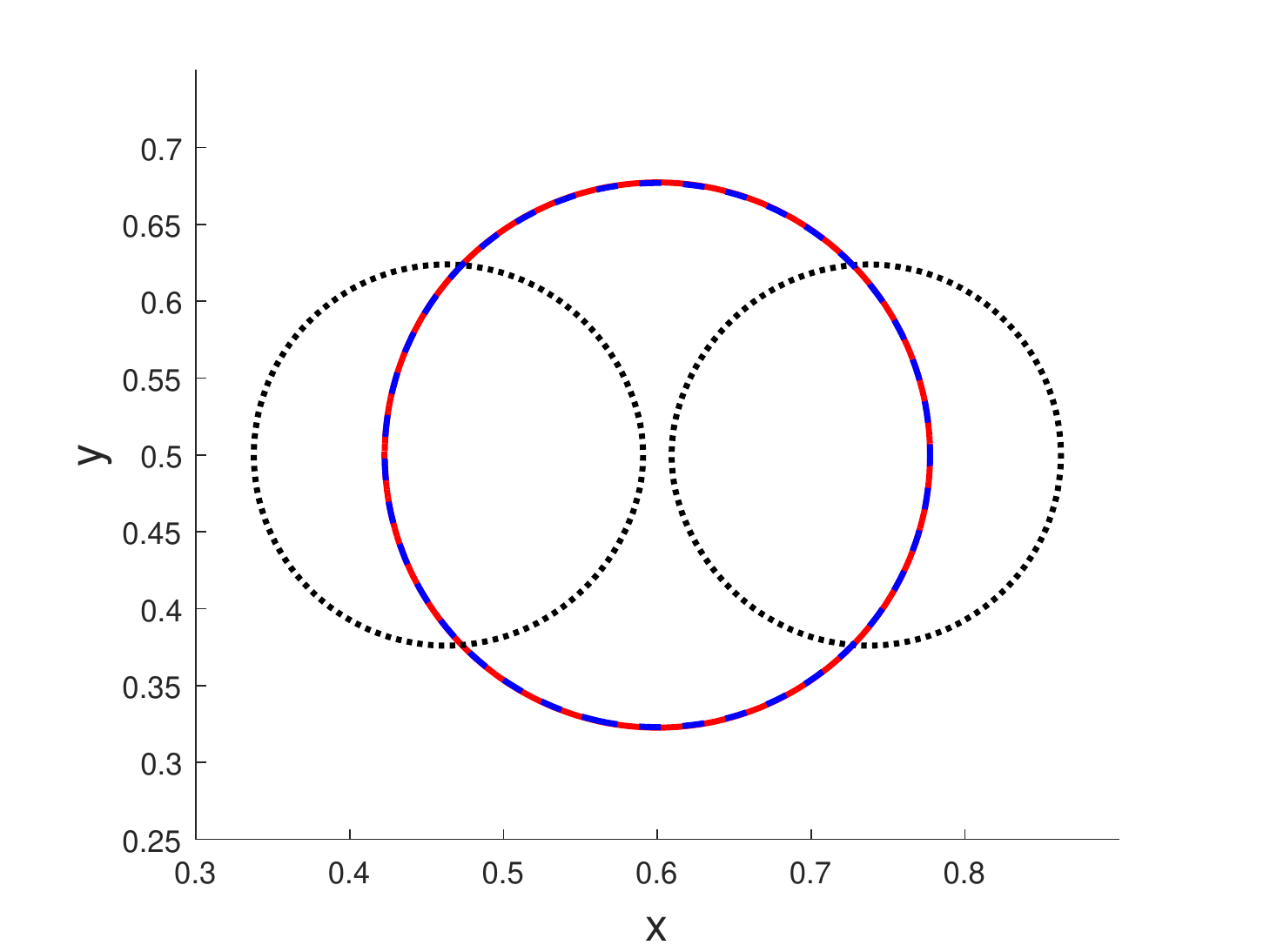}}~
\subfloat[]{\includegraphics[width=0.25\textwidth]{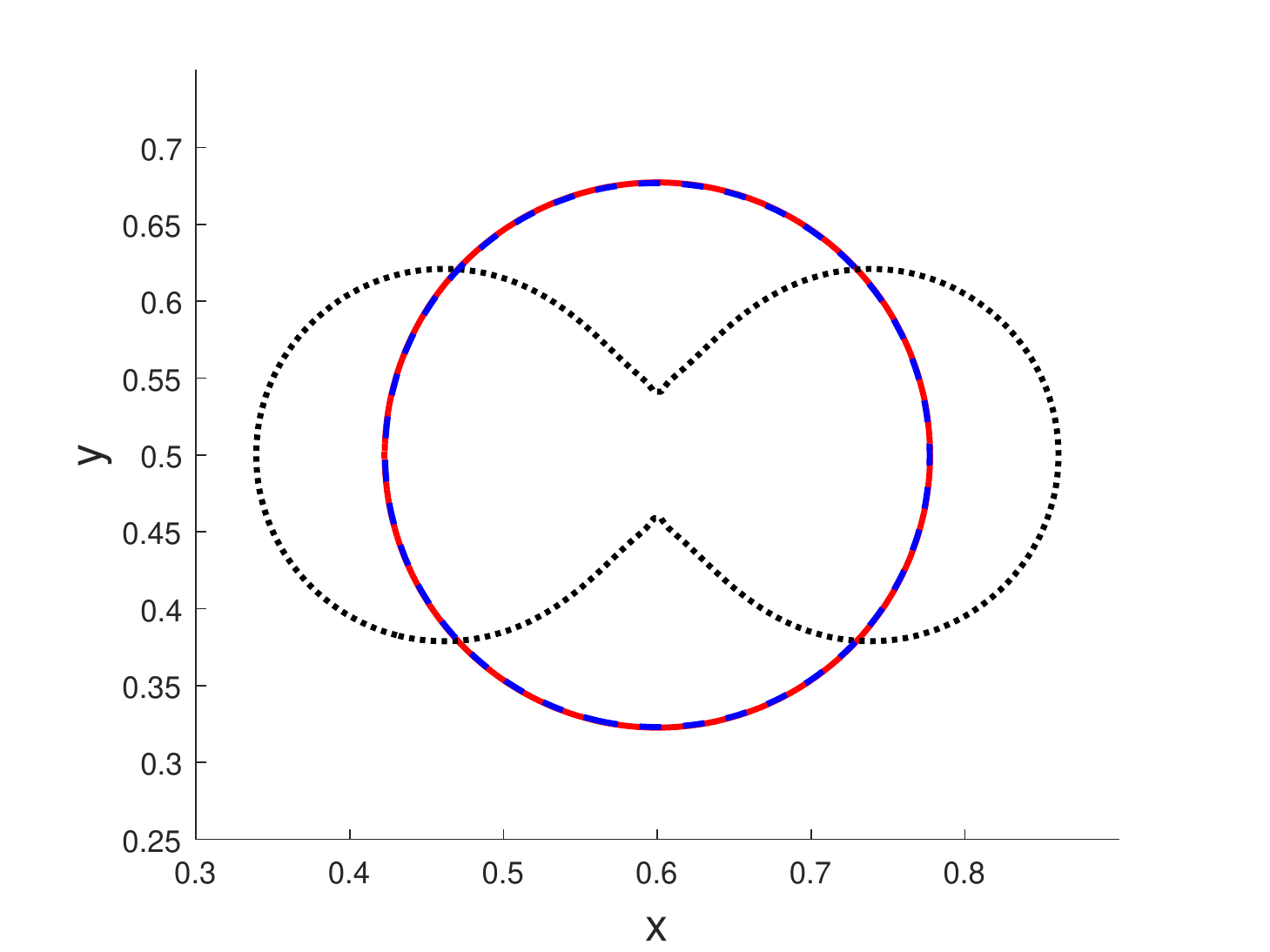}}~
\subfloat[]{\includegraphics[width=0.25\textwidth]{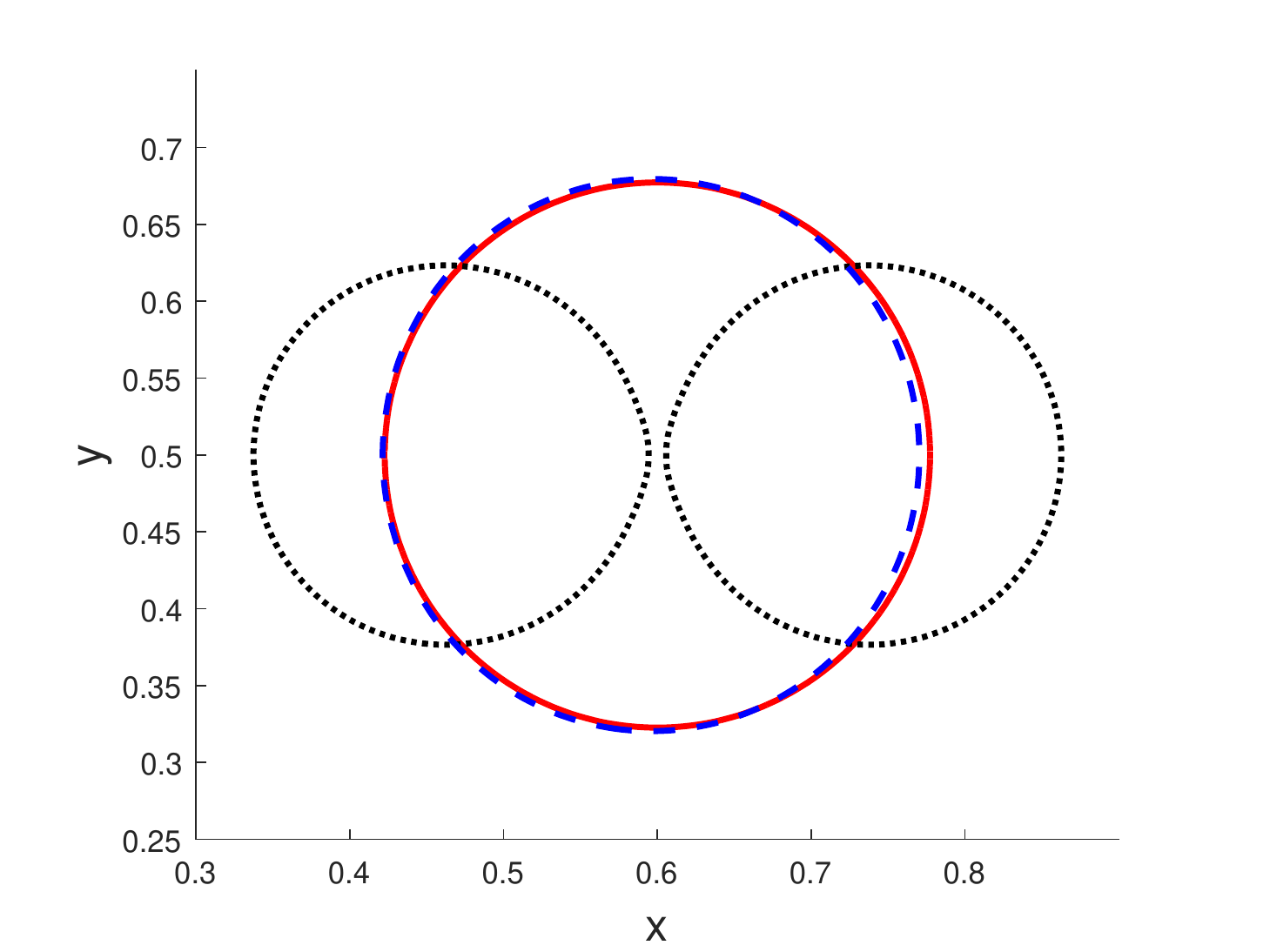}}~
\subfloat[]{\includegraphics[width=0.25\textwidth]{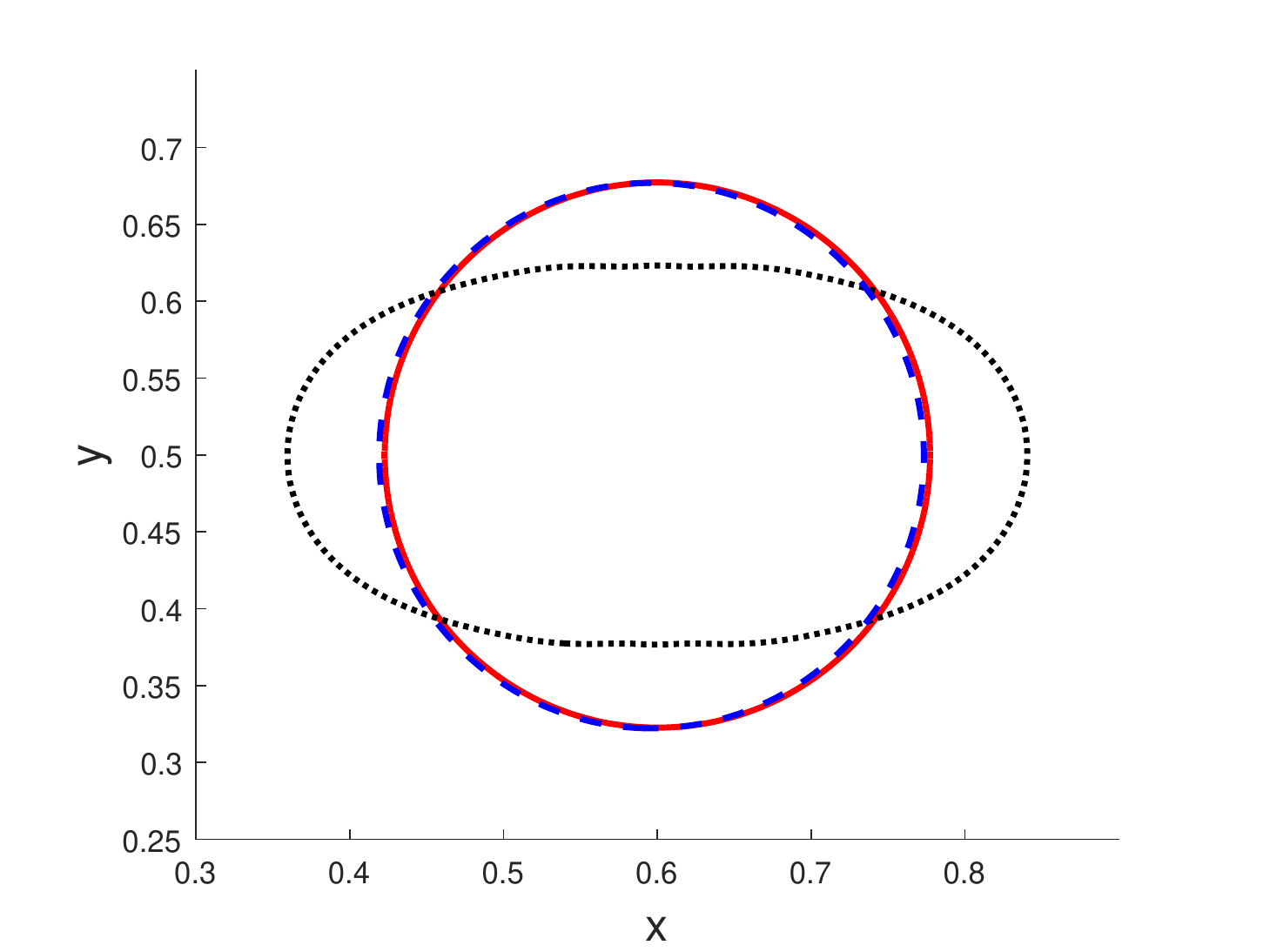}}
\caption{ Interfaces of  two droplets of equal sizes at $t=2T$(dotted line) and $30\text{T}$(dashed line): (a) $\bm F_{stf-1}$,(b) $\bm F_{stf-2}$,(c) $\bm F_{cpf-1}$,(d) $\bm F_{cpf-2}$,(e) $\bm F_{pf-1}$,(f) $\bm F_{pf-2}$,(g) $\bm F_{csf-1}$ and (h) $\bm F_{csf-2}$. Solid line represents the analytical solutions. }
\label{twobubbles-shapes}
\end{figure}

\begin{figure}[htb]
  \centering
\includegraphics[width=0.5\textwidth]{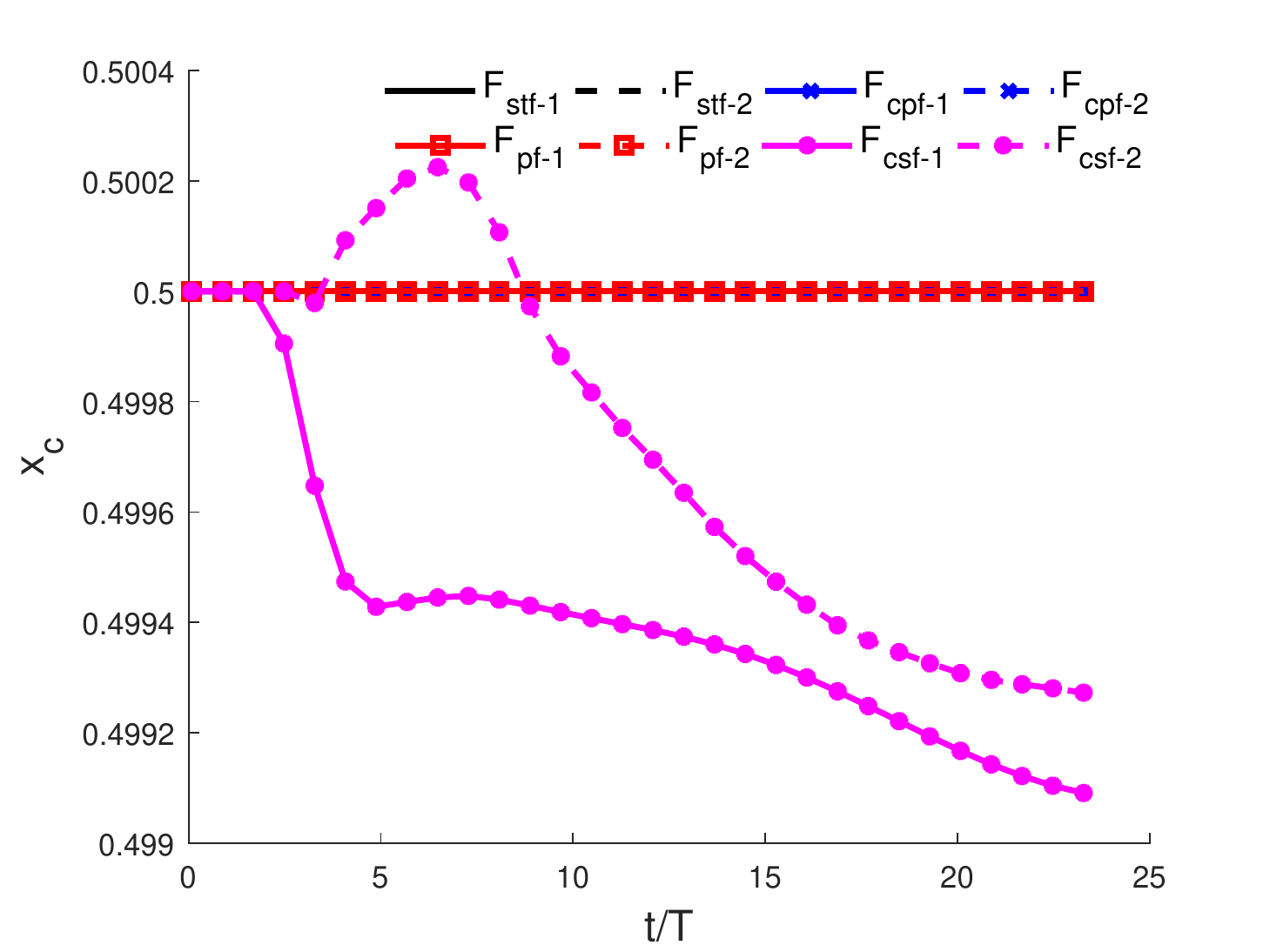}
  \caption{ Time history of  mass center $x_c$ for droplets of equal size.}\label{twobubbles-kineticenergy}
\end{figure}

To test the performance of the LBM with different interfacial force formulations, the merging of two droplets is simulated in this section. Initially, two circular droplets (density $\rho_d$ and viscosity $\nu_d$) are placed in another fluid (density $\rho_s$ and viscosity $\nu_s$) in a rectangle domain of $L_x\times L_y$. When the initial gap $d$ between two droplets  is smaller than $2W$, merging will occur due to the surface tension effect.
The order parameter is initialized to be
\begin{equation}\label{eq}
\phi(x,y)=1+\tanh\left(2\frac{R_1-\sqrt{(x-x_1)^2+(y-y_1)^2}}{W}\right)+ \tanh\left(2\frac{R_2-\sqrt{(x-x_2)^2+(y-y_2)^2}}{W}\right),
\end{equation}
where $(x_1,y_1)=(L_x/2-R_1-d/2, L_y/2)$ and $(x_2,y_2)=(L_x/2+R_2+d/2,L_y/2)
$ are the centers of the two droplets, respectively.  The initial velocity field is zero in the whole domain.
In simulations, the computational domain of $L_x\times L_y=1.2\text{m}\times 1\text{m}$ is discretized by a uniform mesh $240\times 200$.   The initial radius of the two droplets is $R_1=R_2=0.125\text{m}$ and the initial gap  is $d=1.5W$ and $W=0.02 \text{m}$.
The densities of the two phases are $\rho_d=5\text{kg}/\text{m}^3,\rho_s=1\text{kg}/\text{m}^3$ and the viscosities are  $\nu_d=\nu_s=0.01\text{m}^2/\text{s}$. The surface tension coefficient is $\sigma=0.1\text{N}/\text{m}$, and the characteristic velocity is given by $U_c=\sqrt{\sigma \rho_2/R_2}$.
The Peclet number is set as $\text{Pe}=0.1/\text{Cn}$. Periodic boundary conditions are implemented at all boundaries. With these parameters, merging will take place.
Figure~\ref{twobubbles-shapes} shows the interfacial shapes of the droplets at  $t=2\text{T}$ and  $30\text{T}$ with $\text{T}=\sqrt{\rho_1 R^3/\sigma}$.  The interfacial shapes at $t=30\text{T}$ are compared with analytical results. From Fig.~\ref{twobubbles-shapes}, it is observed that the two droplets gradually merge, oscillate and finally form a larger stationary  droplet. Especially, the final interface shapes predicted by all formulations  are in good agreement with the analytical solutions.
However, the interface positions predicted by the LBE models with $\bm F_{cpf-1}$, $\bm F_{pf-2}$, $\bm F_{csf-2}$ at $t=2\text{T}$  are different
from those of the other formulations.
The droplets of the LBE models  with $\bm F_{cpf-1}$, $\bm F_{pf-2}$ and $\bm F_{csf-2}$ have started to merge while the droplets with the other interfacial force formulations  remain at  a distinct distance.
As  no external forces are presented in  the system, the mass centre of the droplets should not change
during coalescence. Figure~\ref{twobubbles-kineticenergy} shows the time development of the position of the mass centre of the droplets. All interfacial forces present similar accuracy.
It's worth pointing out that the computations with $\bm F_{cpf-1}, \bm F_{cpf-2}, \bm F_{pf-1}$ and $\bm F_{pf-2}$ become unstable when $\text{Pe}=1/\text{Cn}$. This implies that  both $\bm F_{stf}$ and $\bm F_{csf}$ have a better numerical stability for this problem.

\begin{figure}[htb]
\centering
\subfloat[]{\includegraphics[width=0.25\textwidth]{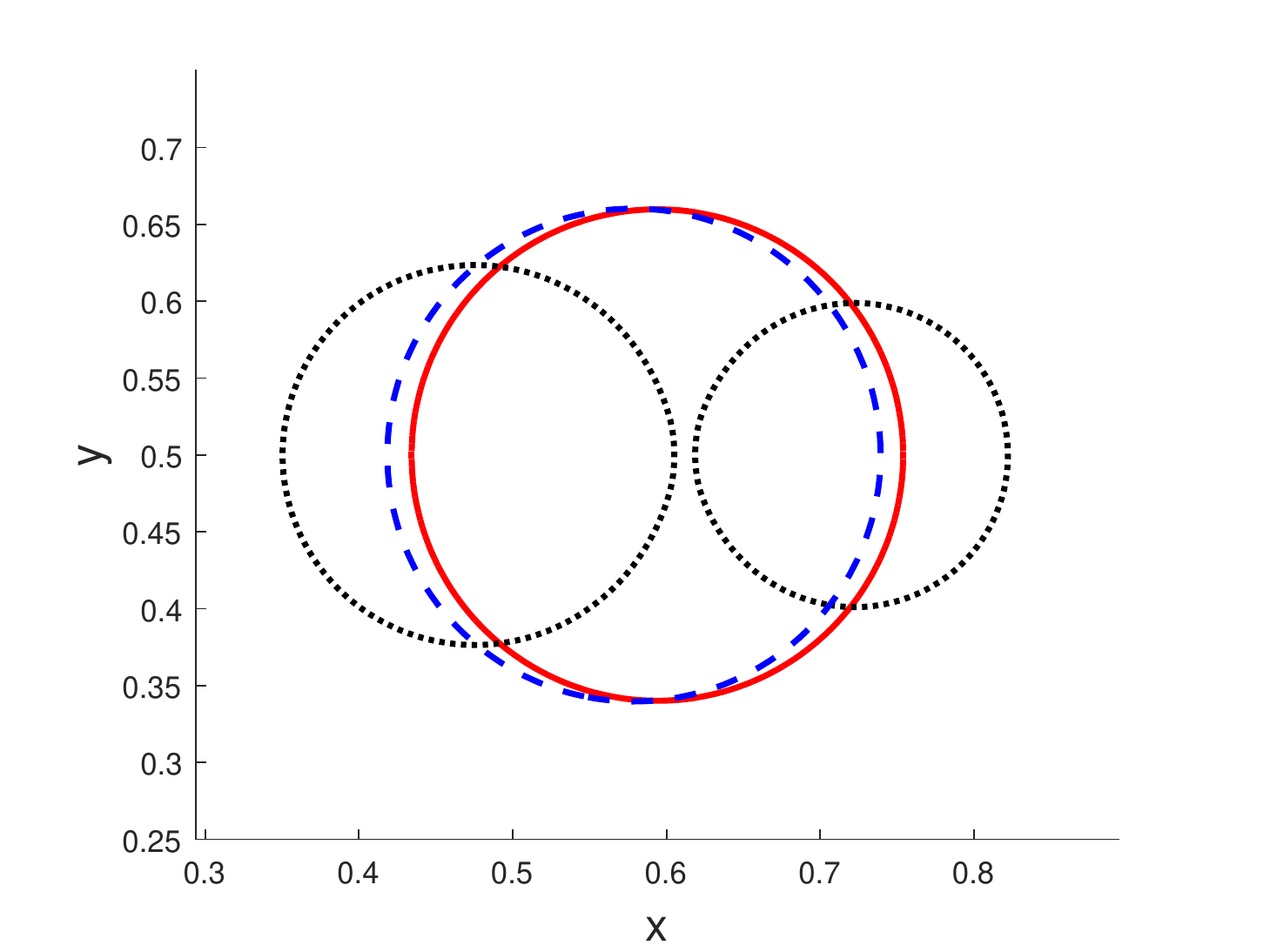}}~
\subfloat[]{\includegraphics[width=0.25\textwidth]{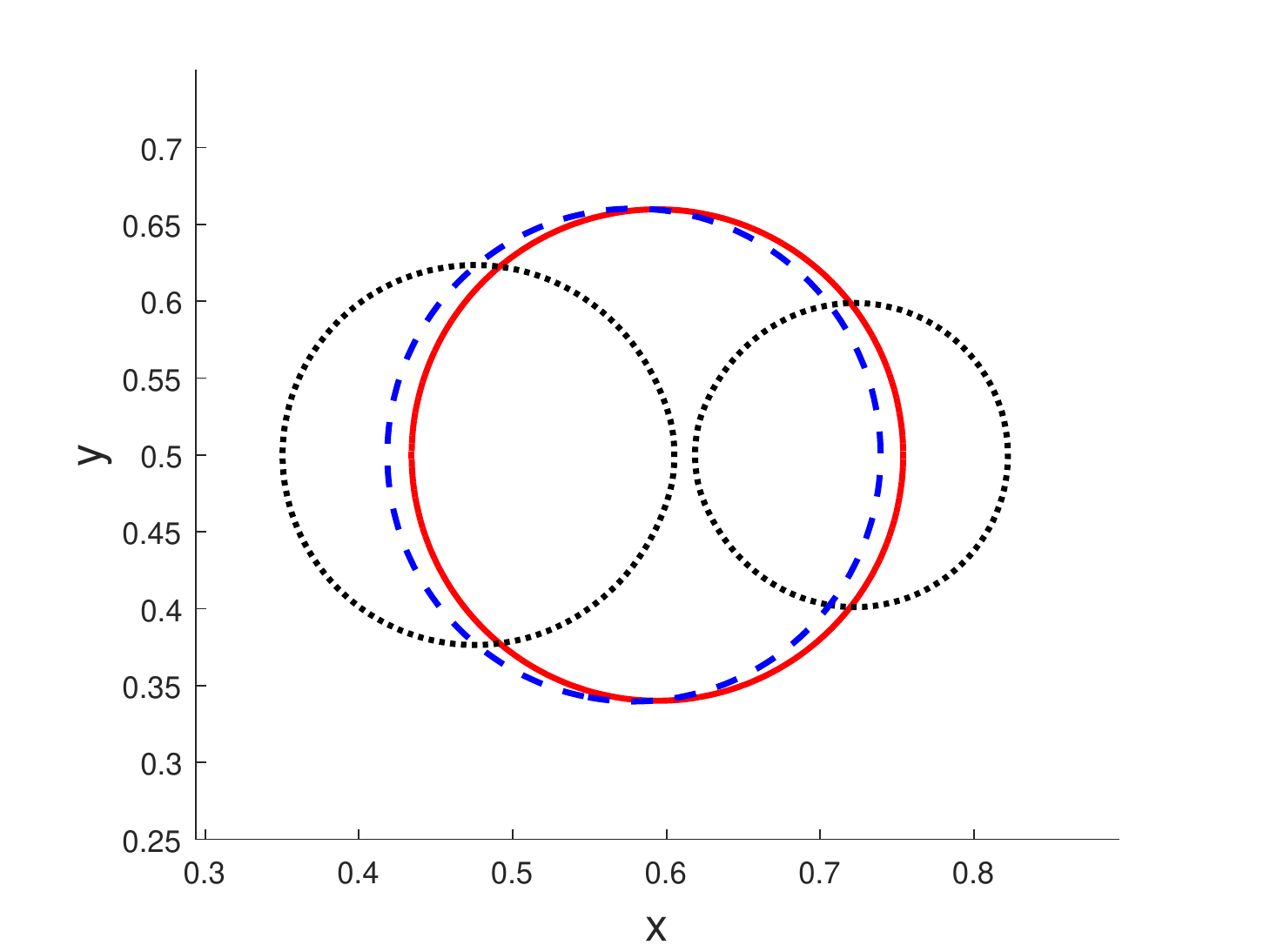}}~
\subfloat[]{\includegraphics[width=0.25\textwidth]{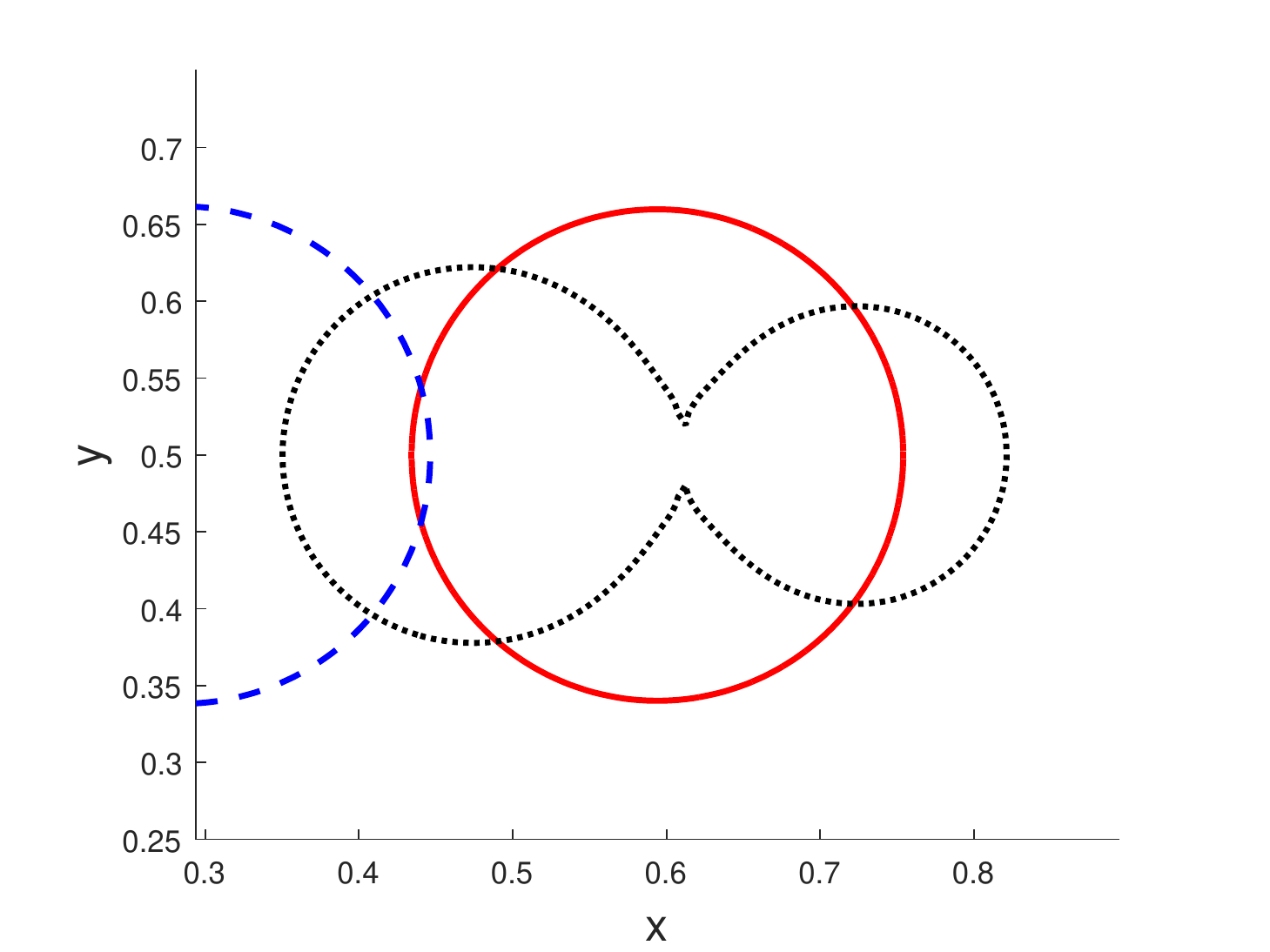}}~
\subfloat[]{\includegraphics[width=0.25\textwidth]{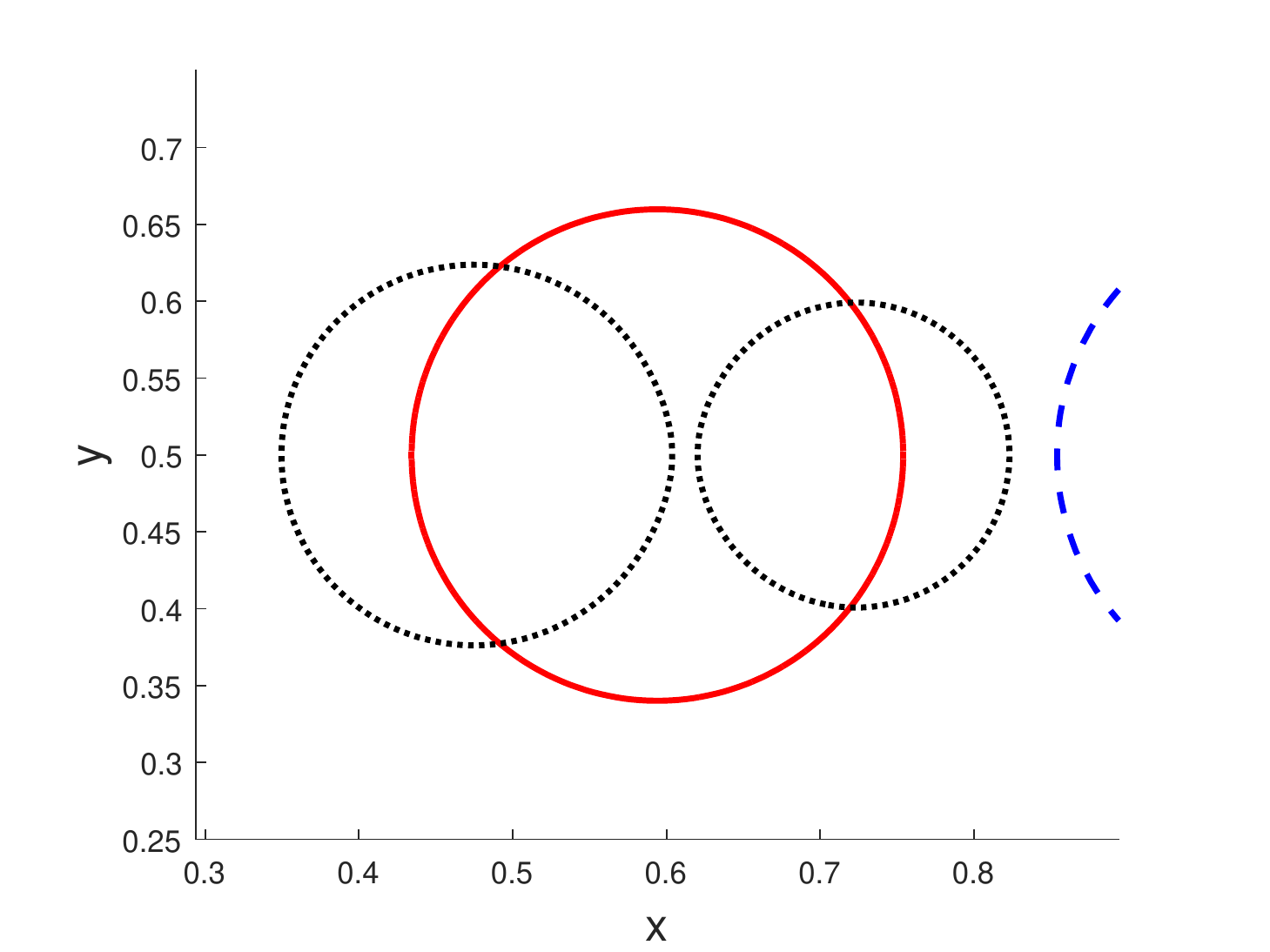}}\\
\subfloat[]{\includegraphics[width=0.25\textwidth]{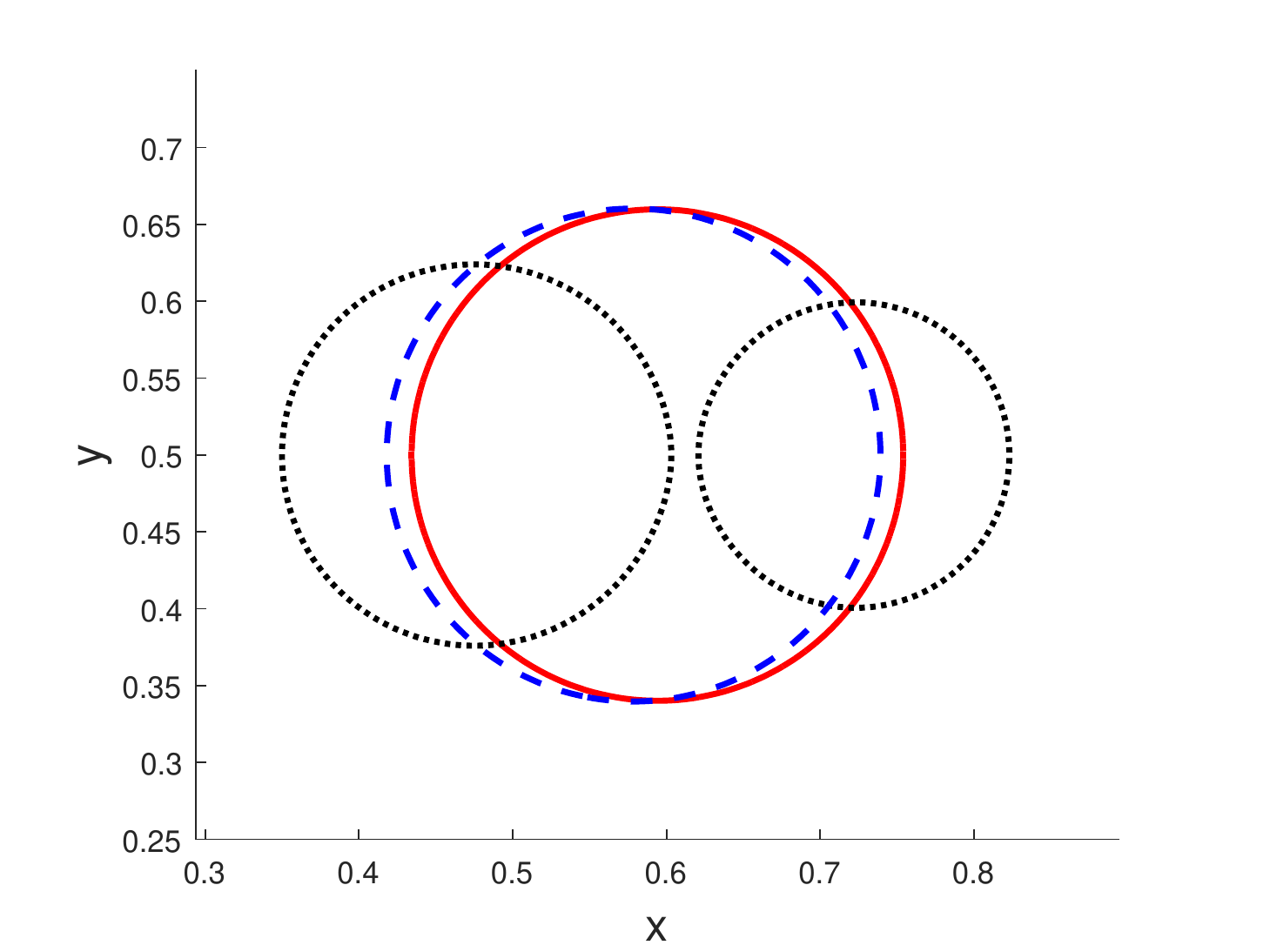}}~
\subfloat[]{\includegraphics[width=0.25\textwidth]{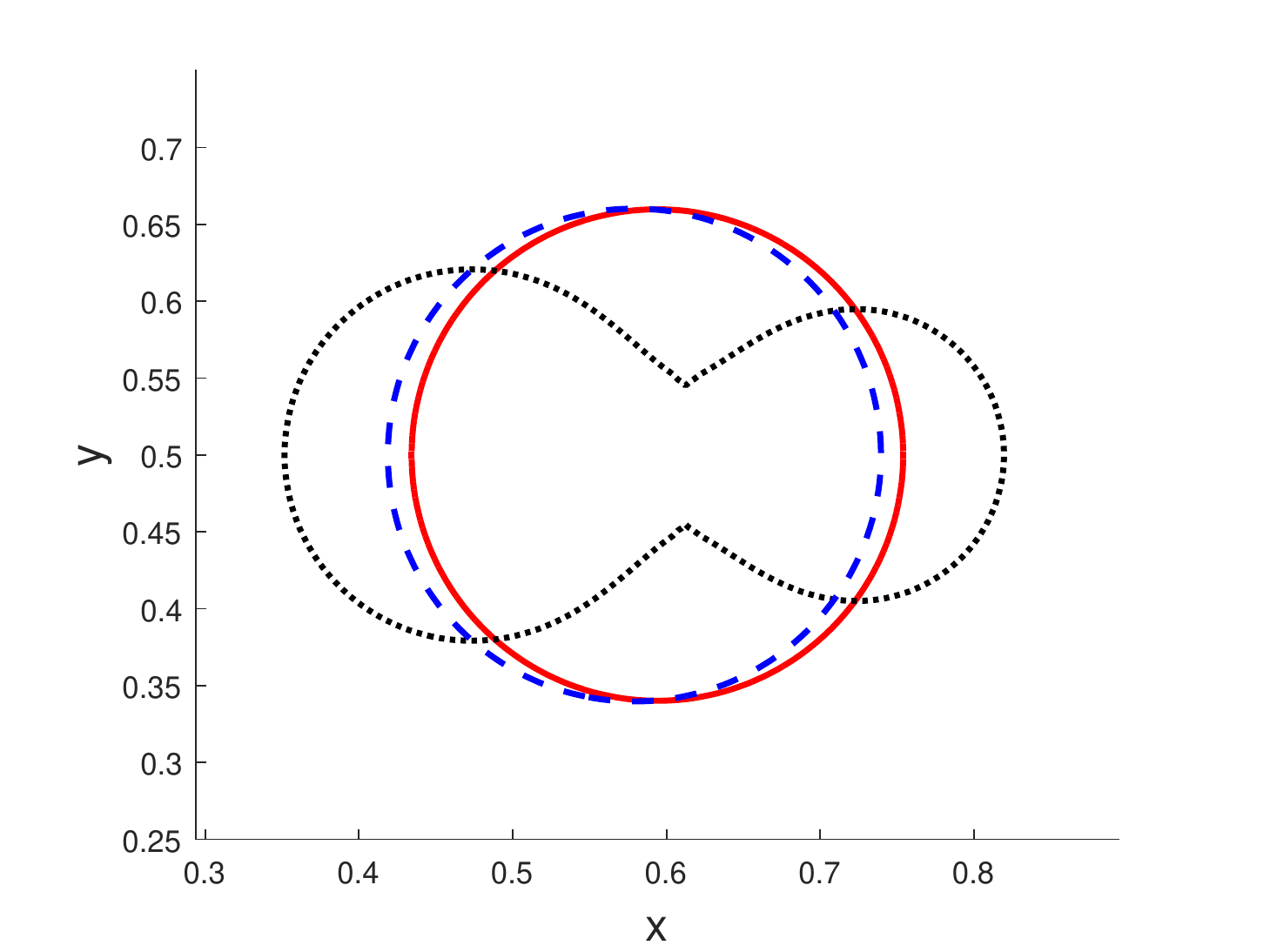}}~
\subfloat[]{\includegraphics[width=0.25\textwidth]{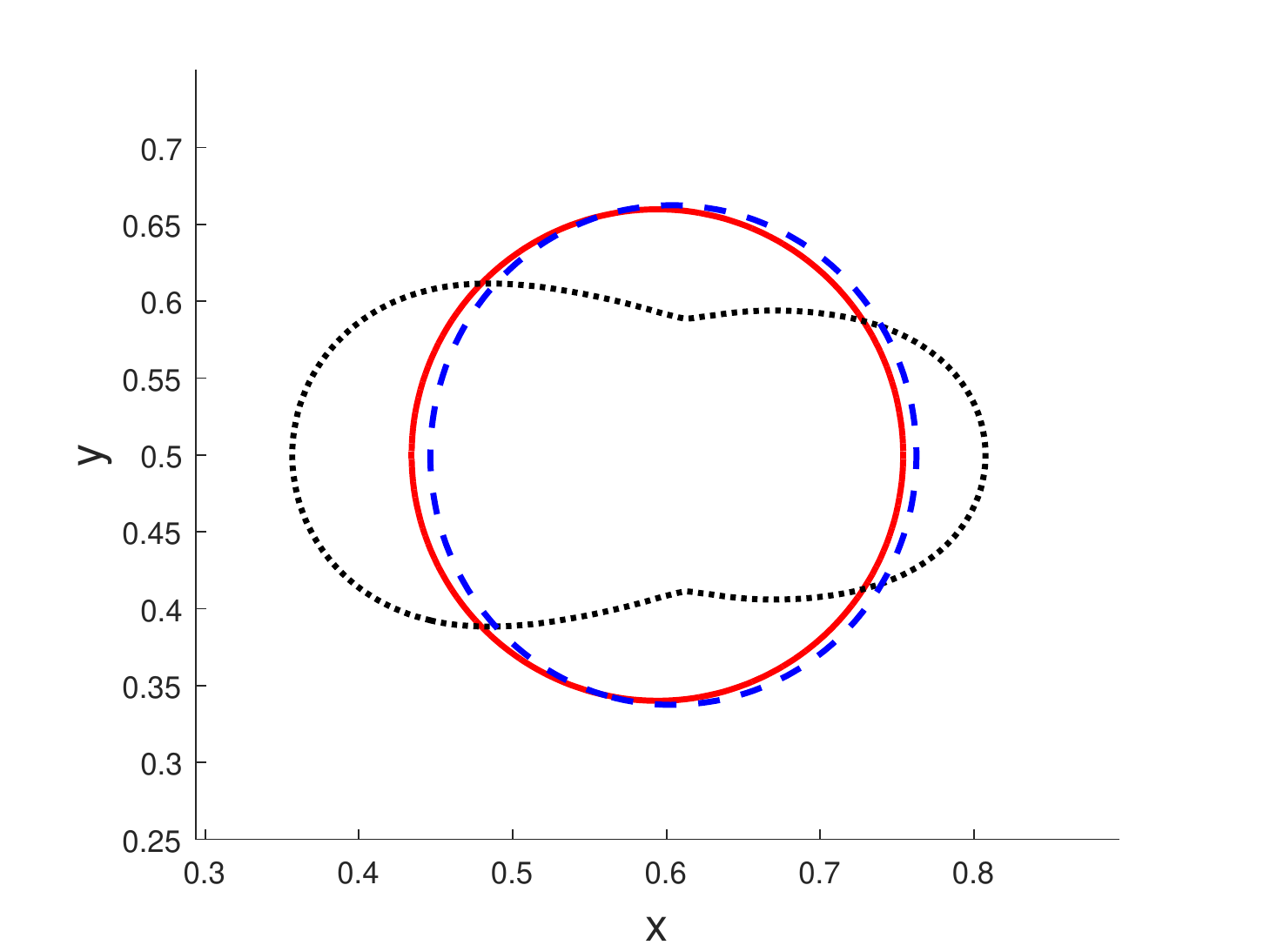}}~
\subfloat[]{\includegraphics[width=0.25\textwidth]{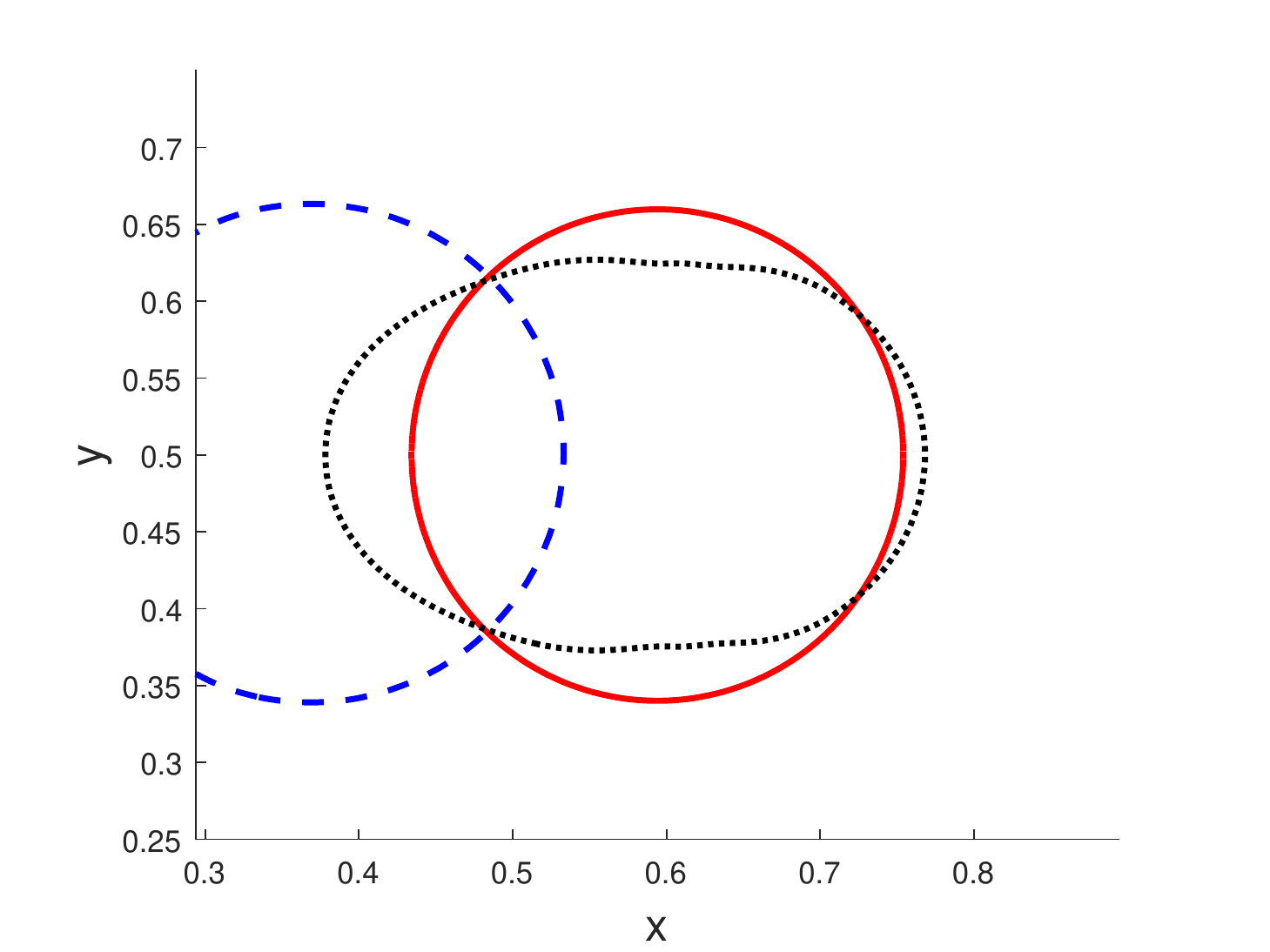}}
\caption{ Interfaces of   two droplets of unequal sizes at $t=2\text{T}$(dotted line) and $30\text{T}$(dashed line): (a) $\bm F_{stf-1}$,(b) $\bm F_{stf-2}$,(c) $\bm F_{cpf-1}$,(d) $\bm F_{cpf-2}$,(e) $\bm F_{pf-1}$,(f) $\bm F_{pf-2}$,(g) $\bm F_{csf-1}$ and (h) $\bm F_{csf-2}$. Solid line represents the analytical solutions.  }
\label{twobubbles-shapes-unequal}
\end{figure}

\begin{figure}[htb]
  \centering
\includegraphics[width=0.5\textwidth]{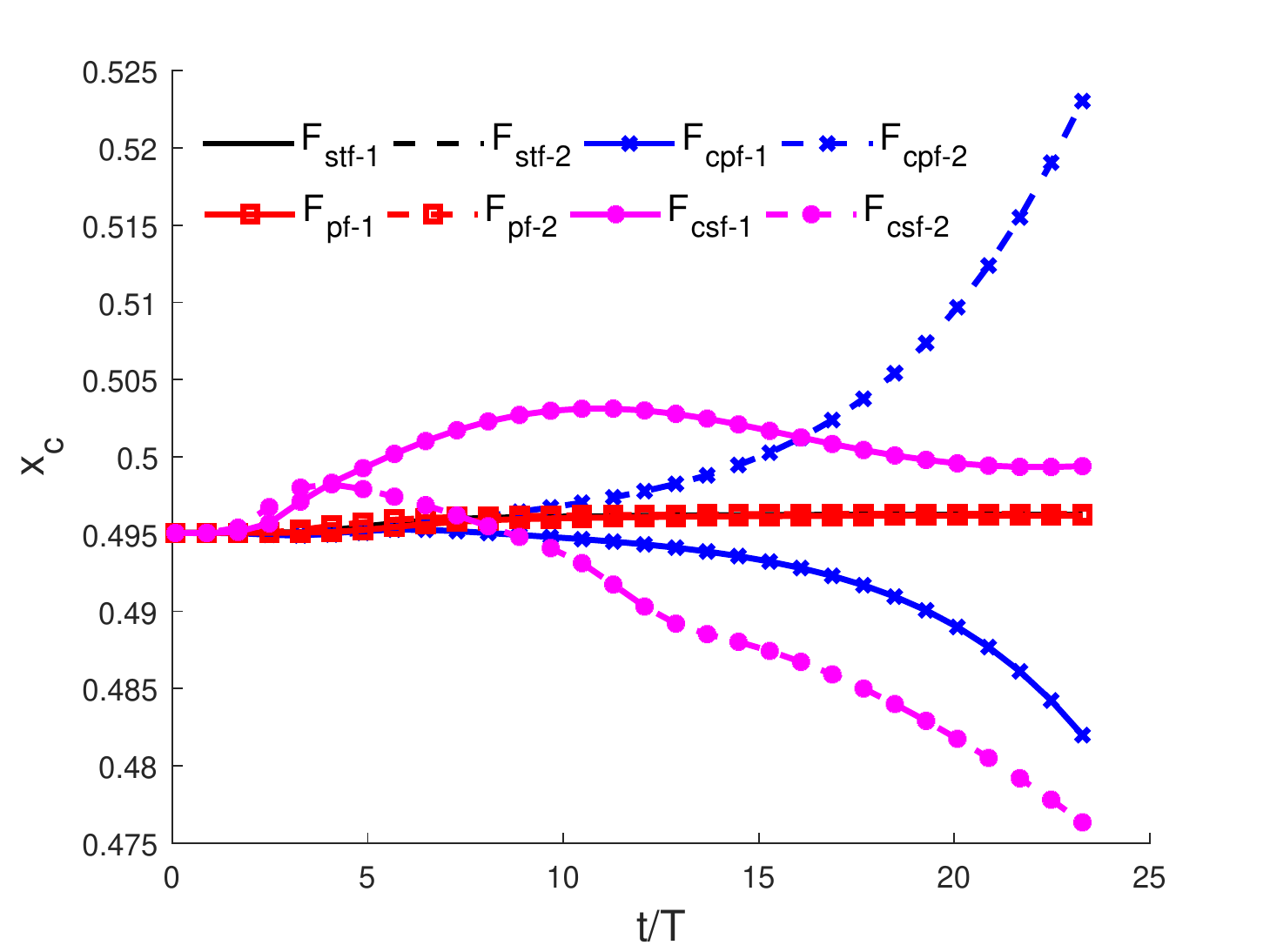}
  \caption{ Time history of mass center $x_c$  for droplets of unequal size.}\label{twobubbles-yc-unequal}
\end{figure}

We further simulate the above system but with two droplets of unequal sizes ($R_1=0.125\text{m}, R_2=0.1\text{m}$).
Figure~\ref{twobubbles-shapes-unequal} shows the interfacial shapes of the droplets at  $t=2\text{T}$ and $30\text{T}$.
The interface positions  are different for  each interfacial force formulation. In particular, the merged  droplets predicted by the BE models with $\bm F_{cpf-1}$, $\bm F_{cpf-2}$ and $\bm F_{csf-2}$ have a distinct movement.
Figure~\ref{twobubbles-yc-unequal} shows the time development of the position of the mass centre of the droplets, which shows that the positions predicted by $\bm F_{cpf}$( $\bm F_{cpf-1},\bm F_{cpf-2}$)  and $\bm F_{csf}$ ($\bm F_{csf-1},\bm F_{csf-2}$) display significant deviations from their initial positions as time increases.

\subsection{Capillary wave}
\begin{figure}[!htb]
\centering
\subfloat[]{\includegraphics[width=0.25\textwidth]{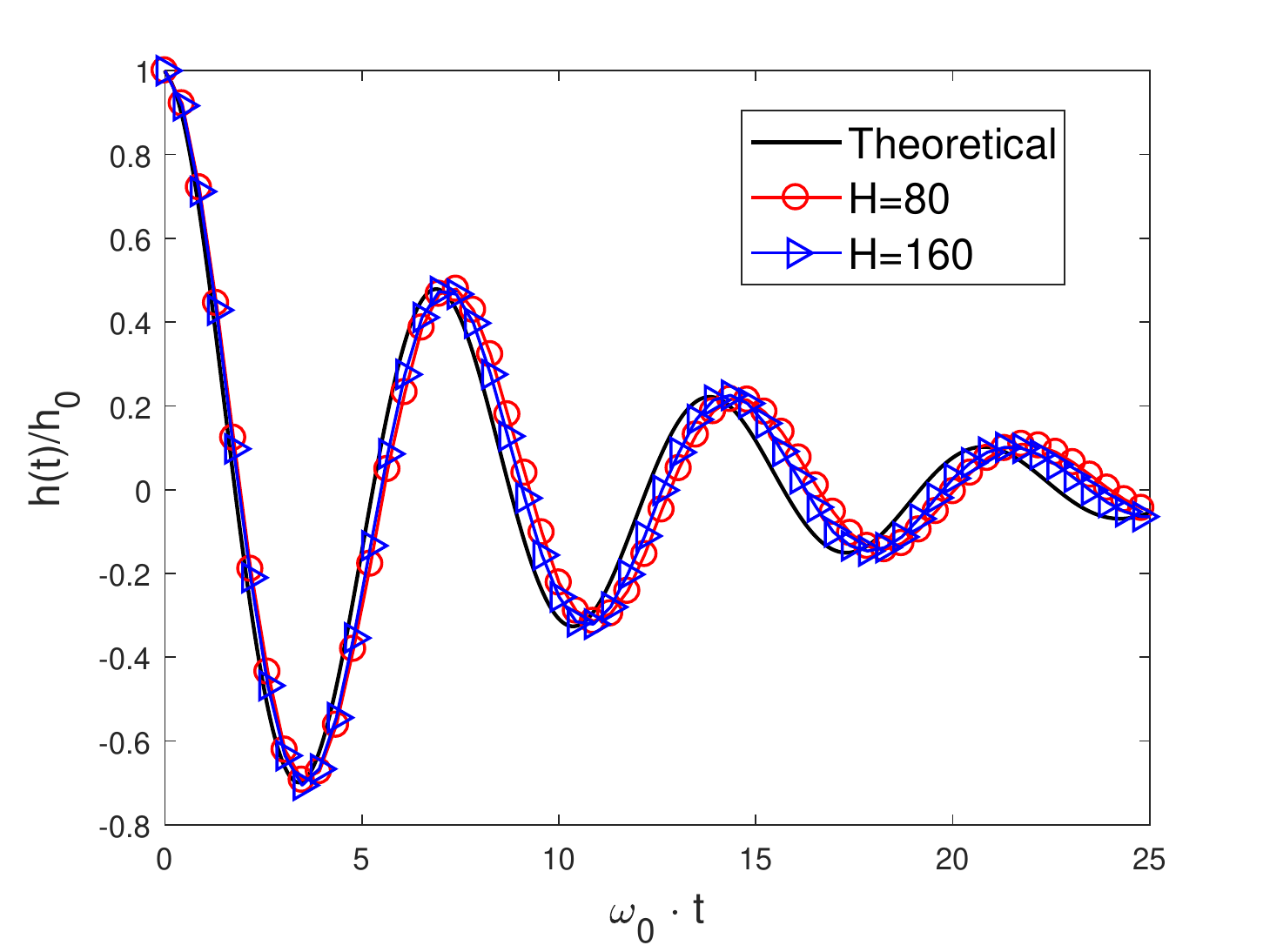}}~
\subfloat[]{\includegraphics[width=0.25\textwidth]{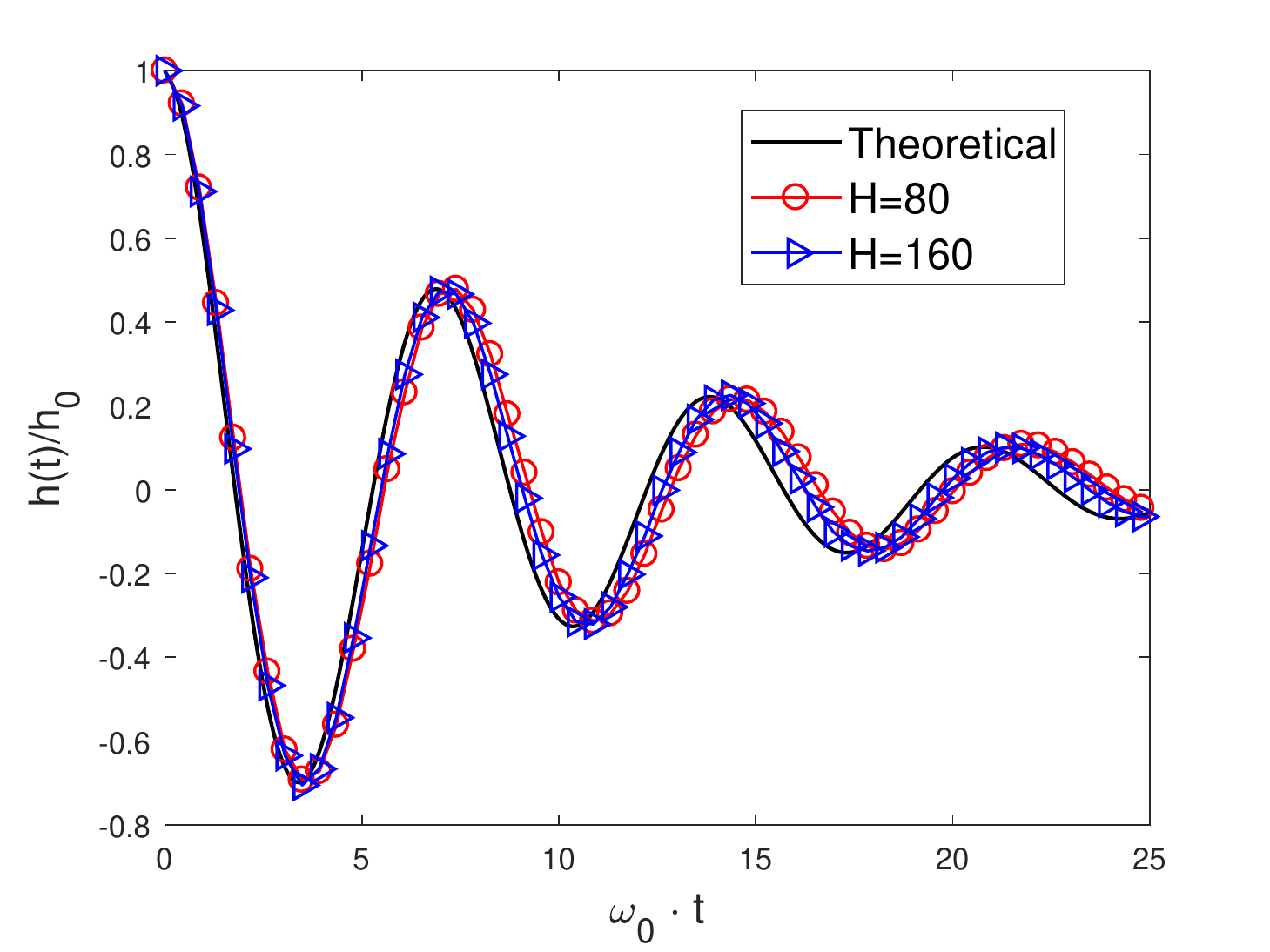}}~
\subfloat[]{\includegraphics[width=0.25\textwidth]{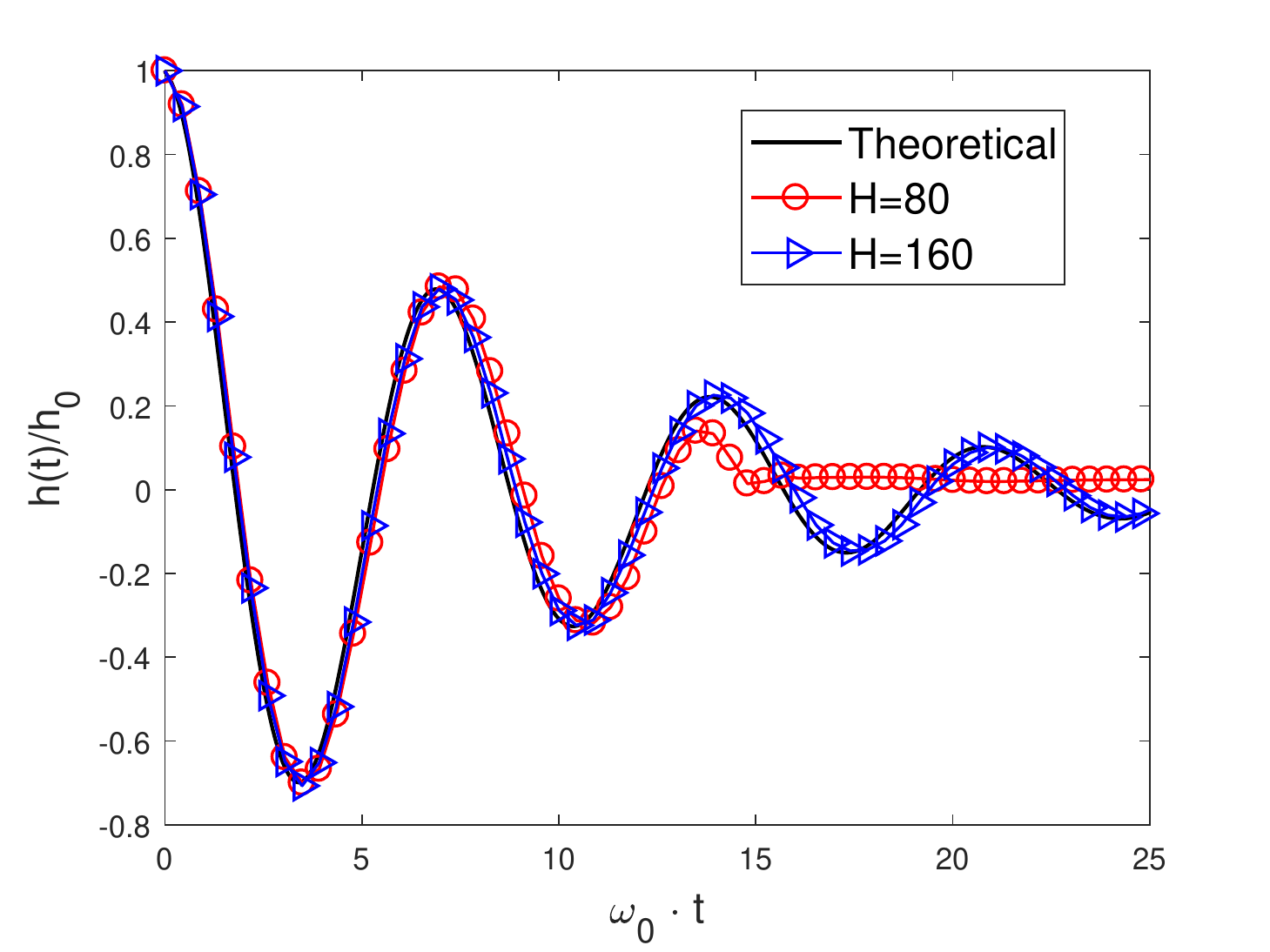}}~
\subfloat[]{\includegraphics[width=0.25\textwidth]{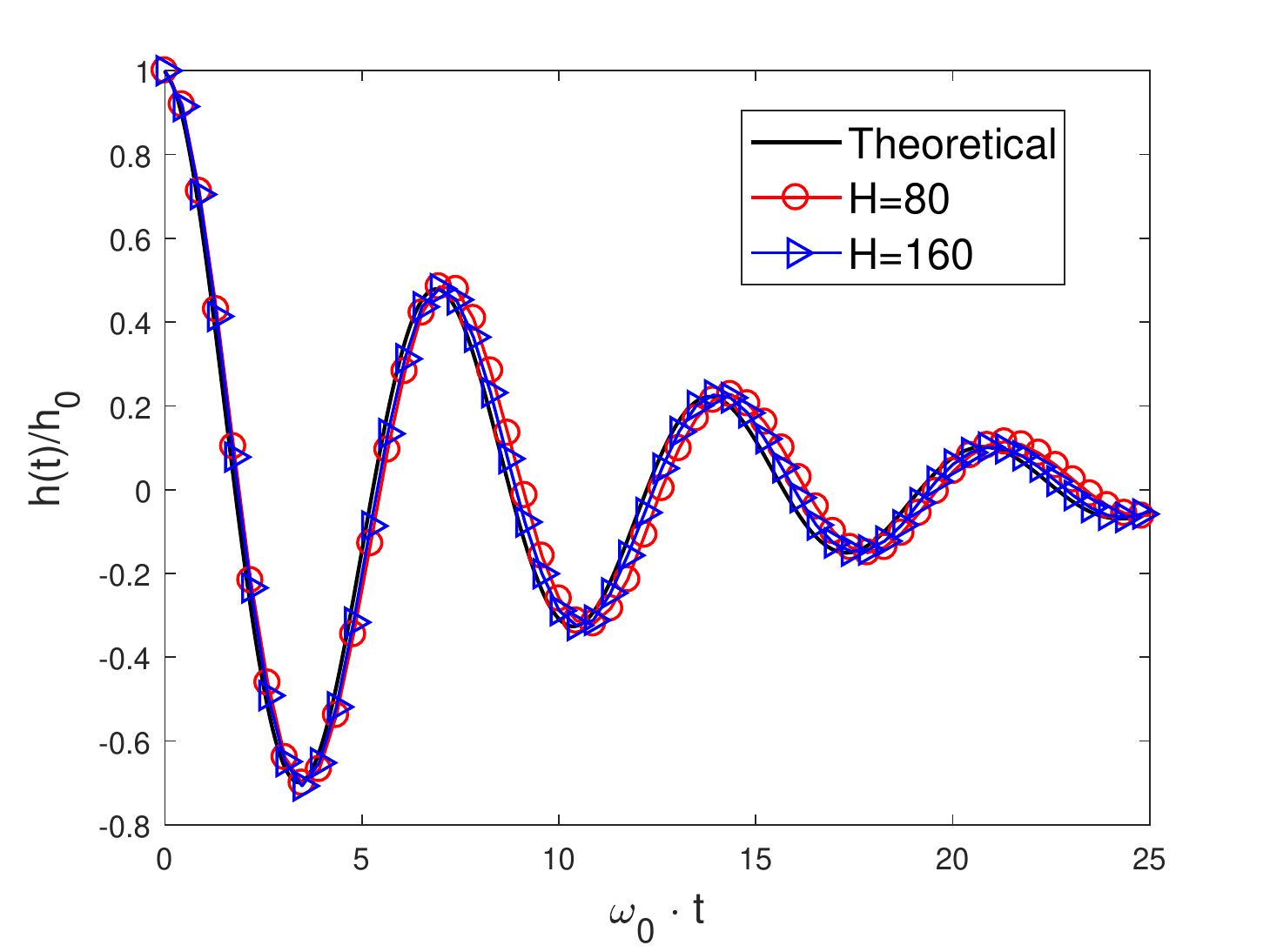}}\\
\subfloat[]{\includegraphics[width=0.25\textwidth]{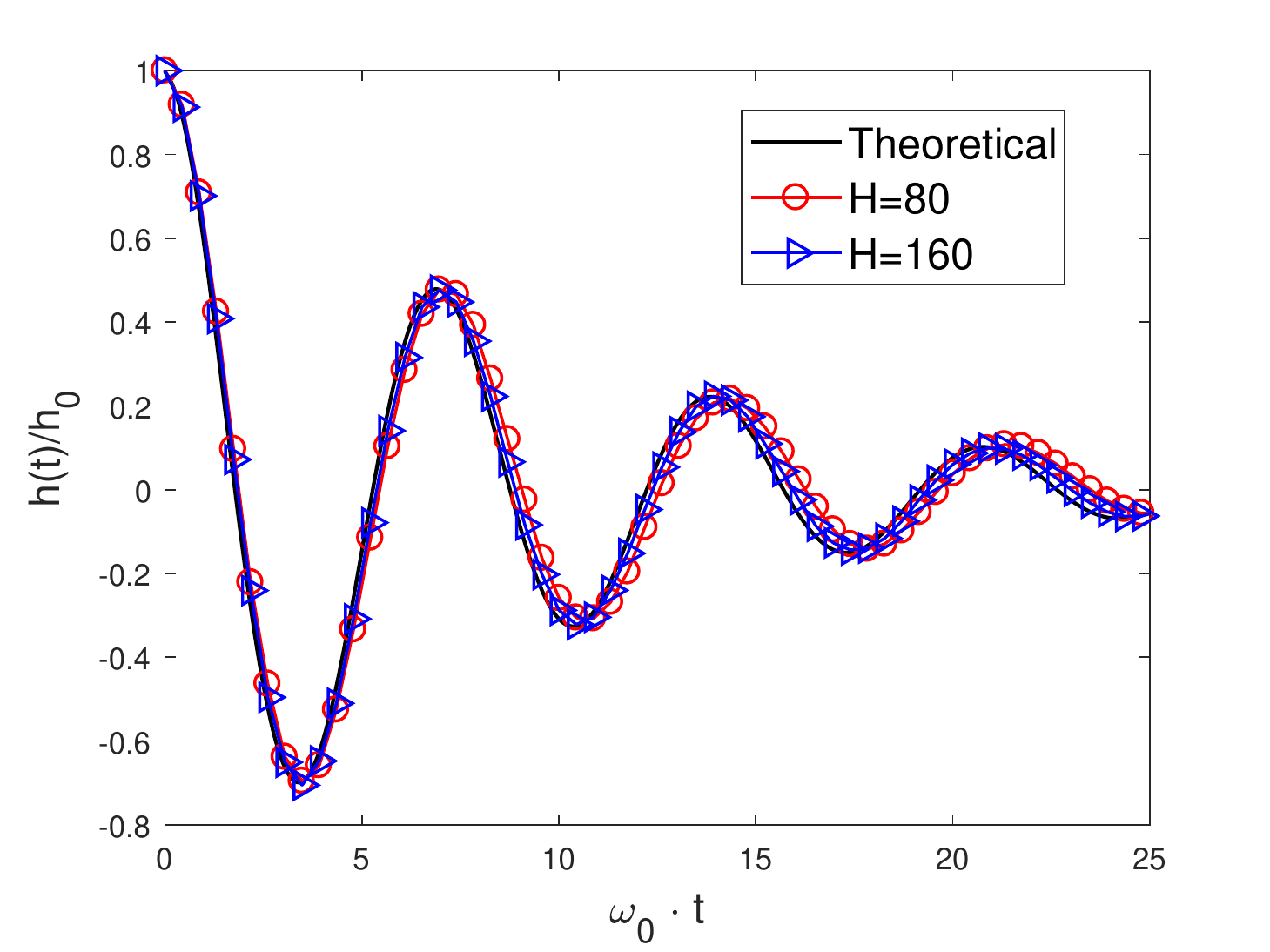}}~
\subfloat[]{\includegraphics[width=0.25\textwidth]{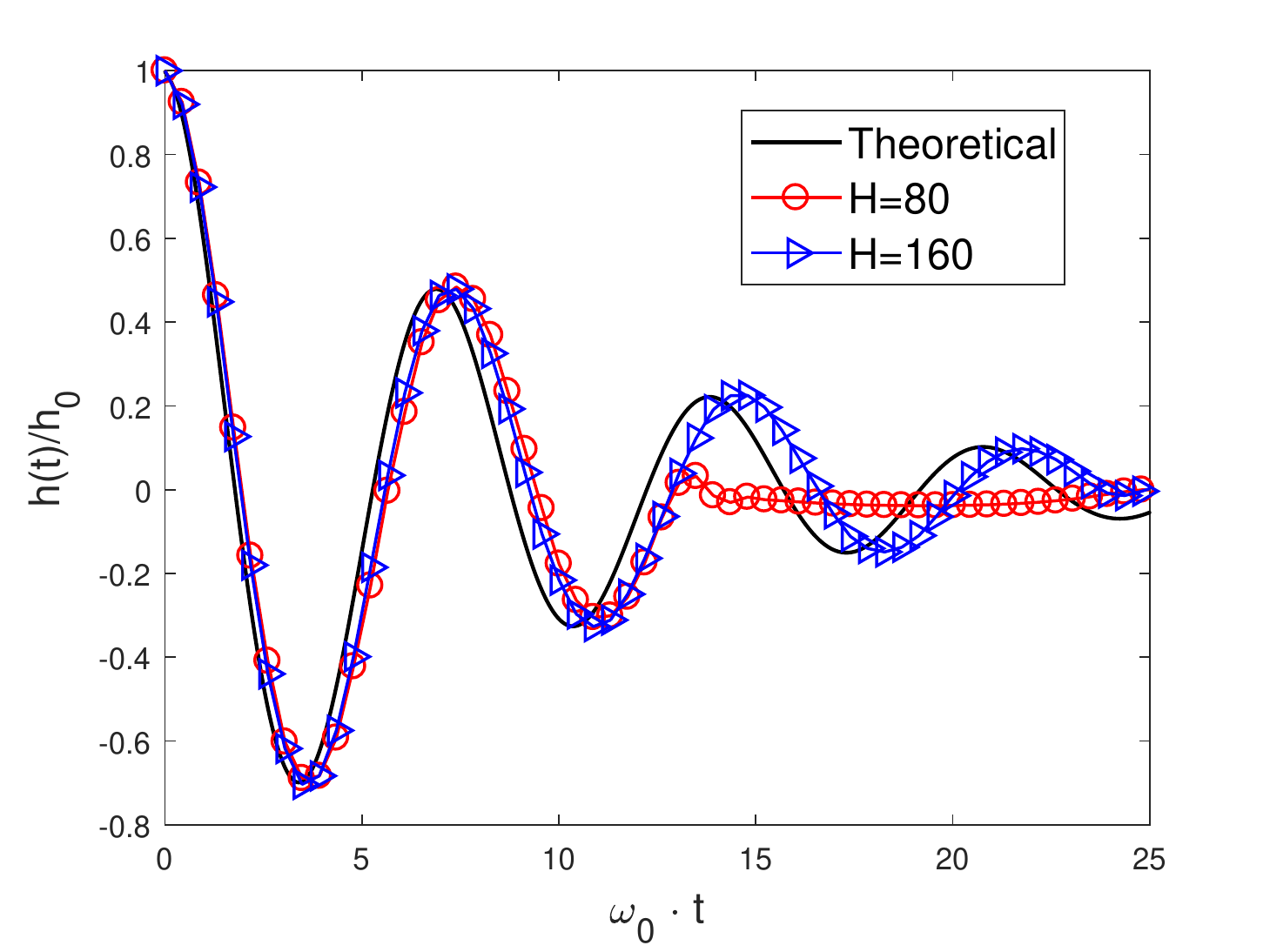}}~
\subfloat[]{\includegraphics[width=0.25\textwidth]{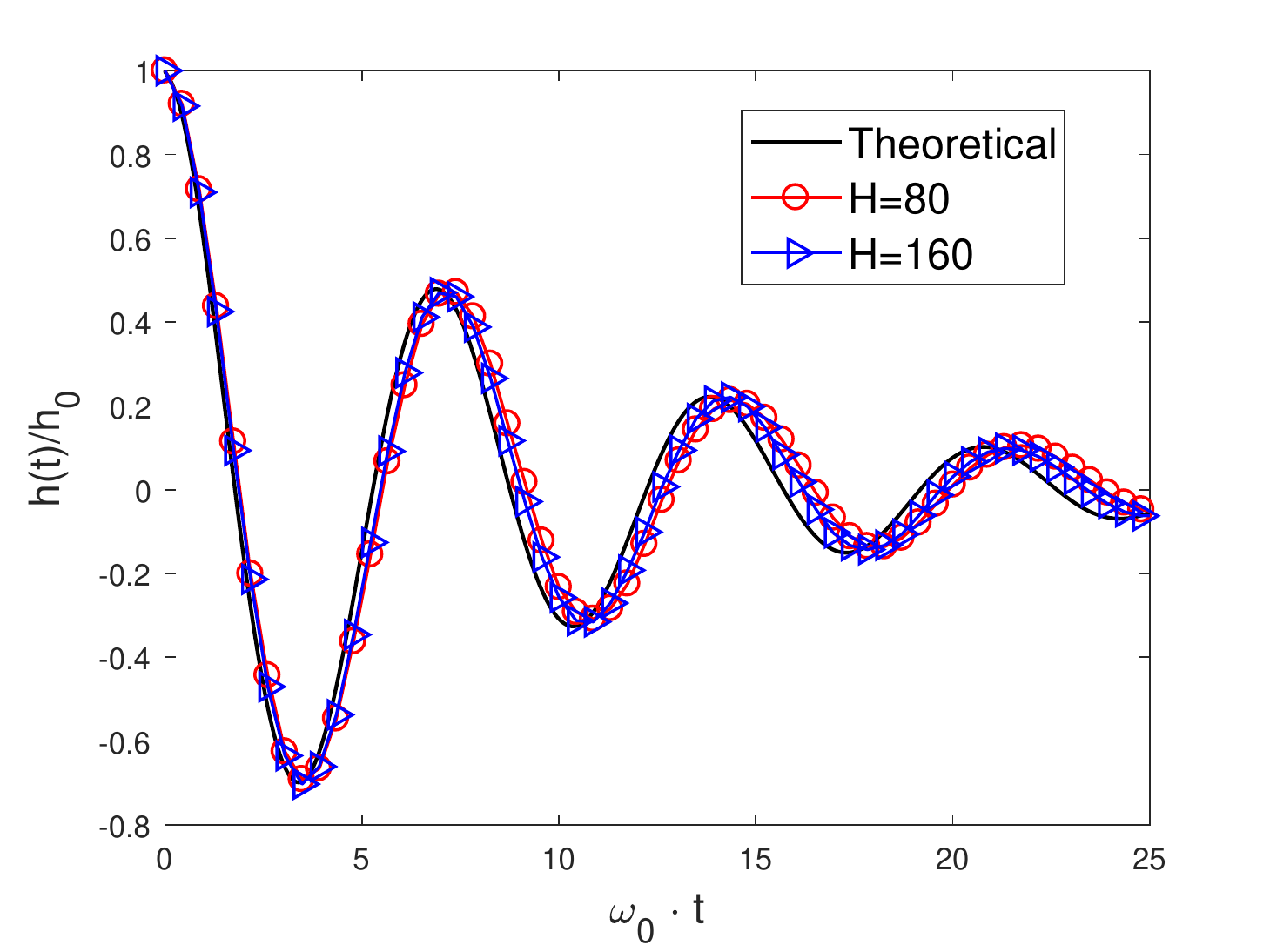}}~
\subfloat[]{\includegraphics[width=0.25\textwidth]{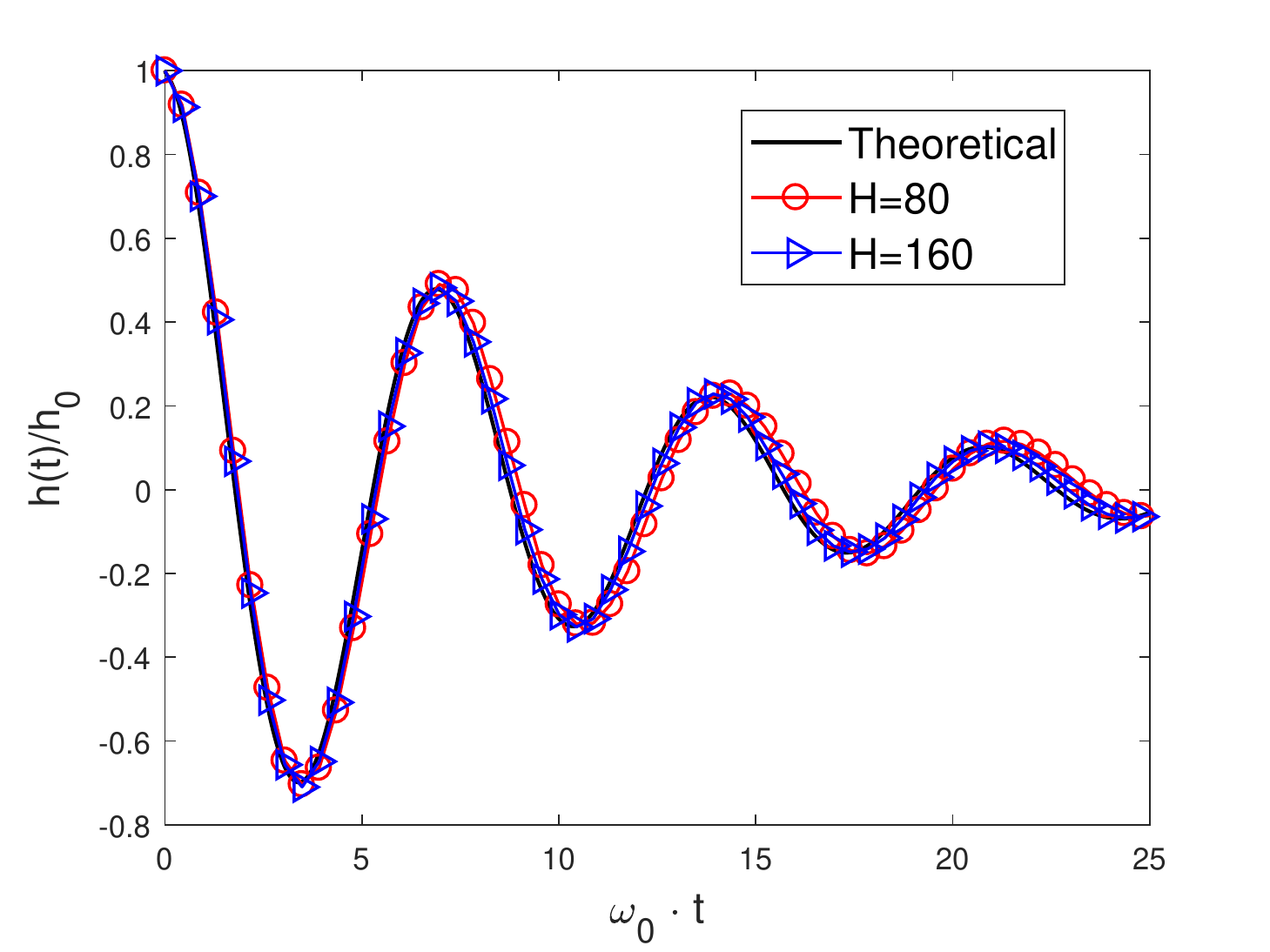}}\\
\caption{Time evolution of capillary wave amplitude with $\rho_1/\rho_2=1$ for (a) $\bm F_{stf-1}$,(b) $\bm F_{stf-2}$,(c) $\bm F_{cpf-1}$,(d) $\bm F_{cpf-2}$,(e) $\bm F_{pf-1}$,(f) $\bm F_{pf-2}$,(g) $\bm F_{csf-1}$ and (h) $\bm F_{csf-2}$. }
\label{fig:Capillary_sigma30}
\end{figure}

We further test the numerical accuracy of the interfacial force formulations by a two-dimensional capillary wave problem.  Initially, a heavier fluid  is placed under a lighter fluid with a small perturbation $y=1.5H+h_0\cos(kx)$ on the interface in a rectangle domain of $H\times 3H$, where $h_0$ is the initial perturbation amplitude and $k=2\pi/H$ is the wave number. The evolution of the interface wave amplitude $h(t)$ is given by~\cite{prosperetti1981motion}
\begin{equation}\label{capillary_wave}
\frac{h(t)}{h_0}=\frac{4(1-4\beta)\nu^2 k^4}{8(1-4\beta)\nu^2 k^4+\omega_0^2}\mbox{erfc}(\sqrt{\nu k^2 t})+
\sum_{i=1}^{4} \frac{z_i}{\bm Z_i}\frac{\omega_0^2}{z_i^2- \nu k^2}e^{( z_i^2-\nu k^2)t} \mbox{erfc}(z_i \sqrt{t}),
\end{equation}
where $\beta=\rho_1 \rho_2/(\rho_1+\rho_2)^2$,  $\omega_0^2=(\sigma k^3)/(\rho_2+\rho_1)$, $\mbox{erfc}(z_i)$ is the complementary error function of a complex variable $z_i$,  $z_i(i=1,\ldots,4)$ are the four roots of the following algebraic equation
\begin{equation}\label{error_function}
z^4-4\beta \sqrt{\nu k^2} z^3+2(1-6\beta)\nu k^2 z^2+4(1-3\beta)(\nu k^2)^{3/2}z+(1-4\beta)\nu^2 k^4+\omega_0^2=0,
\end{equation}
and $ Z_i$ is defined as
\begin{equation}\label{z}
 Z_i=\prod_{j\neq i}(z_j-z_i),\qquad i,j=1,\cdots,4.
\end{equation}
In simulations, periodic boundaries are applied to the left and right sides and no-slip boundaries are imposed on the top and bottom walls~\cite{lallemand2007lattice}.  The physical parameters are set as
 $H=1 \text{m}, \rho_1=\rho_2=1 \text{kg/$\text{m}^3$}$, $\nu_1=\nu_2=0.01 \text{$\text{m}^2$/s}$
 , $\sigma=0.25\text{N/m}$. The characteristic velocity is given by $U_c=\sqrt{\sigma/L_c/\rho_1}$. Hence, the Reynolds number is  $\text{Re}=50$ and the Weber number is $\text{We}=1$.   Two uniform grids of $H=80$ and  $160$ are used.
Figure~\ref{fig:Capillary_sigma30} shows the evolution of the capillary amplitude for each grid. All the numerical results agree well with the theoretical solutions in the initial stage. However, the decaying amplitudes with $\bm F_{cpf-1}$ and $\bm F_{pf-2}$ on $80\times 240$ meshes reach the steady state faster than the other forms as time increases. We found that this behavior can be improved by increasing the Peclet number.

\begin{figure}[!htb]
\centering
\subfloat[]{\includegraphics[width=0.25\textwidth]{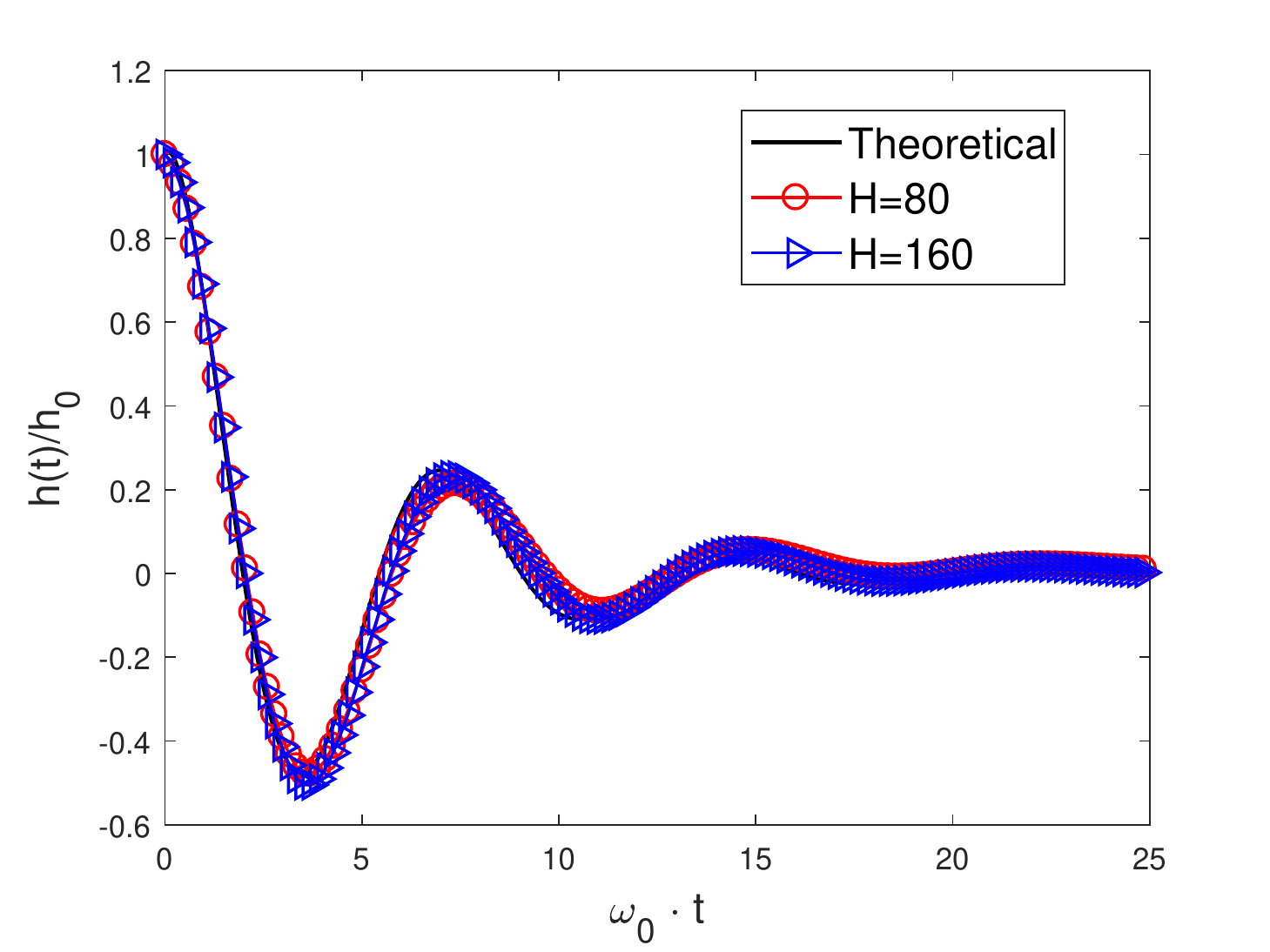}}~
\subfloat[]{\includegraphics[width=0.25\textwidth]{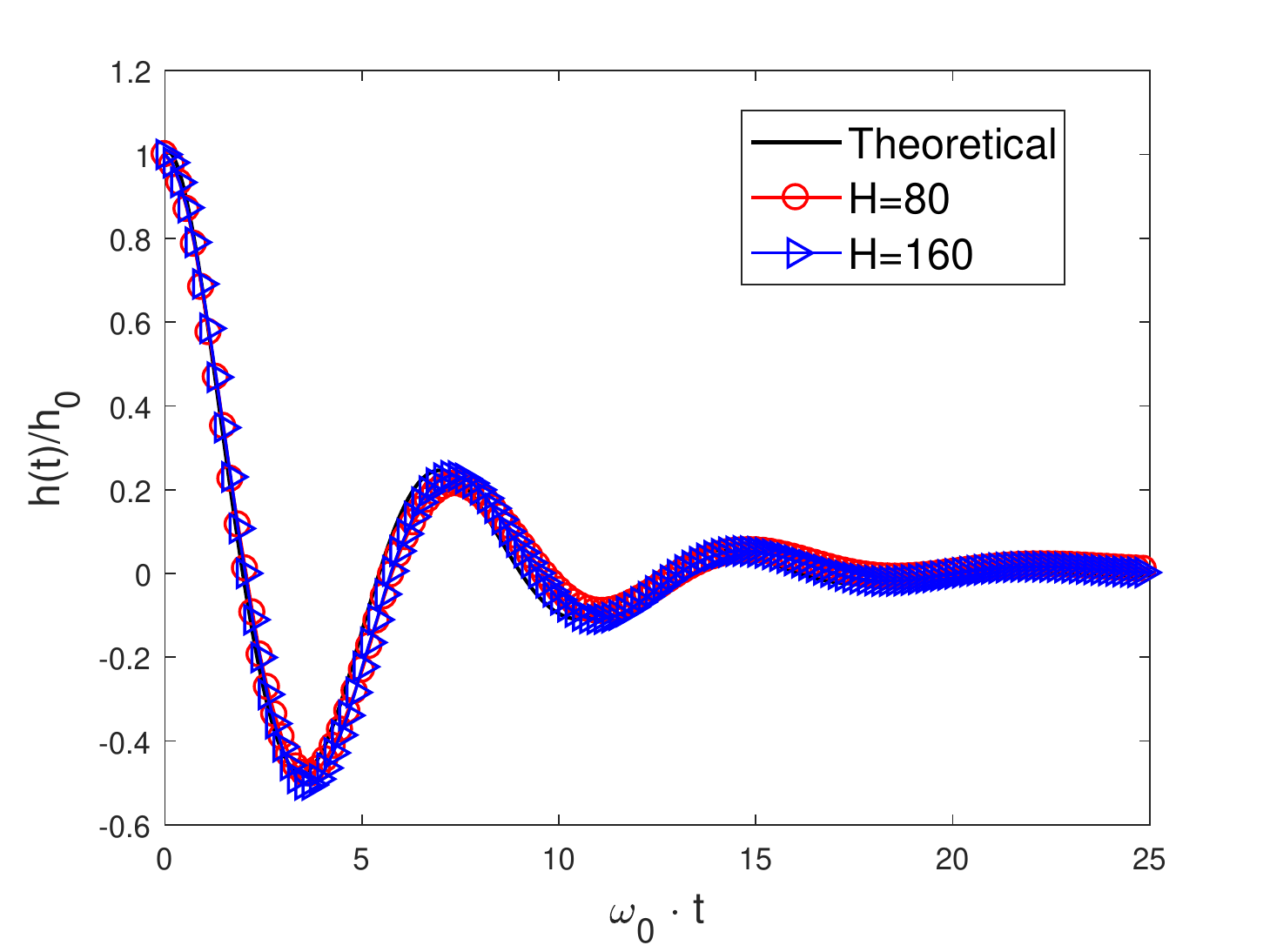}}~
\subfloat[]{\includegraphics[width=0.25\textwidth]{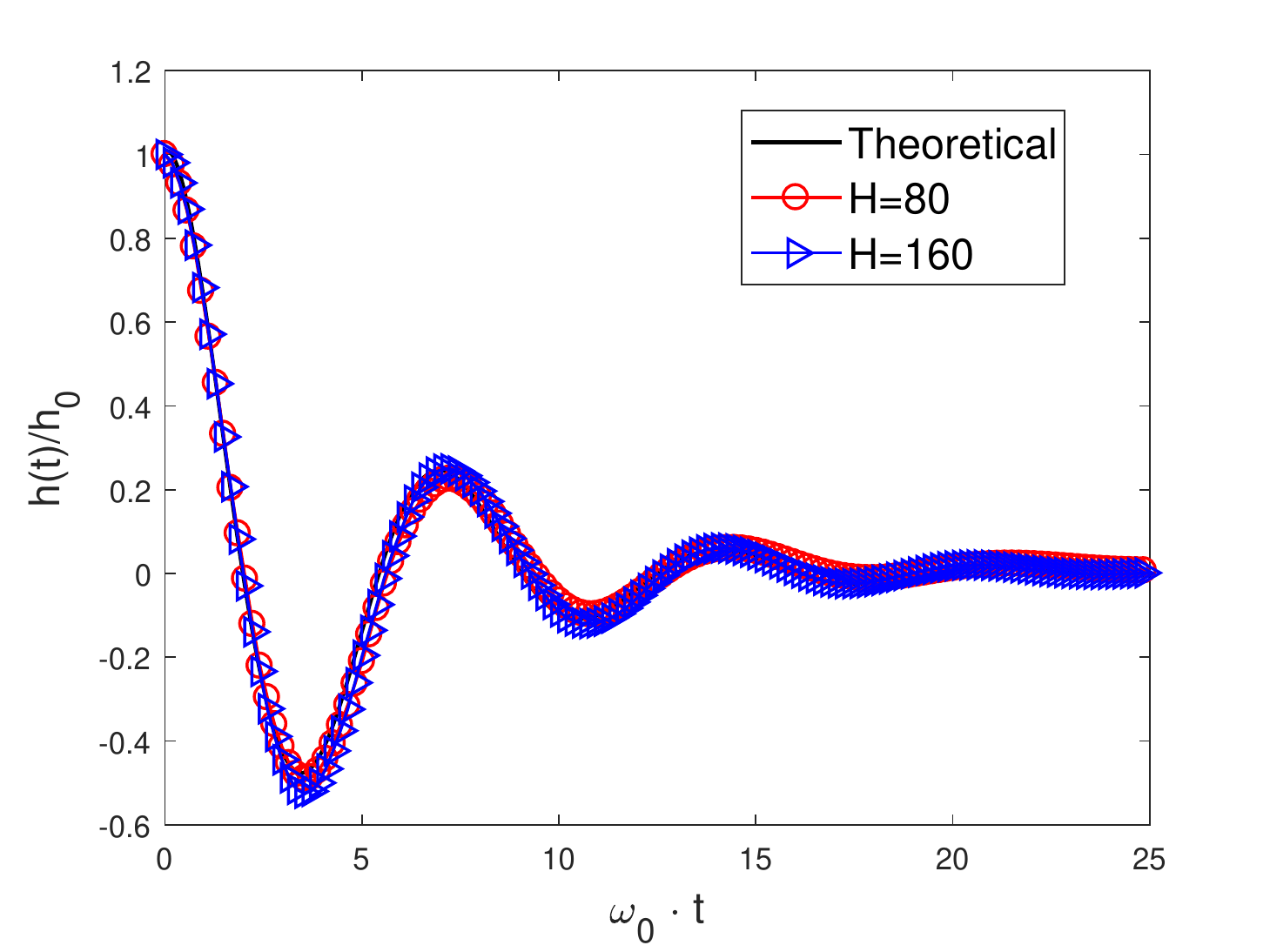}}~
\subfloat[]{\includegraphics[width=0.25\textwidth]{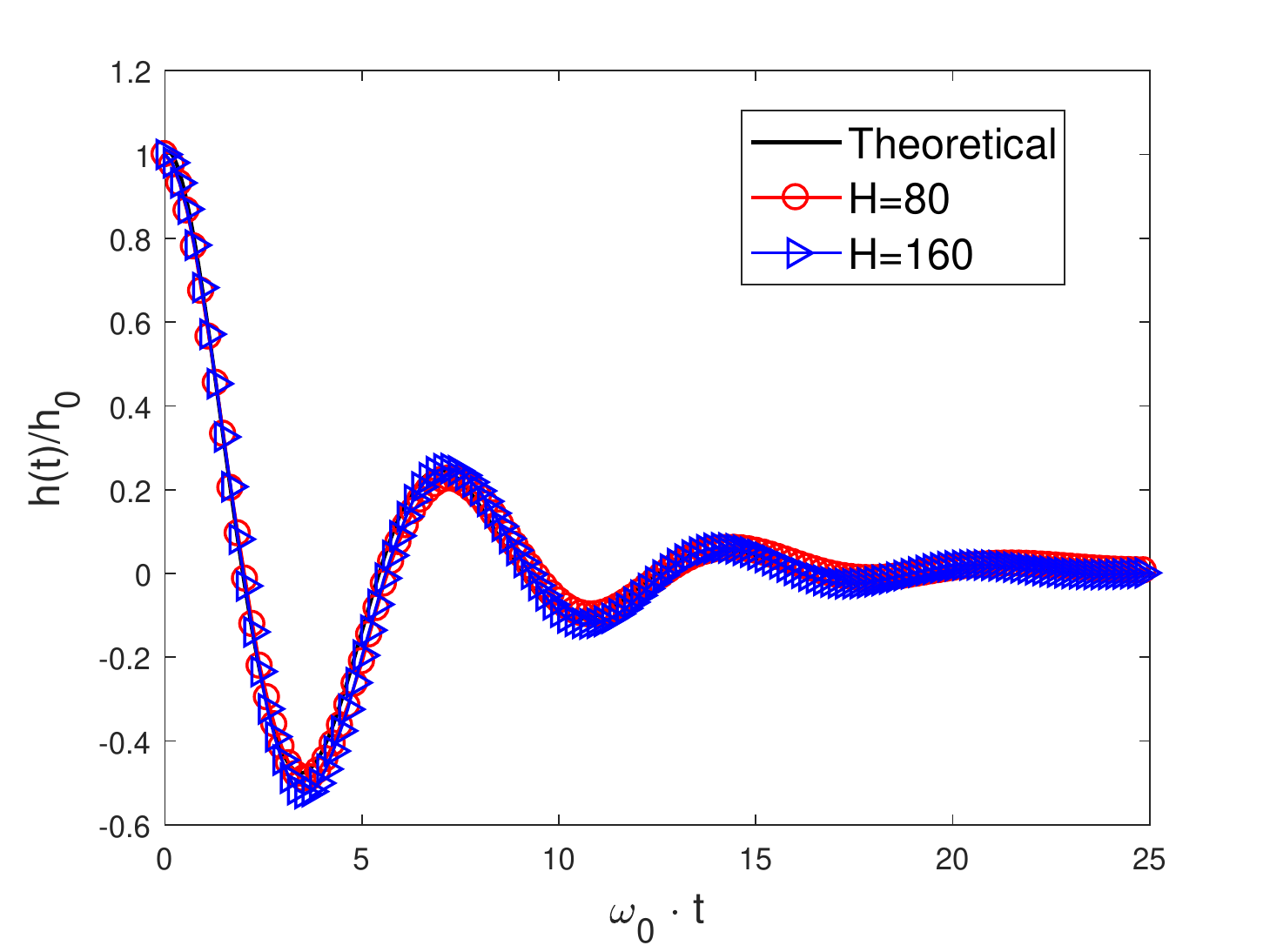}}\\
\subfloat[]{\includegraphics[width=0.25\textwidth]{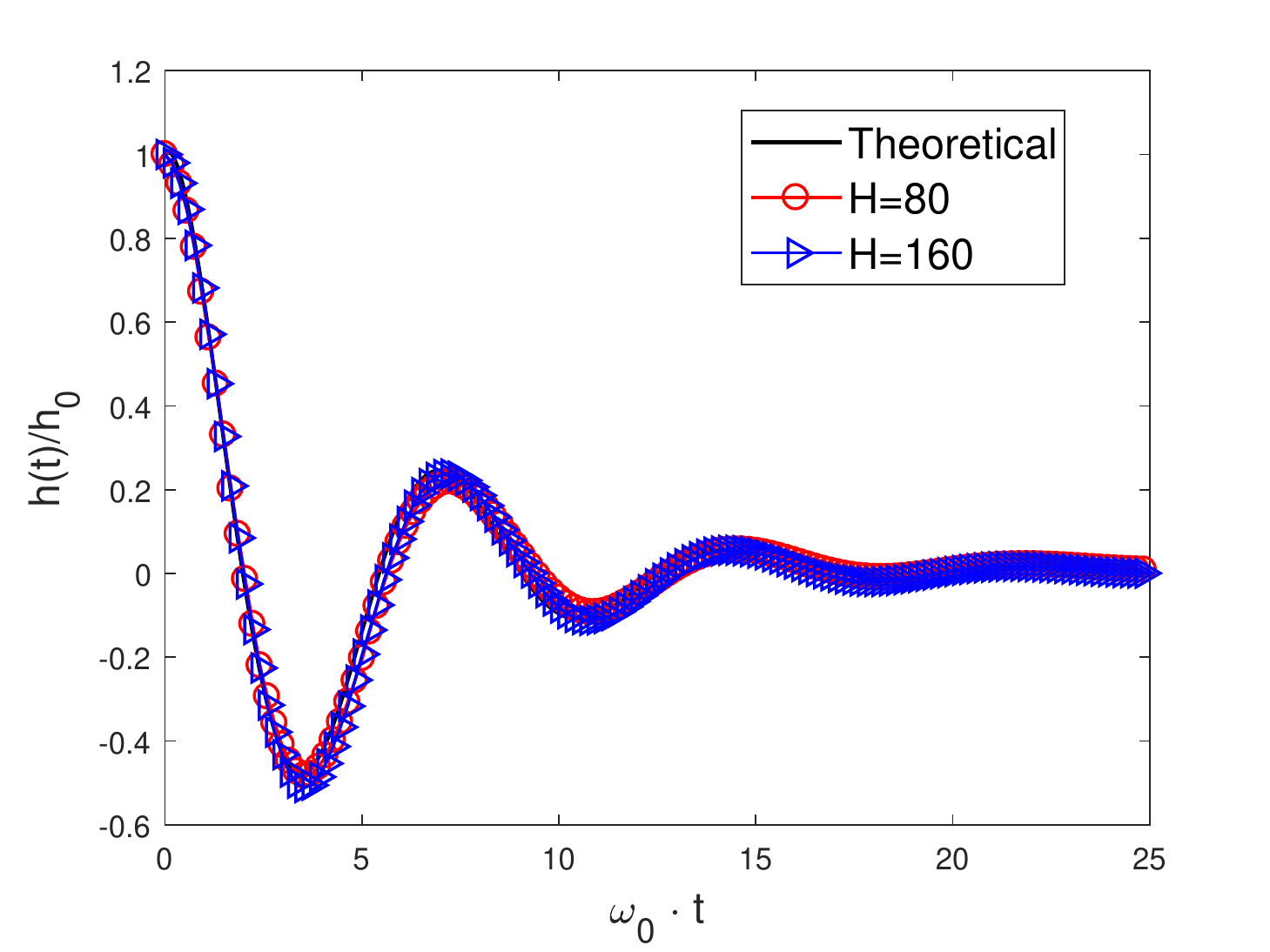}}~
\subfloat[]{\includegraphics[width=0.25\textwidth]{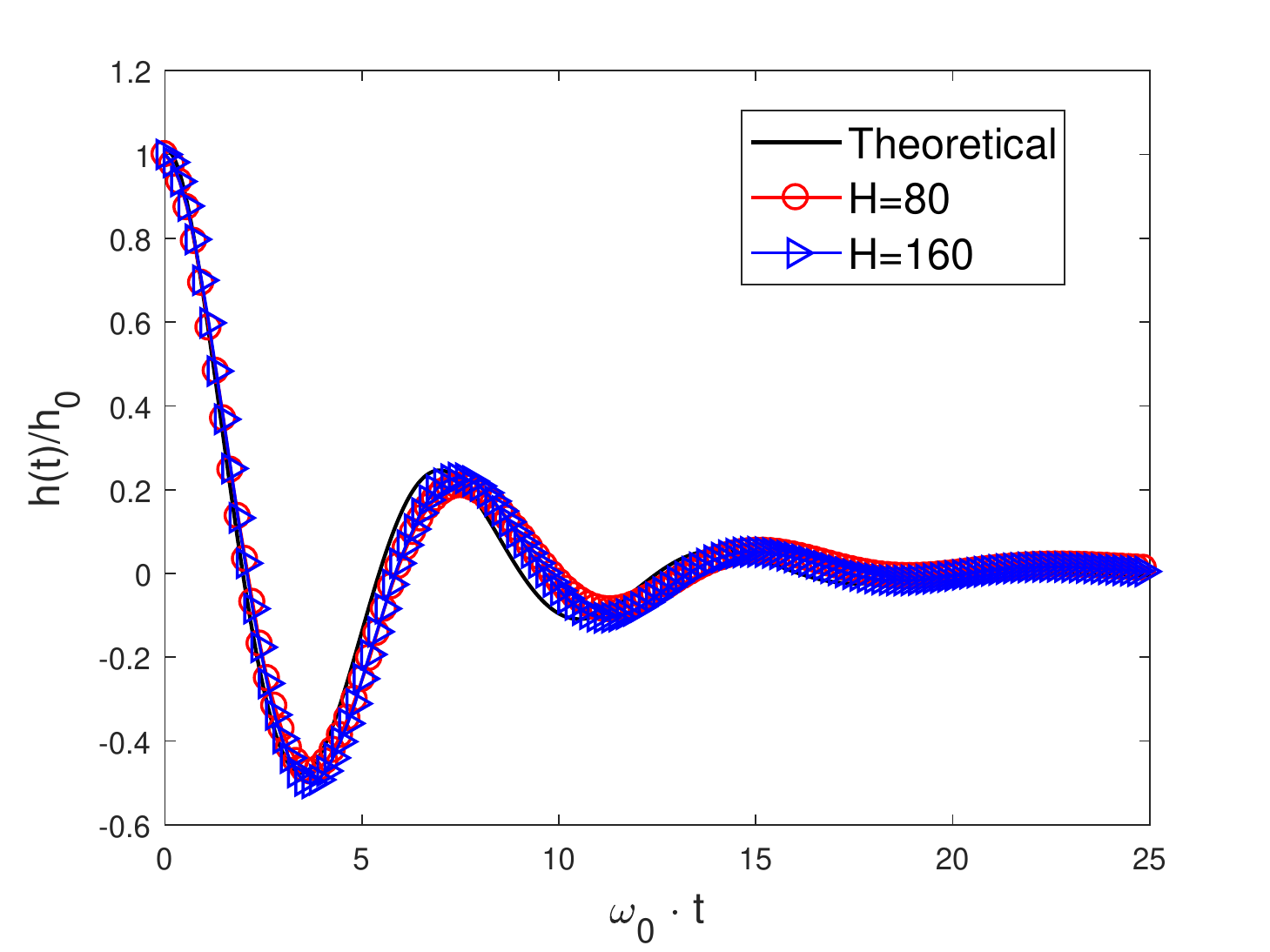}}~
\subfloat[]{\includegraphics[width=0.25\textwidth]{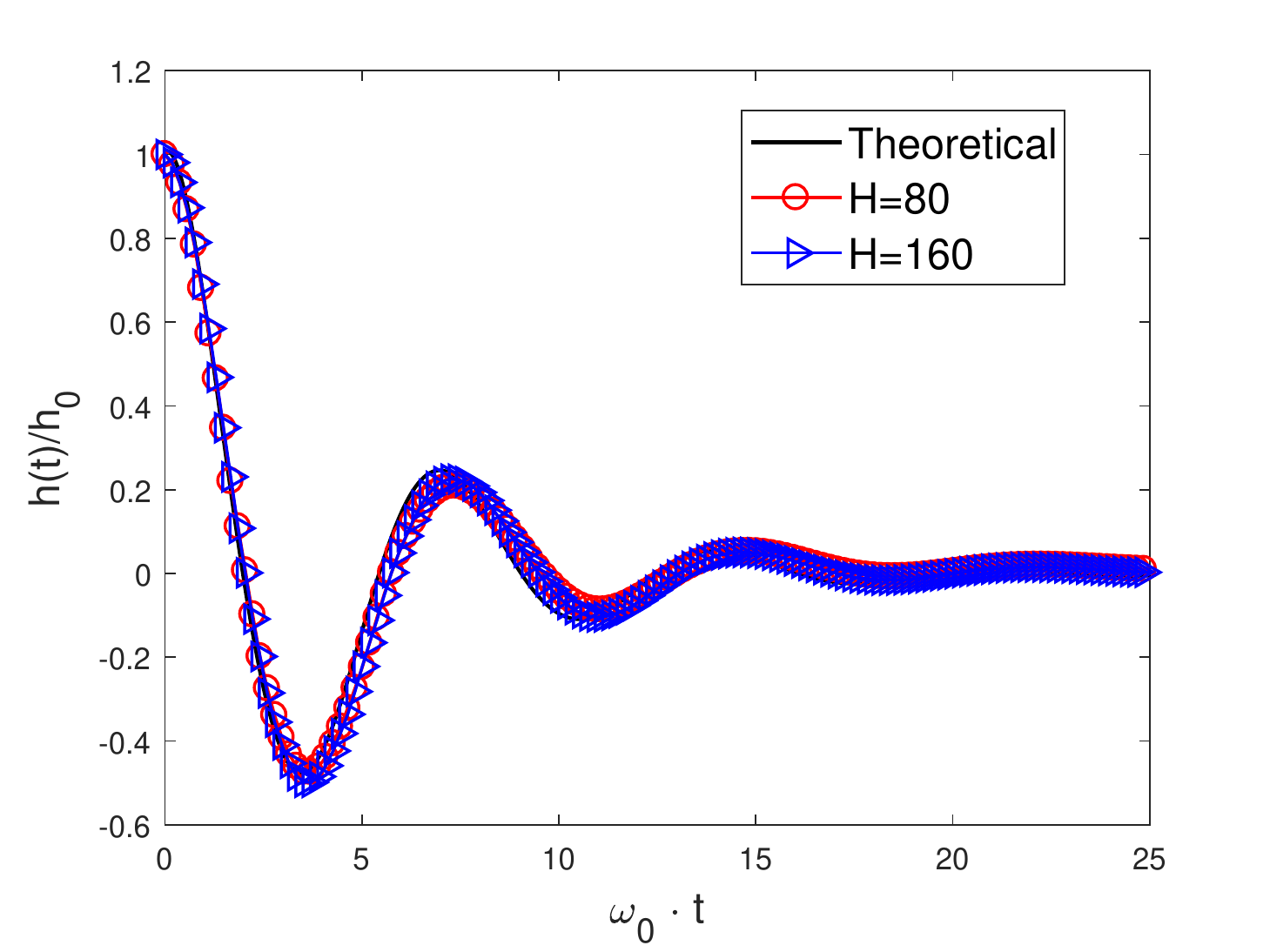}}~
\subfloat[]{\includegraphics[width=0.25\textwidth]{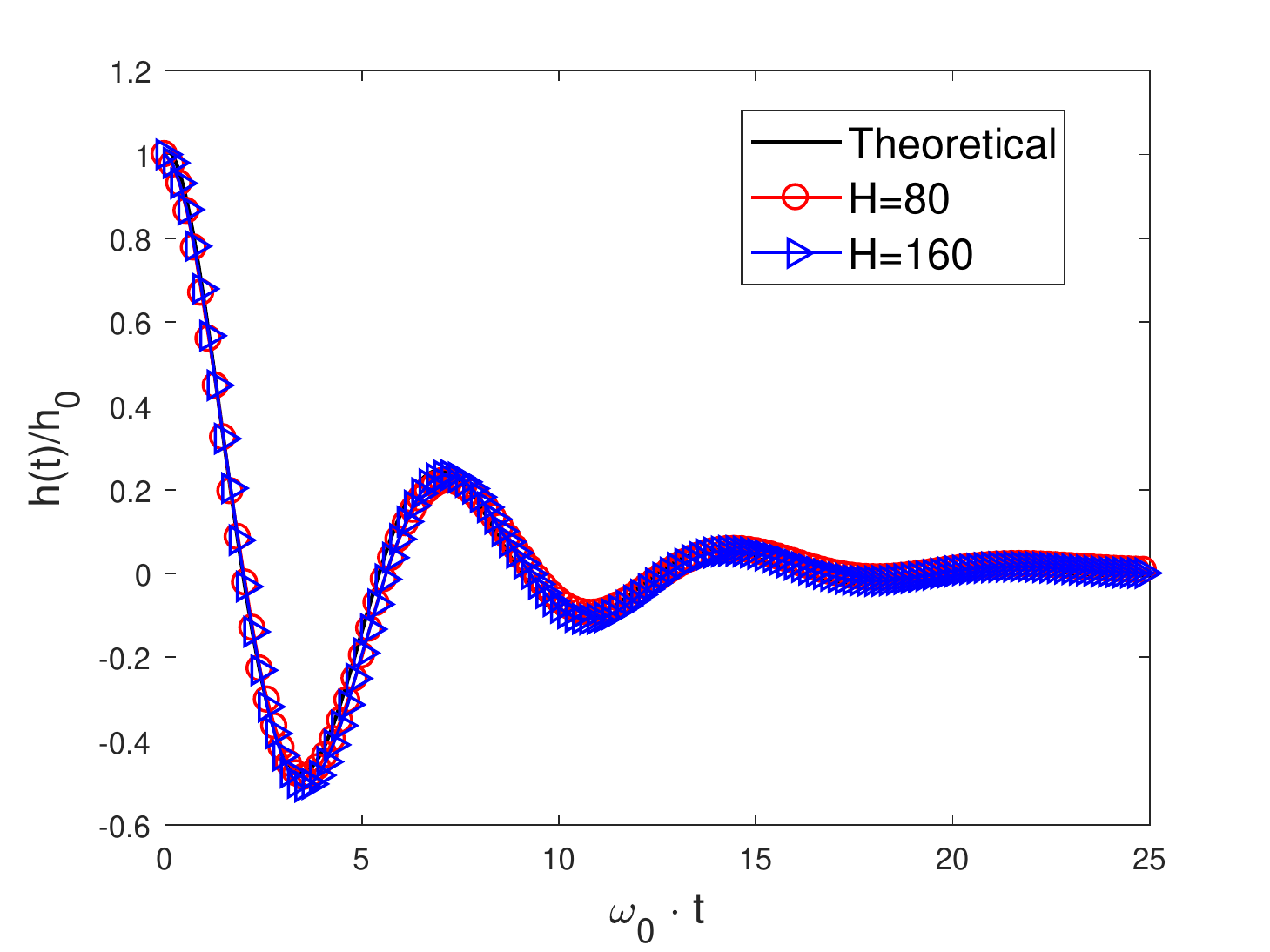}}\\
\caption{Time evolution of capillary wave amplitude with $\rho_1/\rho_2=10$  for (a) $\bm F_{stf-1}$,(b) $\bm F_{stf-2}$,(c) $\bm F_{cpf-1}$,(d) $\bm F_{cpf-2}$,(e) $\bm F_{pf-1}$,(f) $\bm F_{pf-2}$,(g) $\bm F_{csf-1}$ and (h) $\bm F_{csf-2}$. }
\label{fig:Capillary_sigma3}
\end{figure}

We further repeated the above simulations  with $\rho_1/\rho_2=10$. The results are shown in Fig.~\ref{fig:Capillary_sigma3}. In this case, all the results give a good agreement with the theoretical solutions.
  For quantitative comparison, the time averaged $L_2$-norm error for the wave amplitude is measured, which is defined as
\begin{equation}\label{eq:capilary_error}
E_2(h)=\sqrt{\frac{\omega_0}{25}\int_{0}^{\frac{25}{\omega_0}}|\bar{h}(t)-\bar{h}_{exact}(t)|^2dt}.
\end{equation}
Table~\ref{tab:time_normed_error} presents the time averaged $L_2$-norm error of wave amplitude, from which we can observe that  all the averaged errors  monotonically decrease as the numerical grid increases. Among the results, it can be found that the results given by $\bm F_{pf-1}$ and $\bm F_{csf-2}$ are closer to the analytical solutions.

\begin{table}[!htb]
  \centering
  \caption{The time averaged $L_2$-norm error for capillary wave time evolution. }\label{tab:time_normed_error}
\setlength{\tabcolsep}{3mm}{%
\begin{tabular}{cccccccccc}
\hline
 $\frac{\rho_1}{\rho_2}$& $grid$& $\bm F_{stf-1}$ &$\bm F_{stf-2}$& $\bm F_{cpf-1}$& $\bm F_{cpf-2}$& $\bm F_{pf-1}$&$\bm F_{pf-2}$ &$\bm F_{csf-1}$&$\bm F_{csf-2}$ \\
\hline
1    & H=80    &0.0905     &0.0906   &0.0795    &0.0609
               &0.055      &0.1220   &0.0775    &0.0503   \\

1    & H=160   &0.0612     &0.0613   &0.0310    &0.0314
               &0.0275     &0.0950   &0.0559    &0.0231    \\

10    &H=80   &0.0384     &0.0384   &0.0258    &0.0259
              &0.0261     &0.0498   &0.0359    &0.0232    \\

10    & H=160   &0.0341     &0.0341   &0.0233    &0.0233
                &0.0220     &0.0466   &0.0343    &0.0209    \\
\hline
\end{tabular}}
\end{table}

\subsection{Rising bubble}
\begin{figure}
\centering
\subfloat{\includegraphics[width=0.4\textwidth,trim=0 0 0 5,clip]{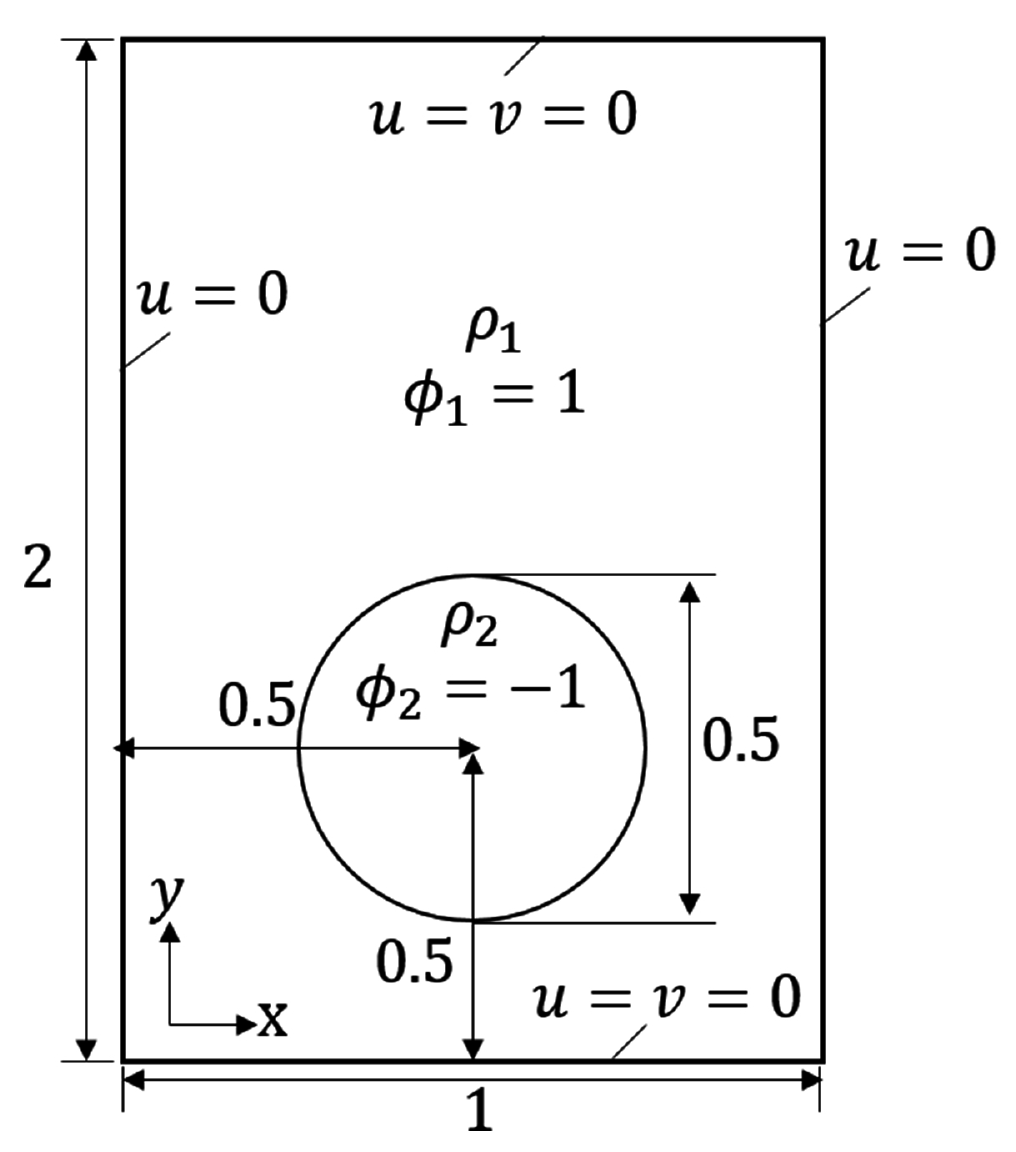}}
\caption{Initial configuration for the rising bubble.}
\label{fig:Rising_initial}
\end{figure}

We now examine the performance of different interfacial force formulations by simulating a bubble rising in a two-dimensional domain, which was also simulated by Hysing \emph{et al.}~\cite{hysing2009quantitative} and S. Aland \emph{et al}.~\cite{aland2012benchmark}.
Although  no analytical solution is available for this problem, some numerical results were presented in ~\cite{hysing2009quantitative}.
The results from group 3 on the finest grids in \cite{hysing2009quantitative} are taken as the reference solutions.
The schematic of the domain is shown in Fig~\ref{fig:Rising_initial}. Initially, a bubble with radius $R=0.25\text{m}$ is placed at $(0.5\text{m}, 0.5\text{m})$ in a rectangle domain of  $2\text{m}\times 1\text{m}$. For the velocities, no-slip velocity boundary conditions are applied to the top and bottom boundaries and free-slip boundary conditions are imposed on the side boundaries. The gravitational force acts in the opposite direction of the vertical direction. Two uniform grids of $120 \times 240$
and $240\times 480$  are employed. The fluid parameters are listed in Table~\ref{tab:test_Paras}.
The  related non-dimensional numbers are given by
\begin{equation}
\text{Re}=\frac{\rho_1 U_g L}{\mu_1},\quad \text{Eo}=\frac{\rho_1 U_g^2 L}{\sigma}, \quad \text{Mo}=\frac{\text{Eo}^3}{\text{Re}^4}=\frac{U_g^2\mu_1^4}{2\rho_1 \sigma^3 R}
\end{equation}
where  $U_g=\sqrt{2Rg}$ and $L$ are  the reference velocity and length, respectively.

For comparison, the benchmark quantities, including bubble shape at $t=3s$, rising velocity, center of mass and circularity are measured  by
\begin{equation}\label{eq:quantify-yc}
  y_c=\frac{\int_{\Omega}(1-\phi)y d\bm x}{\int_{\Omega} (1-\phi) d\bm x},
\end{equation}
\begin{equation}\label{eq:quantify-uc}
  v_c=\frac{\int_{\Omega}(1-\phi) v d\bm x}{\int_{\Omega} (1-\phi ) d\bm x},
\end{equation}
\begin{equation}\label{eq:circularity}
C=\frac{\text{perimenter of area-equivalent circle}}{\text{perimeter of bubble} }=\frac{2\sqrt{\int_{\phi<0} \pi d x  } }{P_b}
\end{equation}
where $v$ is the velocity component in the vertical direction and $P_b$ is obtained by integration over the contour line at $\phi=0$ in Matlab.

\begin{figure}[!htb]
\centering
\subfloat[]{\includegraphics[width=0.25\textwidth]{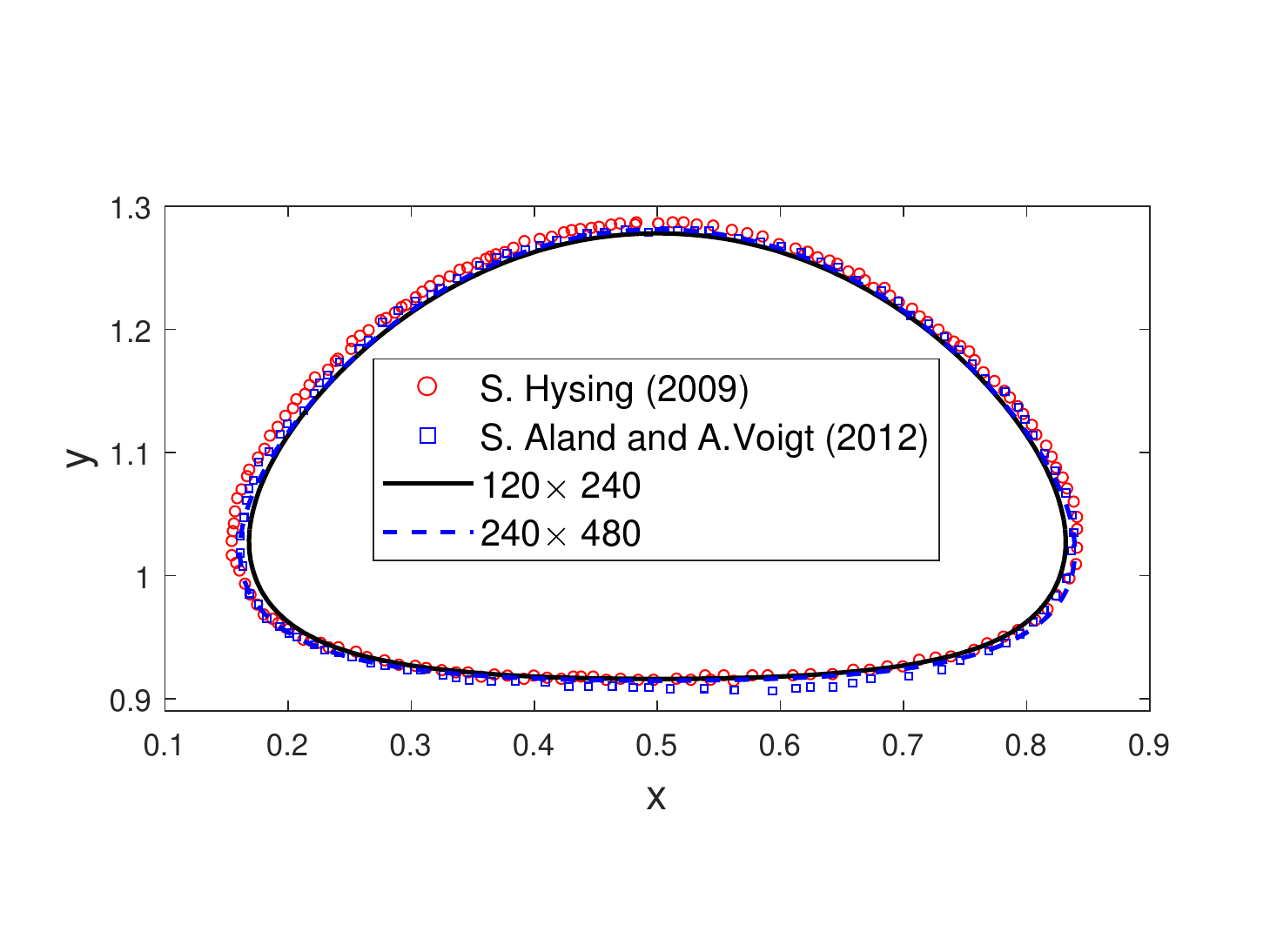}}~
\subfloat[]{\includegraphics[width=0.25\textwidth]{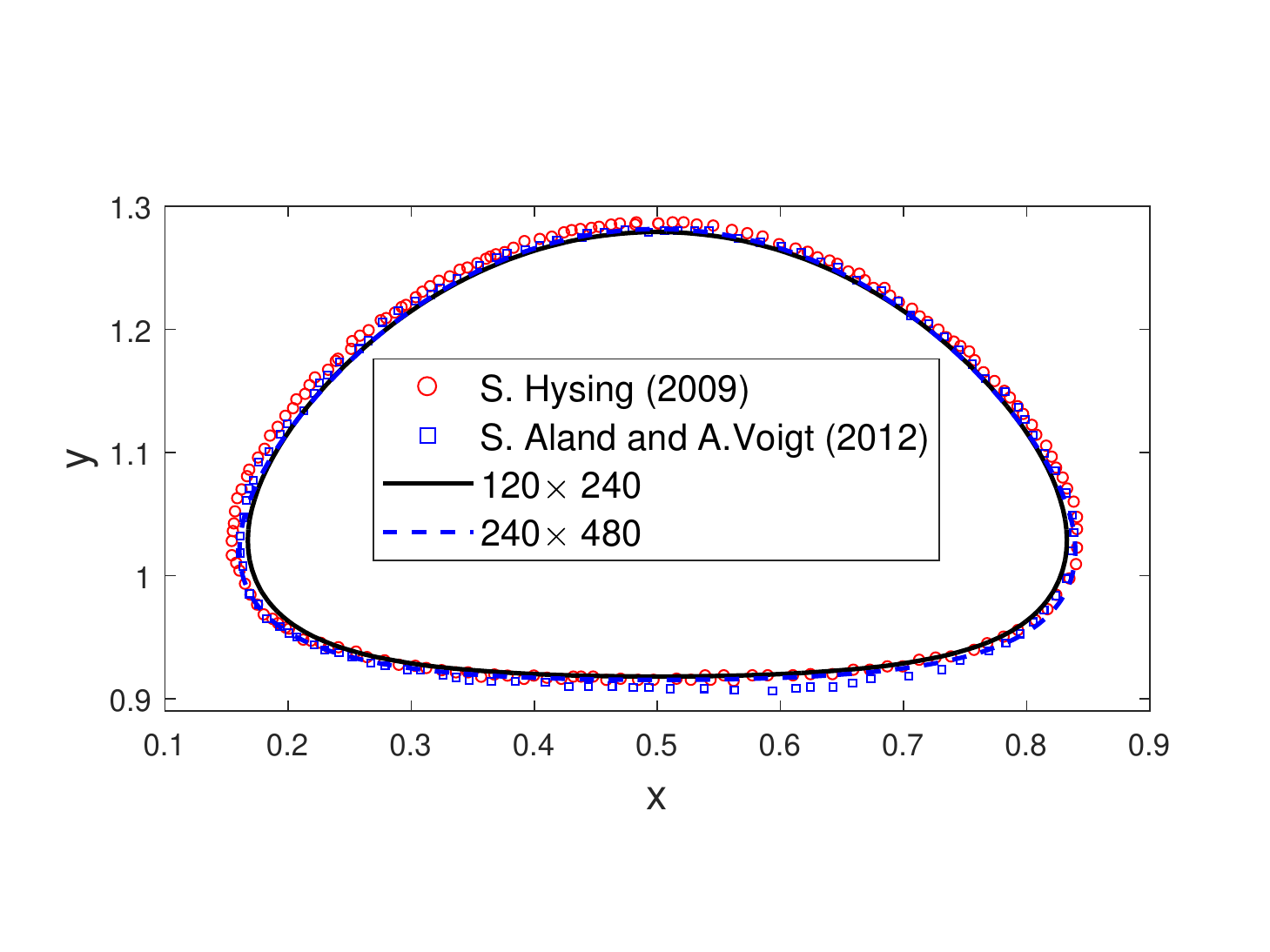}}~
\subfloat[]{\includegraphics[width=0.25\textwidth]{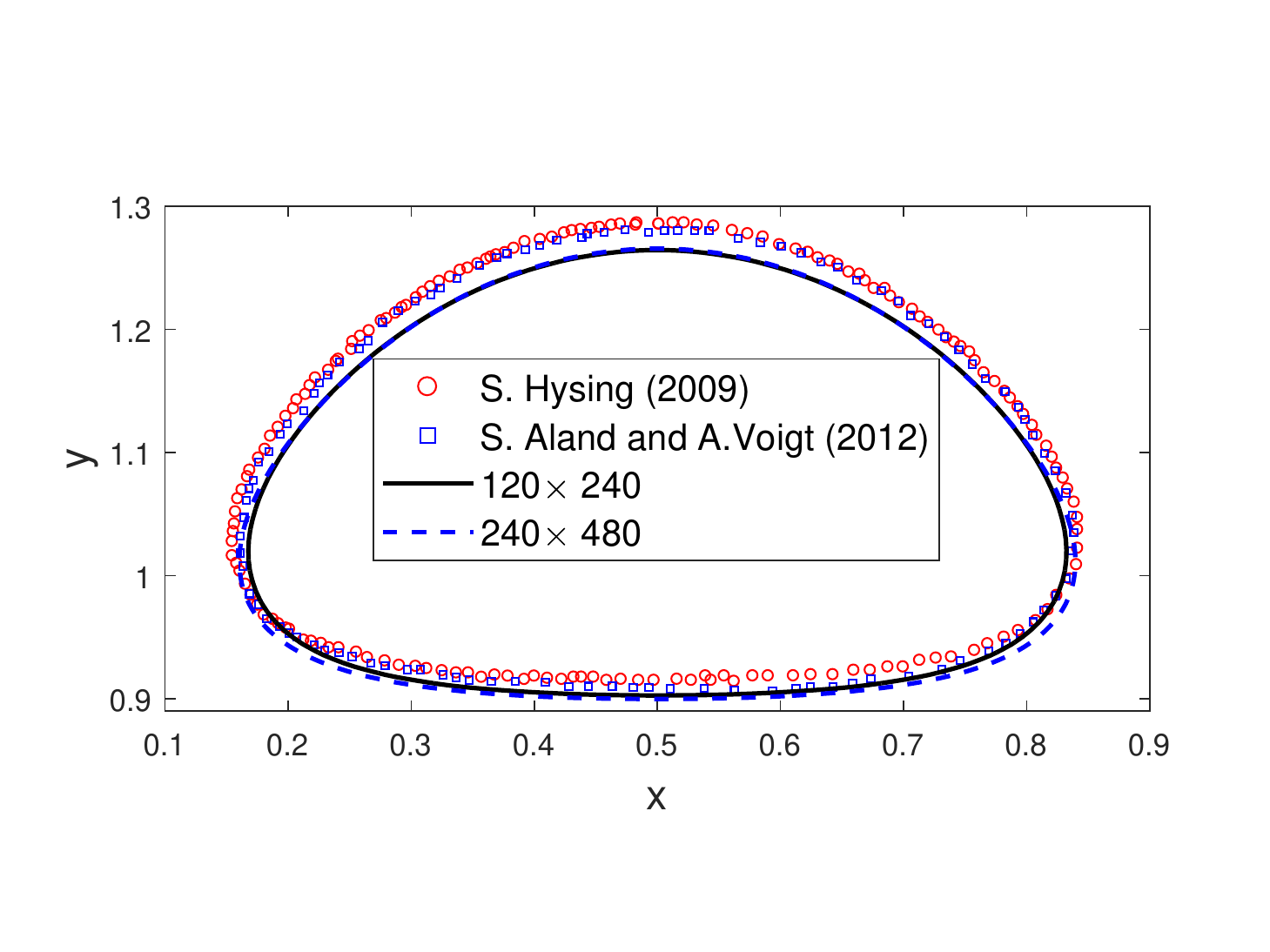}}~
\subfloat[]{\includegraphics[width=0.25\textwidth]{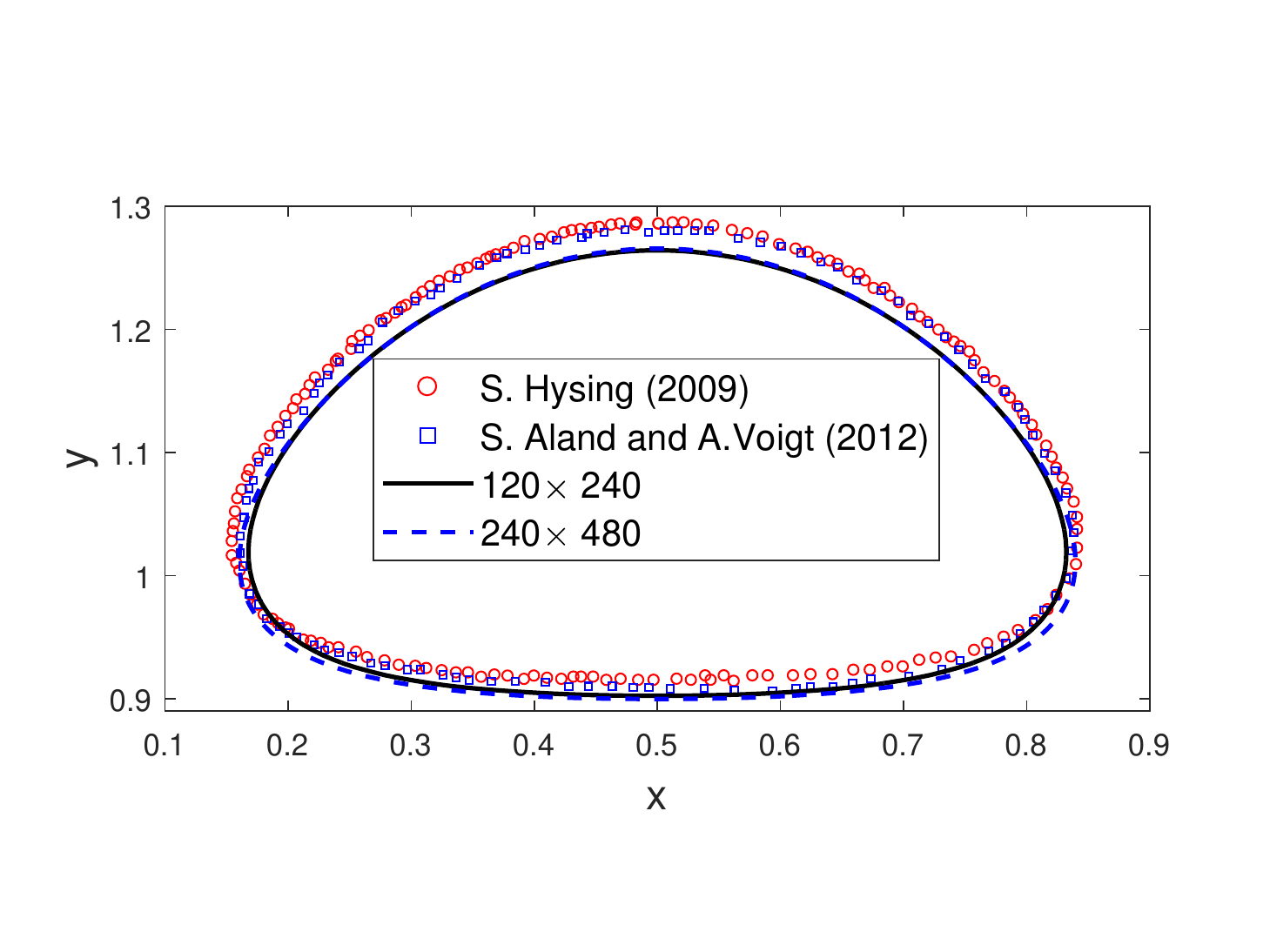}} \\
\subfloat[]{\includegraphics[width=0.25\textwidth]{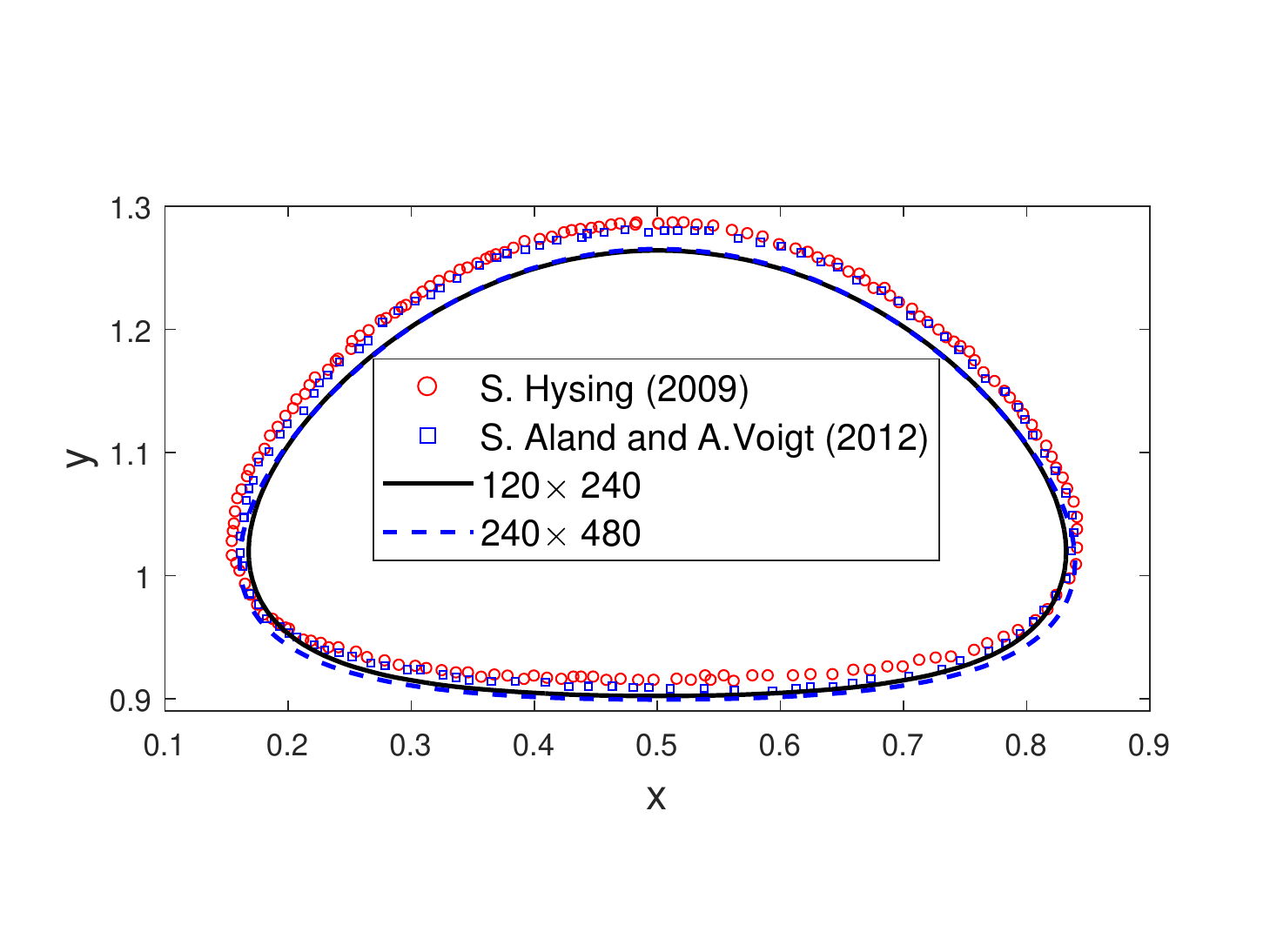}}~
\subfloat[]{\includegraphics[width=0.25\textwidth]{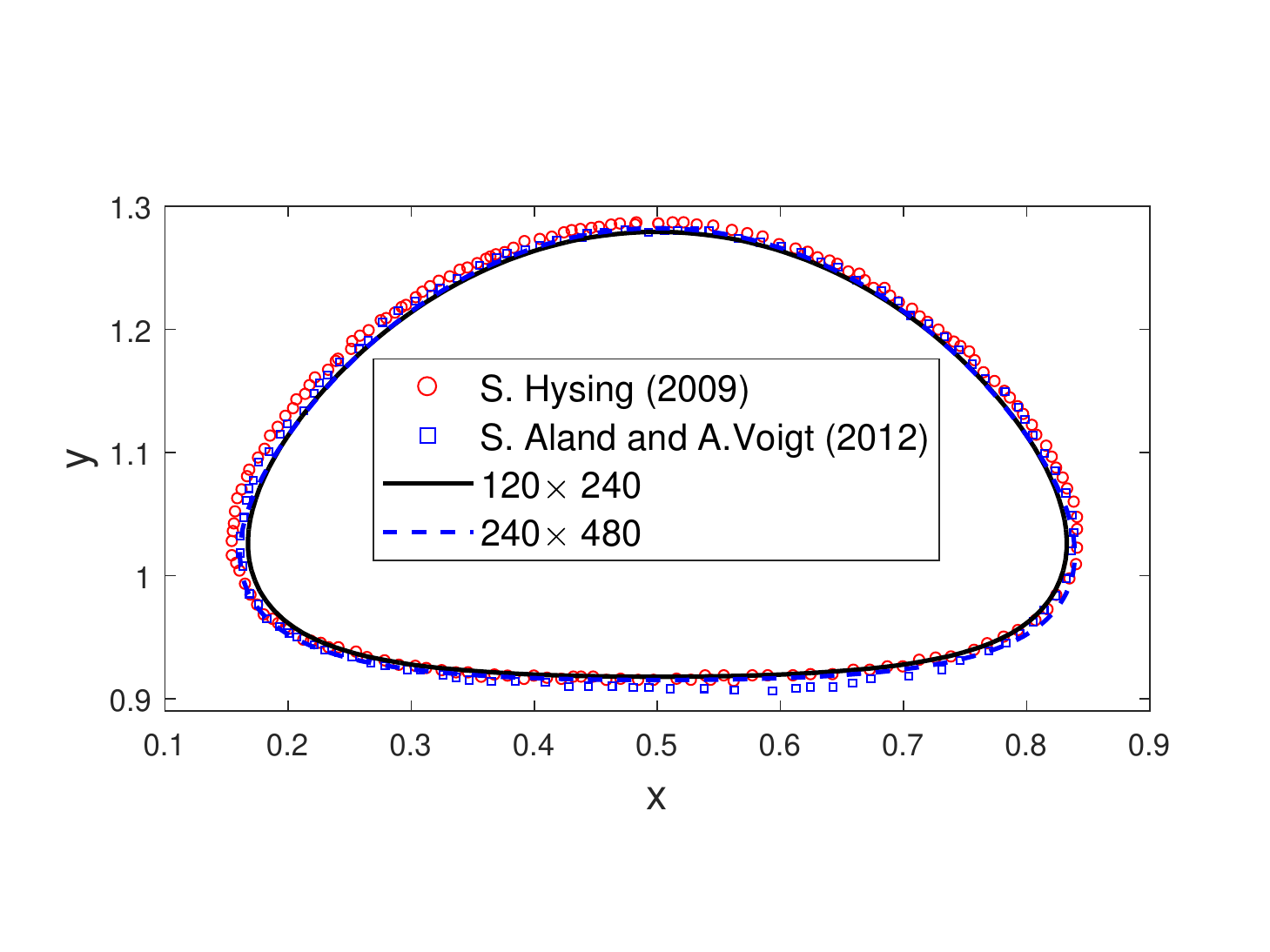}}~
\subfloat[]{\includegraphics[width=0.25\textwidth]{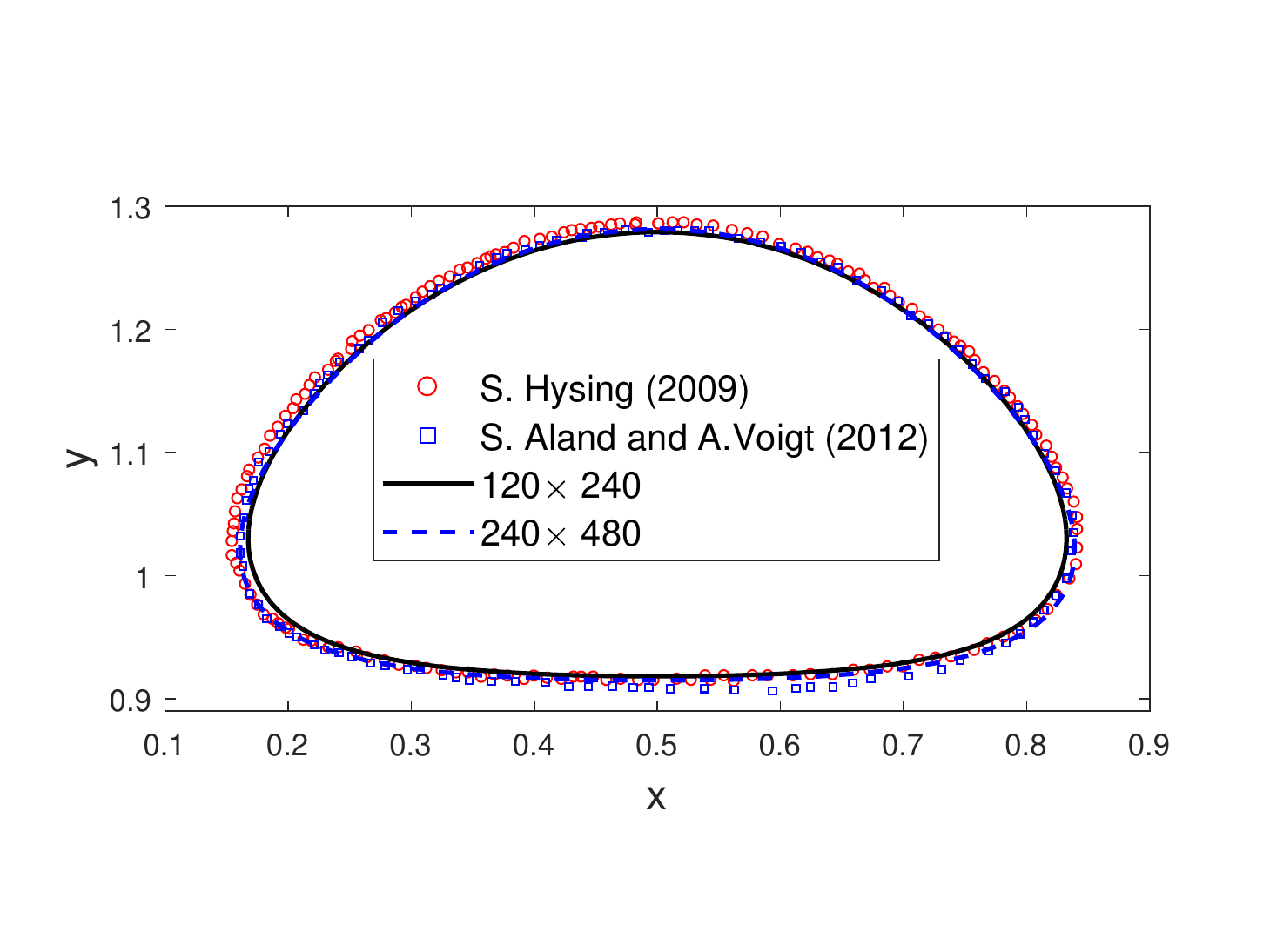}}~
\subfloat[]{\includegraphics[width=0.25\textwidth]{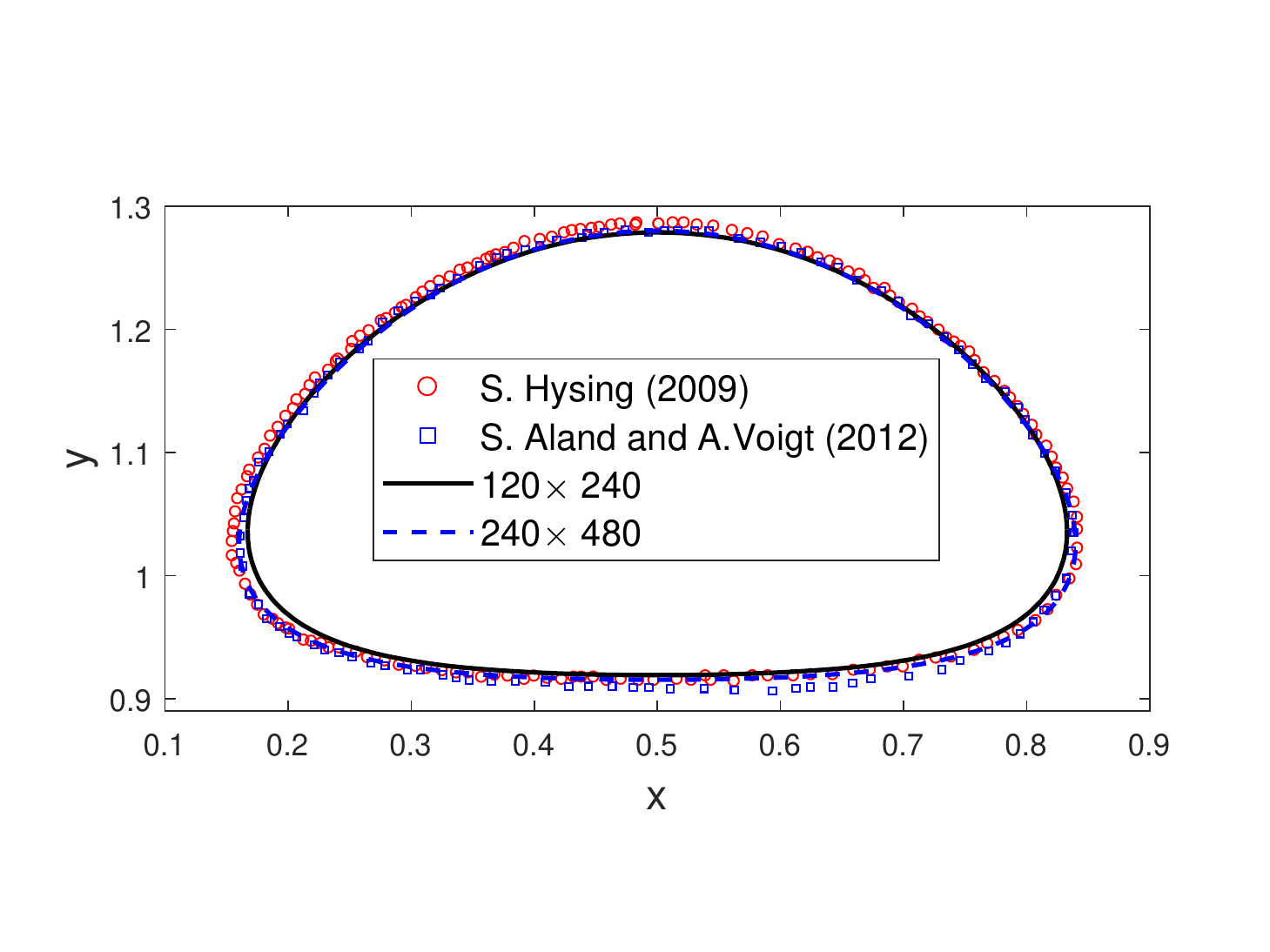}}~
\caption{ Bubble shapes at time $t=3s$ for (a) $\bm F_{stf-1}$,(b) $\bm F_{stf-2}$,(c) $\bm F_{cpf-1}$,(d) $\bm F_{cpf-2}$,(e) $\bm F_{pf-1}$,(f) $\bm F_{pf-2}$,(g) $\bm F_{csf-1}$ and (h) $\bm F_{csf-2}$. }
\label{fig:bubbleshape}
\end{figure}

\begin{figure}[!htb]
\centering
\subfloat[]{\includegraphics[width=0.25\textwidth]{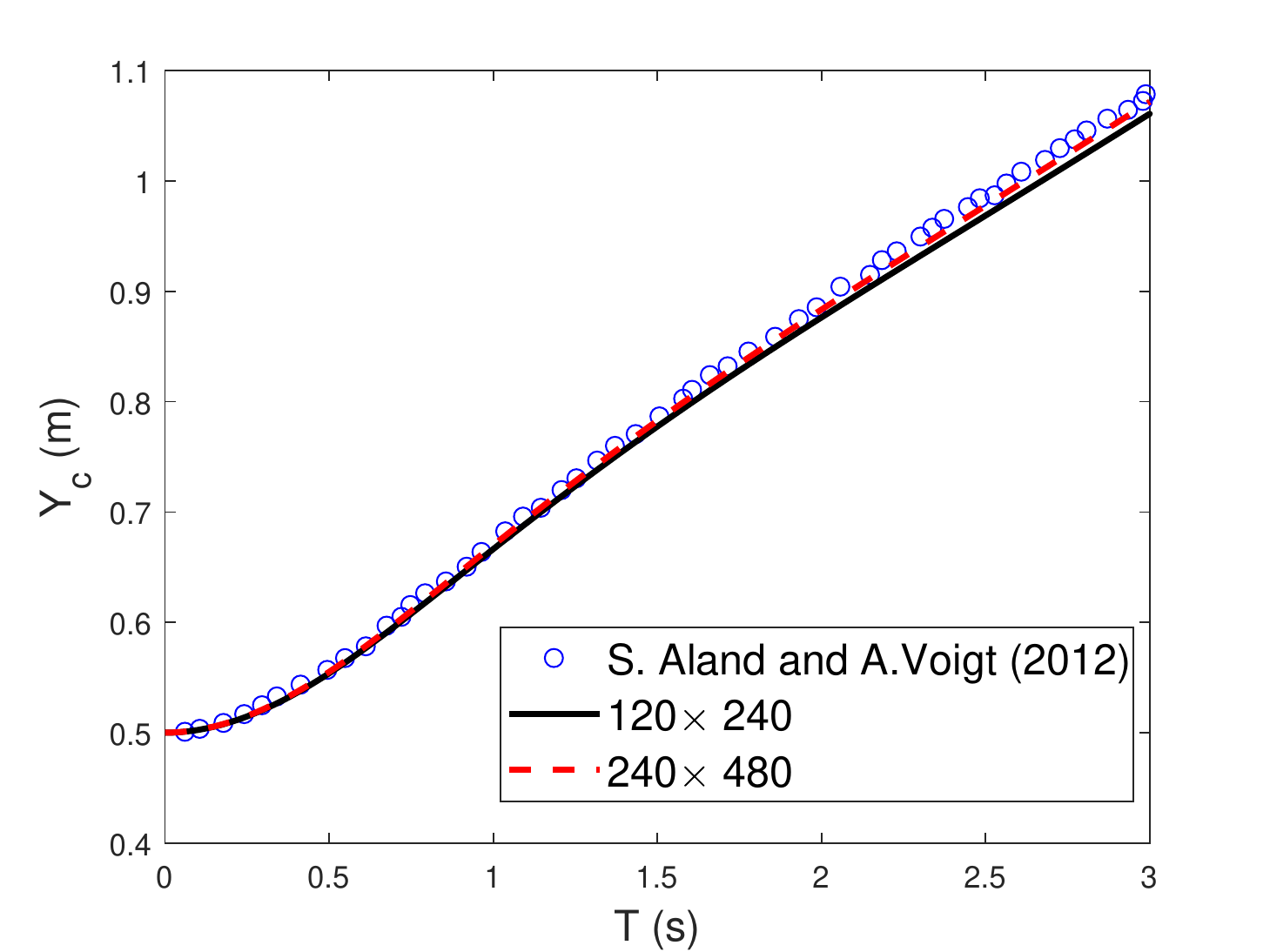}}~
\subfloat[]{\includegraphics[width=0.25\textwidth]{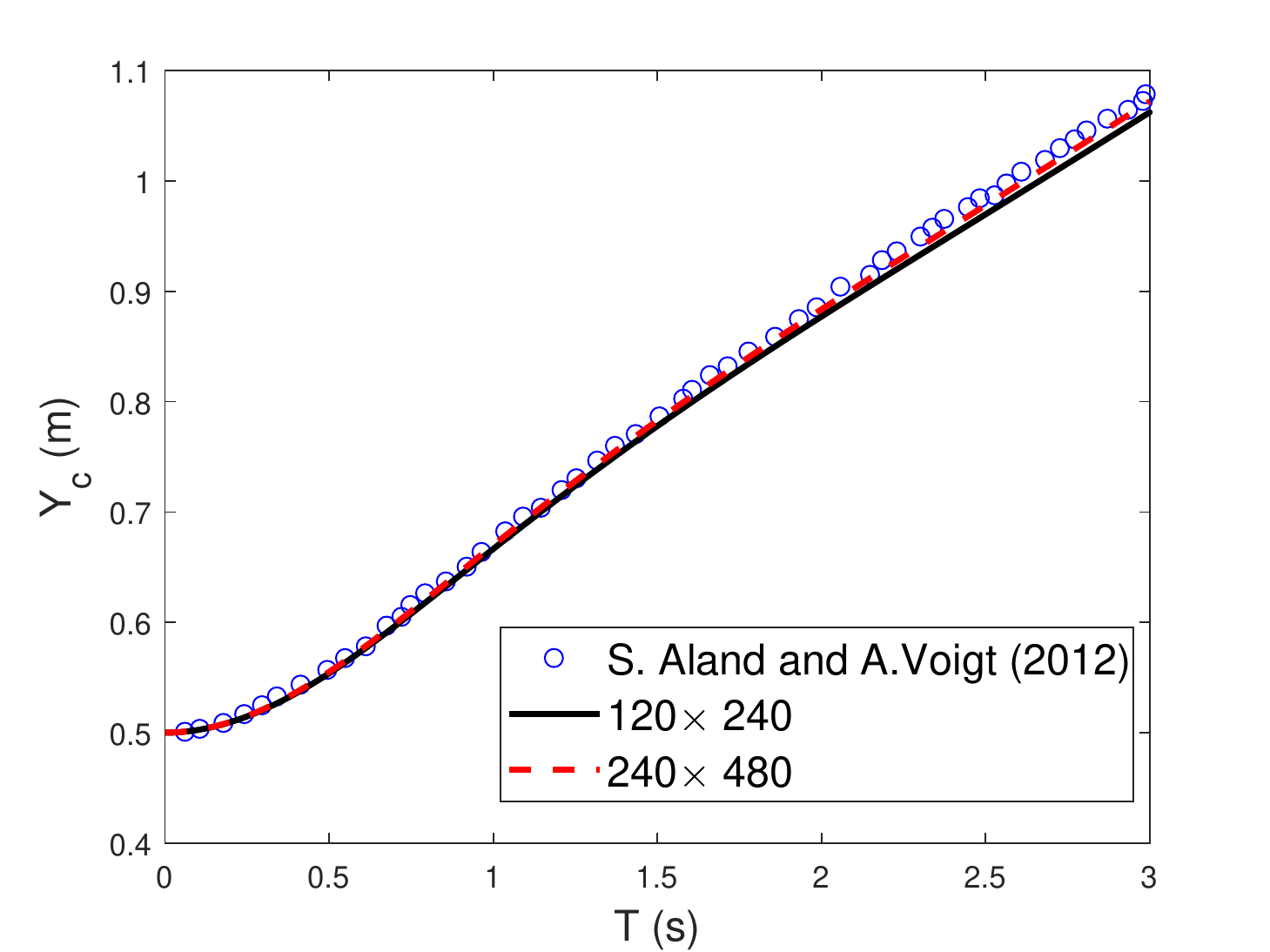}}~
\subfloat[]{\includegraphics[width=0.25\textwidth]{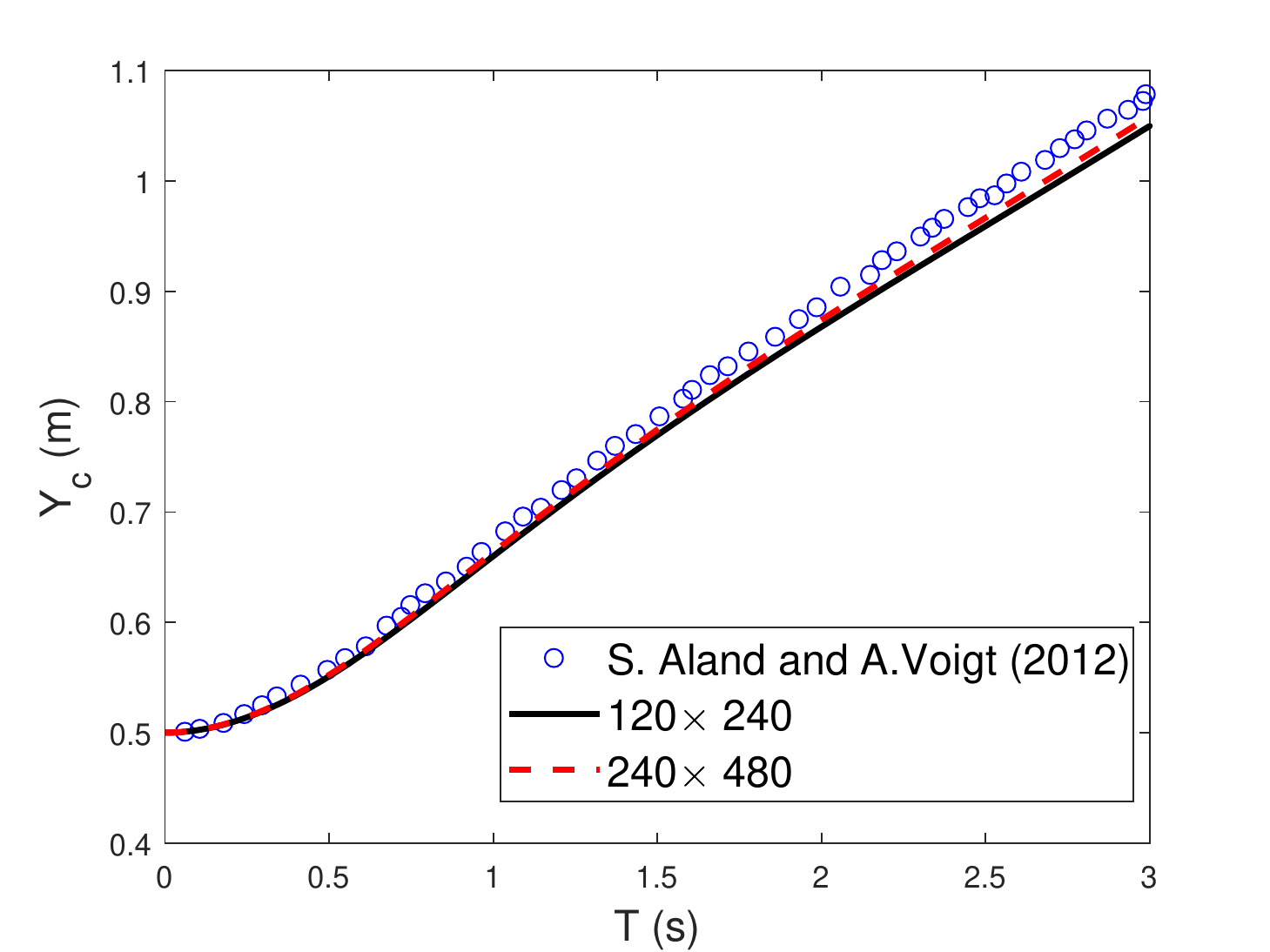}}~
\subfloat[]{\includegraphics[width=0.25\textwidth]{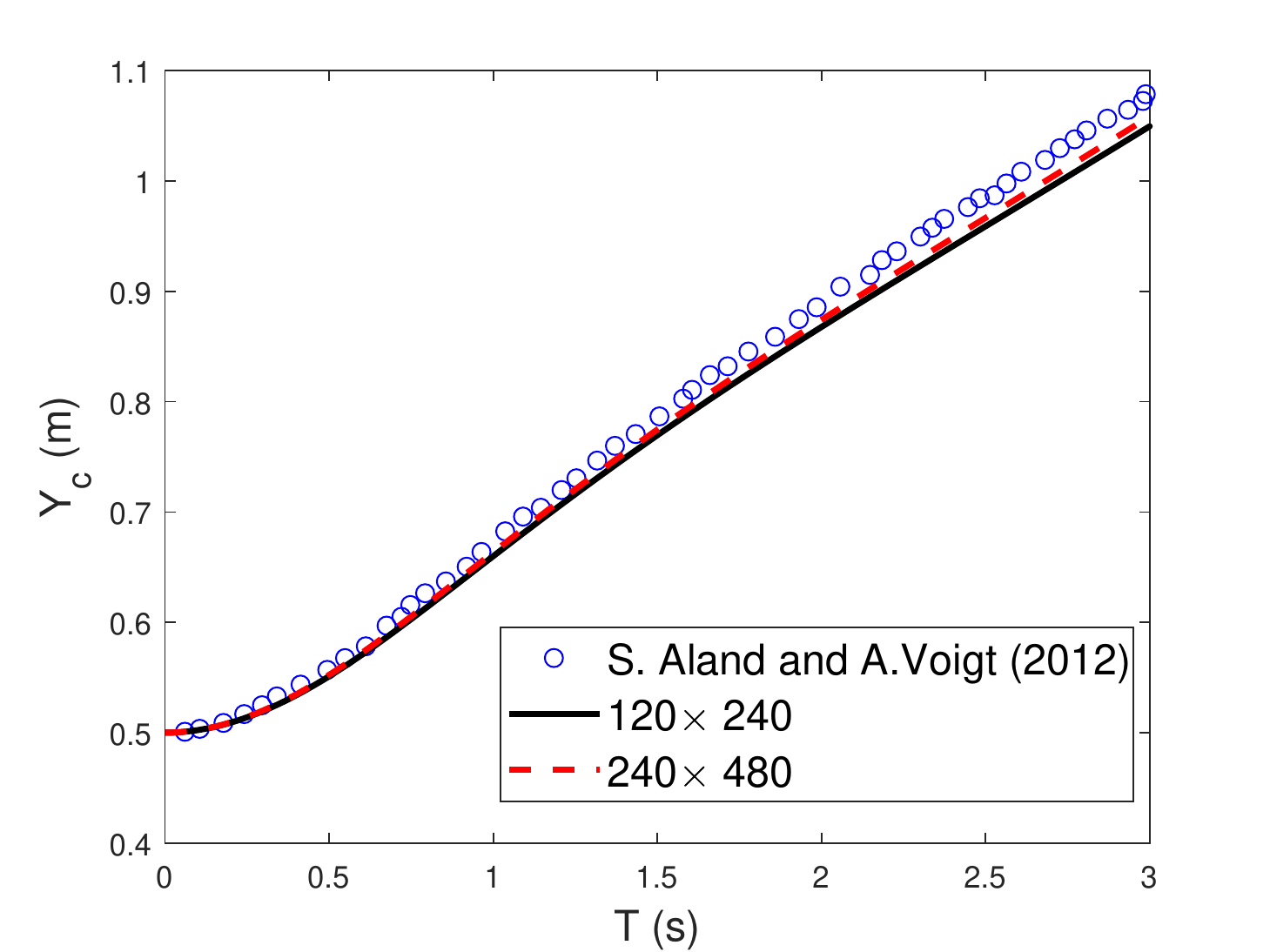}}\\
\subfloat[]{\includegraphics[width=0.25\textwidth]{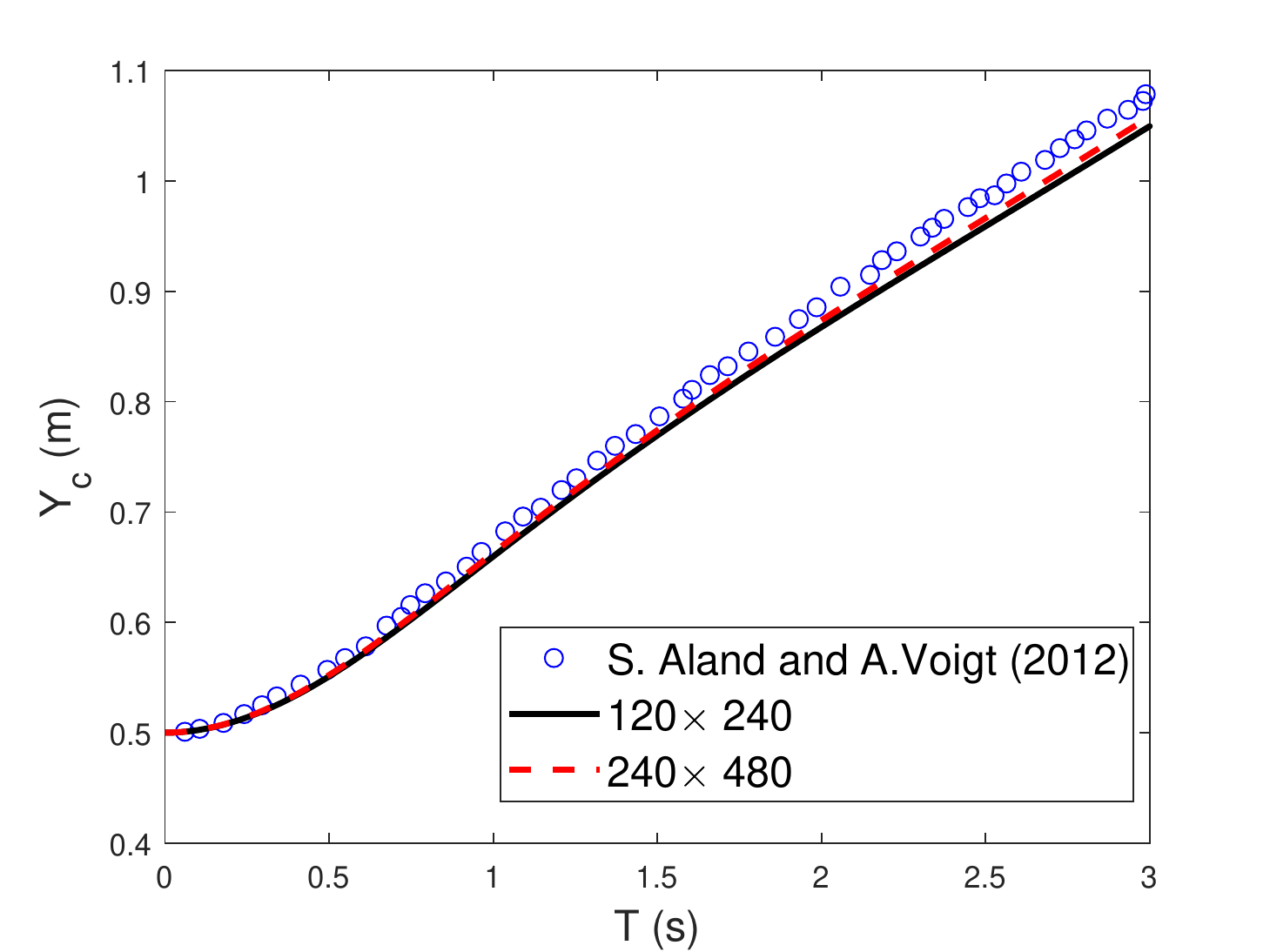}}~
\subfloat[]{\includegraphics[width=0.25\textwidth]{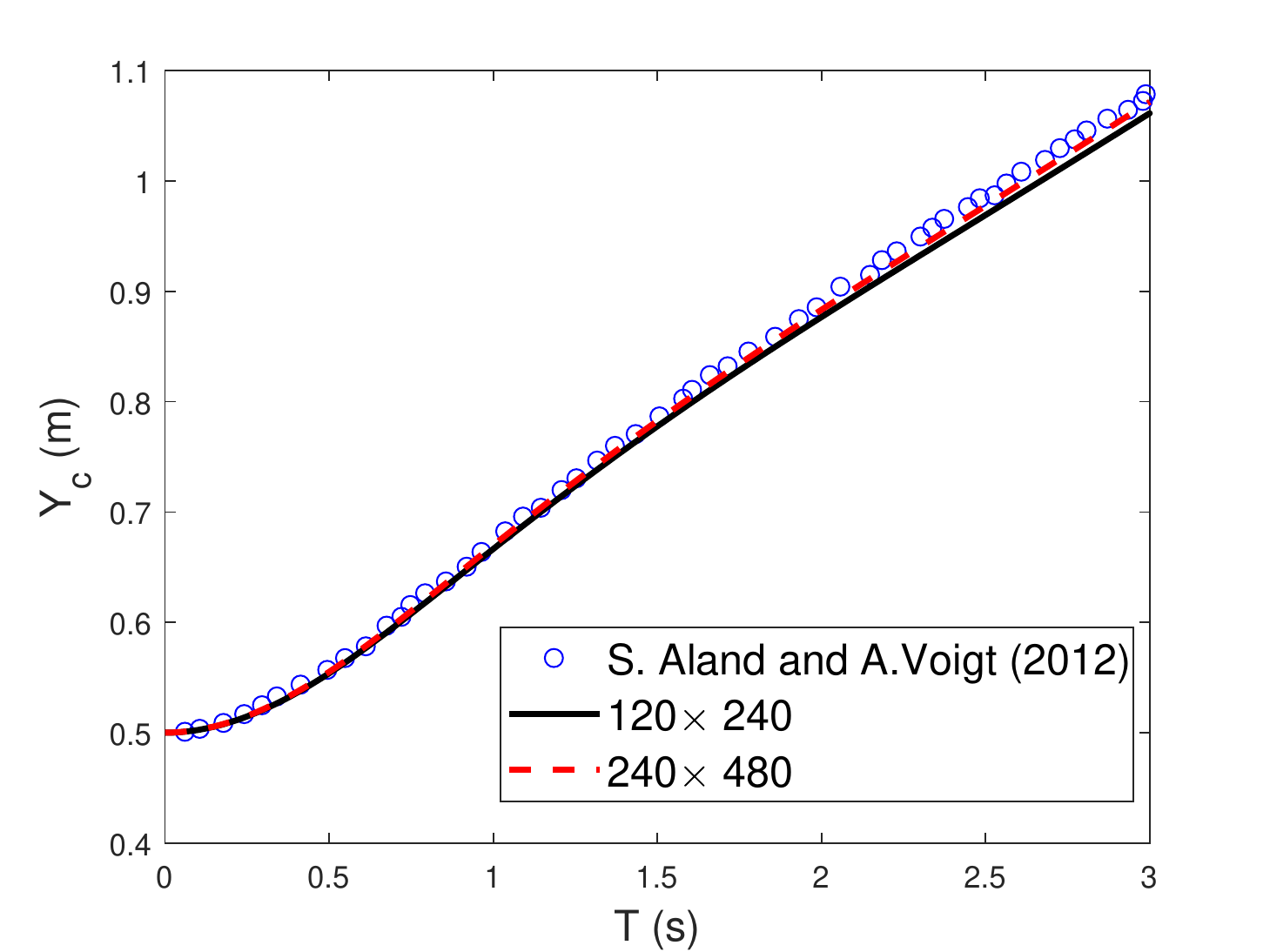}}~
\subfloat[]{\includegraphics[width=0.25\textwidth]{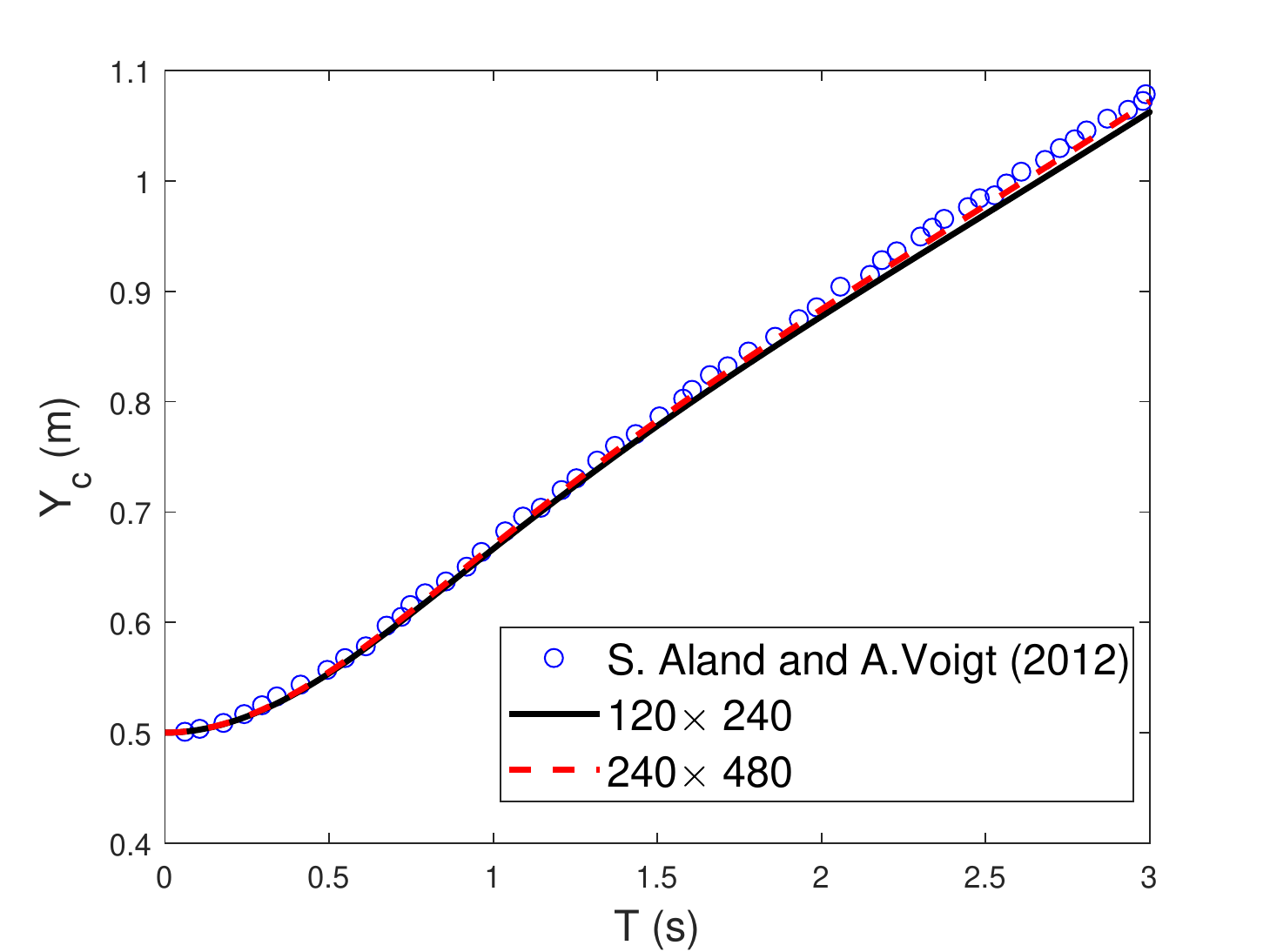}}~
\subfloat[]{\includegraphics[width=0.25\textwidth]{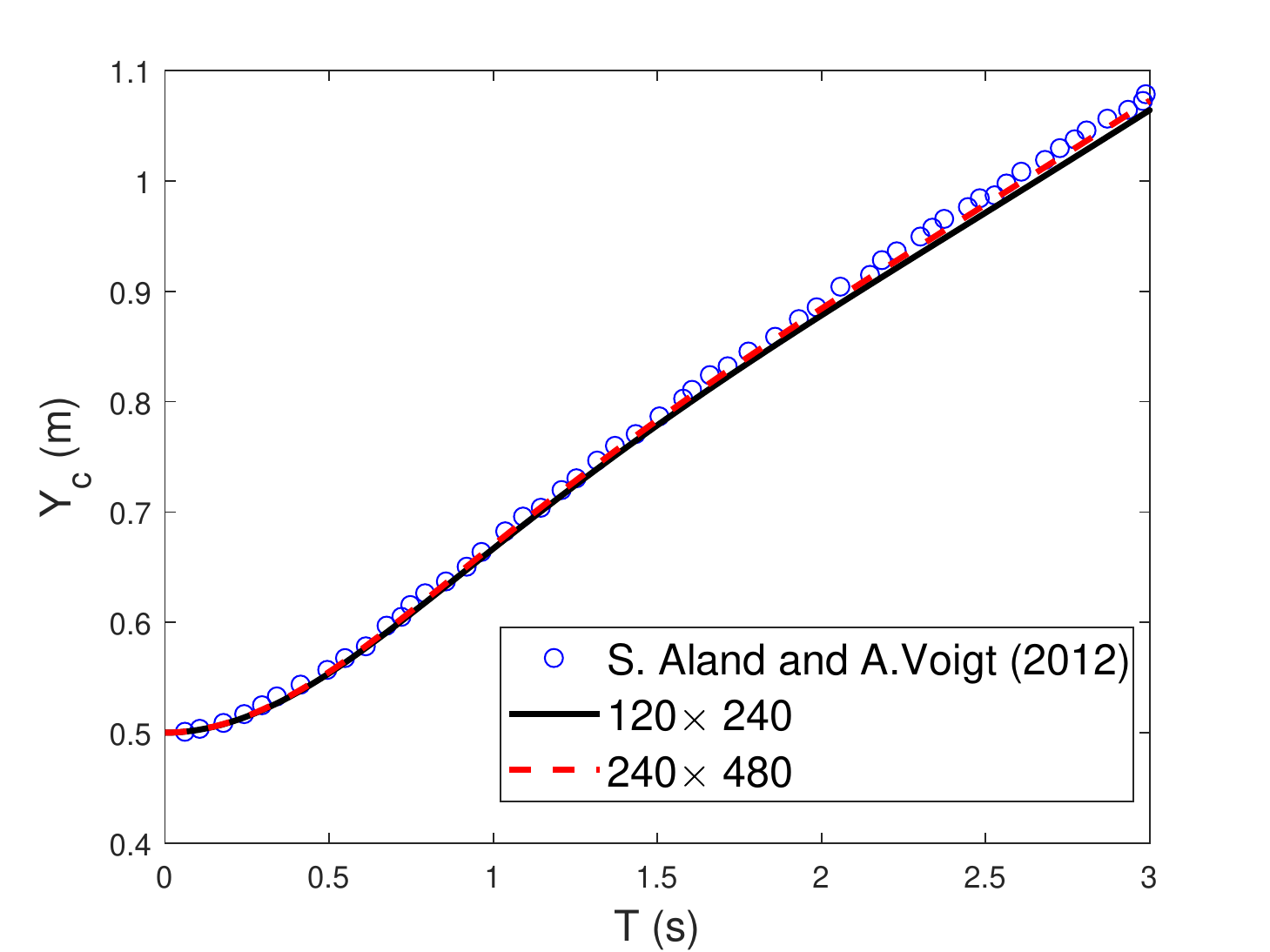}}~
\caption{The evolution of the center of mass for  (a) $\bm F_{stf-1}$,(b) $\bm F_{stf-2}$,(c) $\bm F_{cpf-1}$,(d) $\bm F_{cpf-2}$,(e) $\bm F_{pf-1}$,(f) $\bm F_{pf-2}$,(g) $\bm F_{csf-1}$ and (h) $\bm F_{csf-2}$. }
\label{fig:bubbleYc}
\end{figure}

\begin{figure}[!htb]
\centering
\subfloat[]{\includegraphics[width=0.25\textwidth]{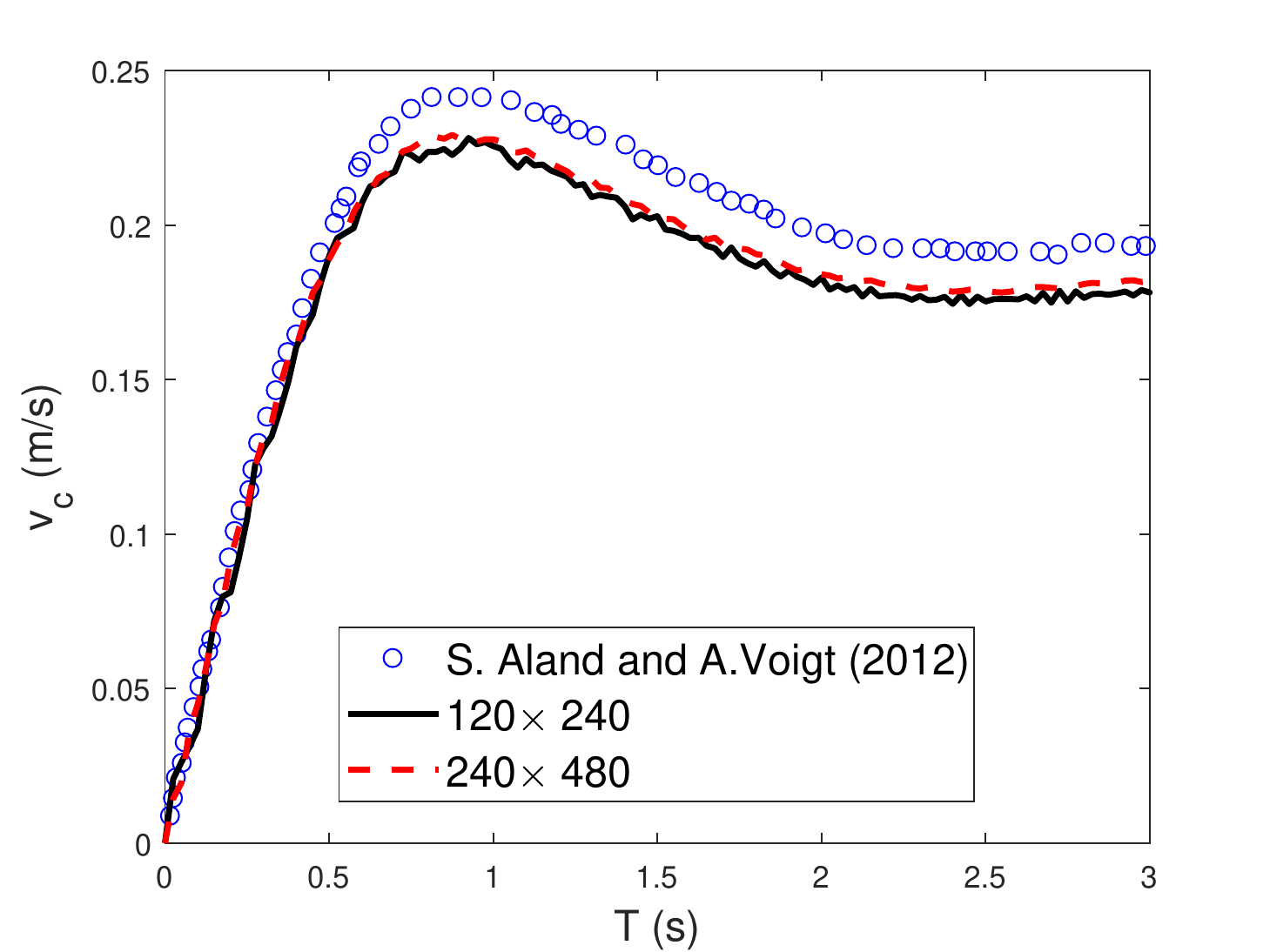}}~
\subfloat[]{\includegraphics[width=0.25\textwidth]{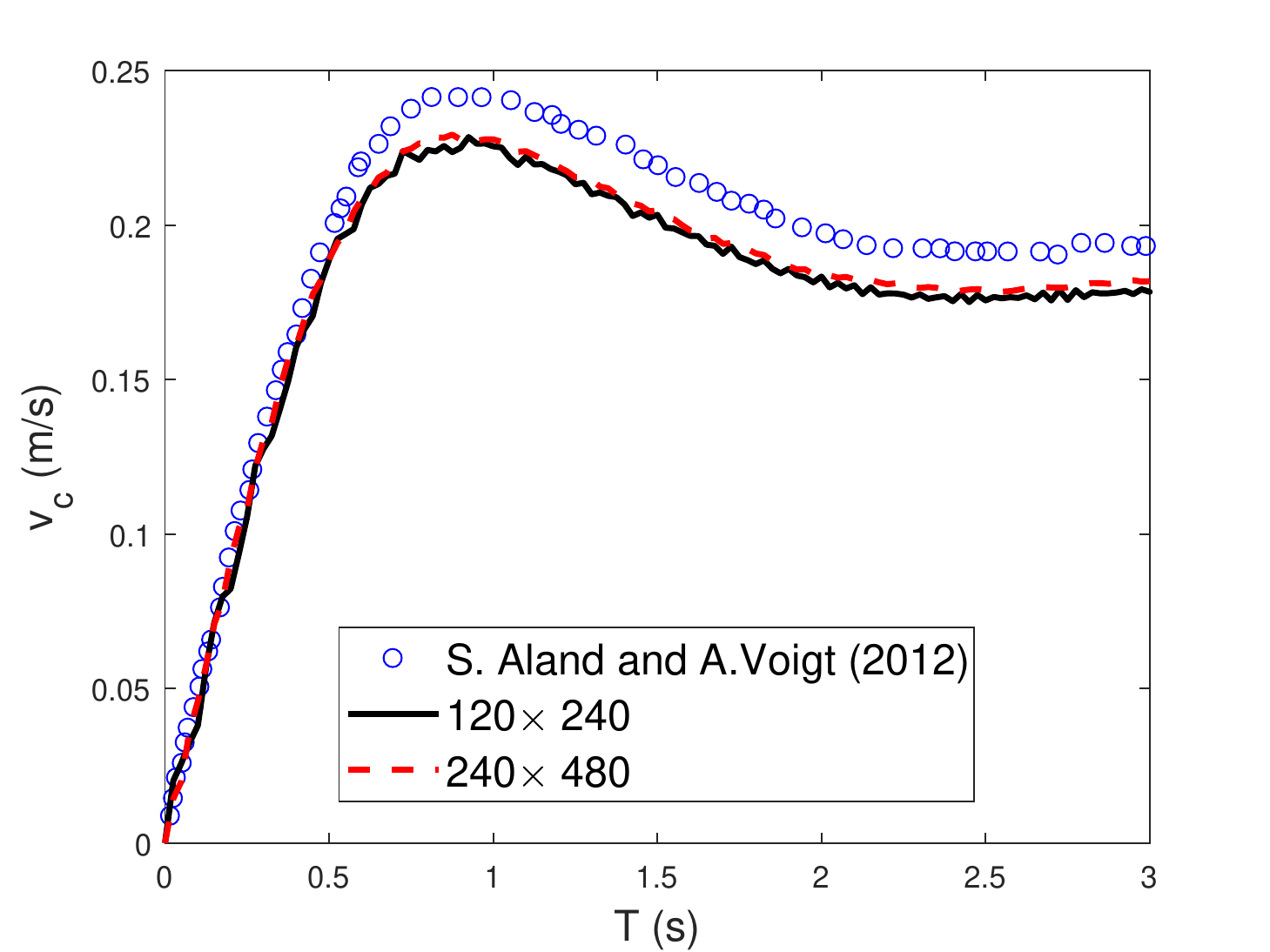}}~
\subfloat[]{\includegraphics[width=0.25\textwidth]{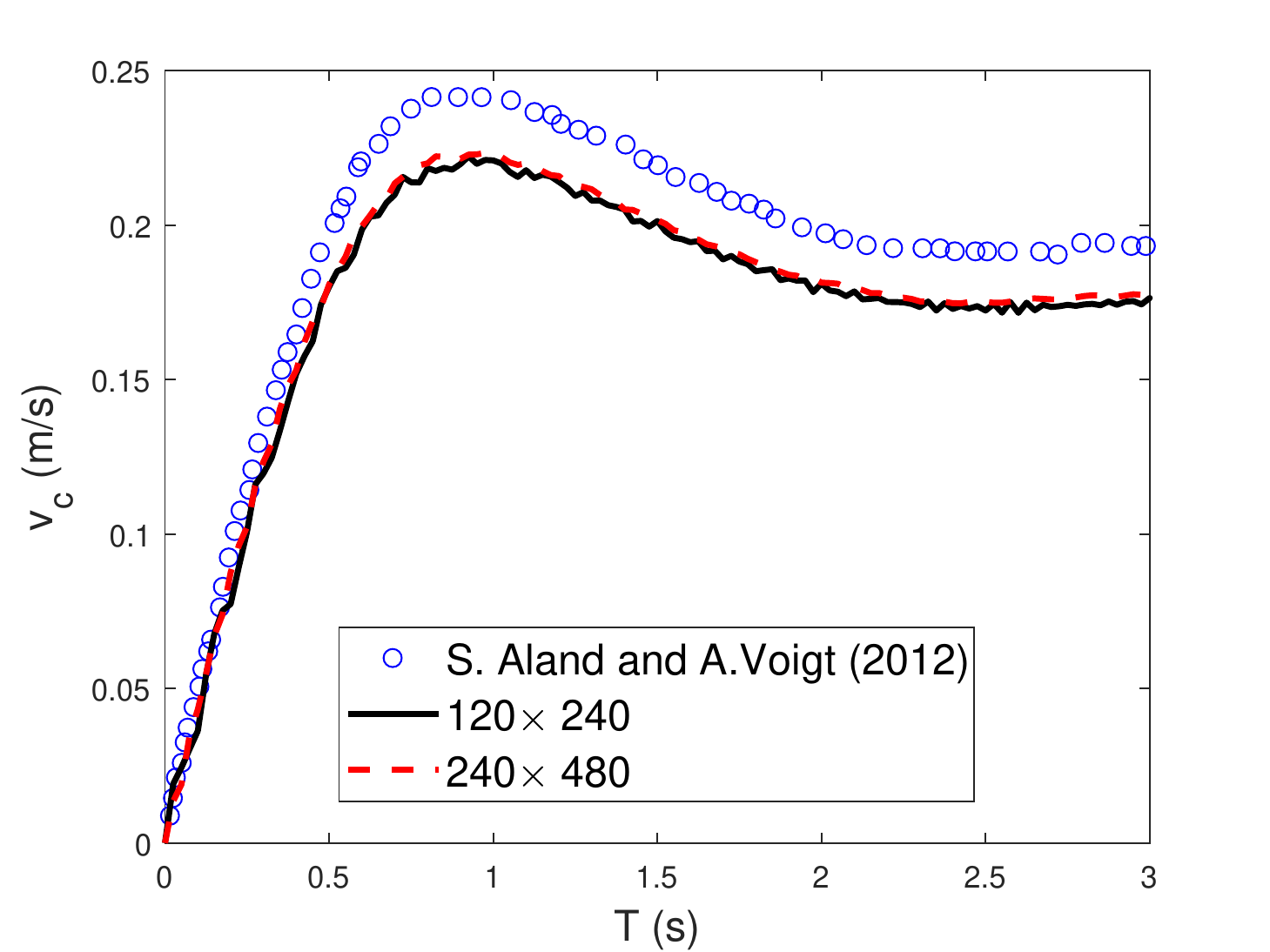}}~
\subfloat[]{\includegraphics[width=0.25\textwidth]{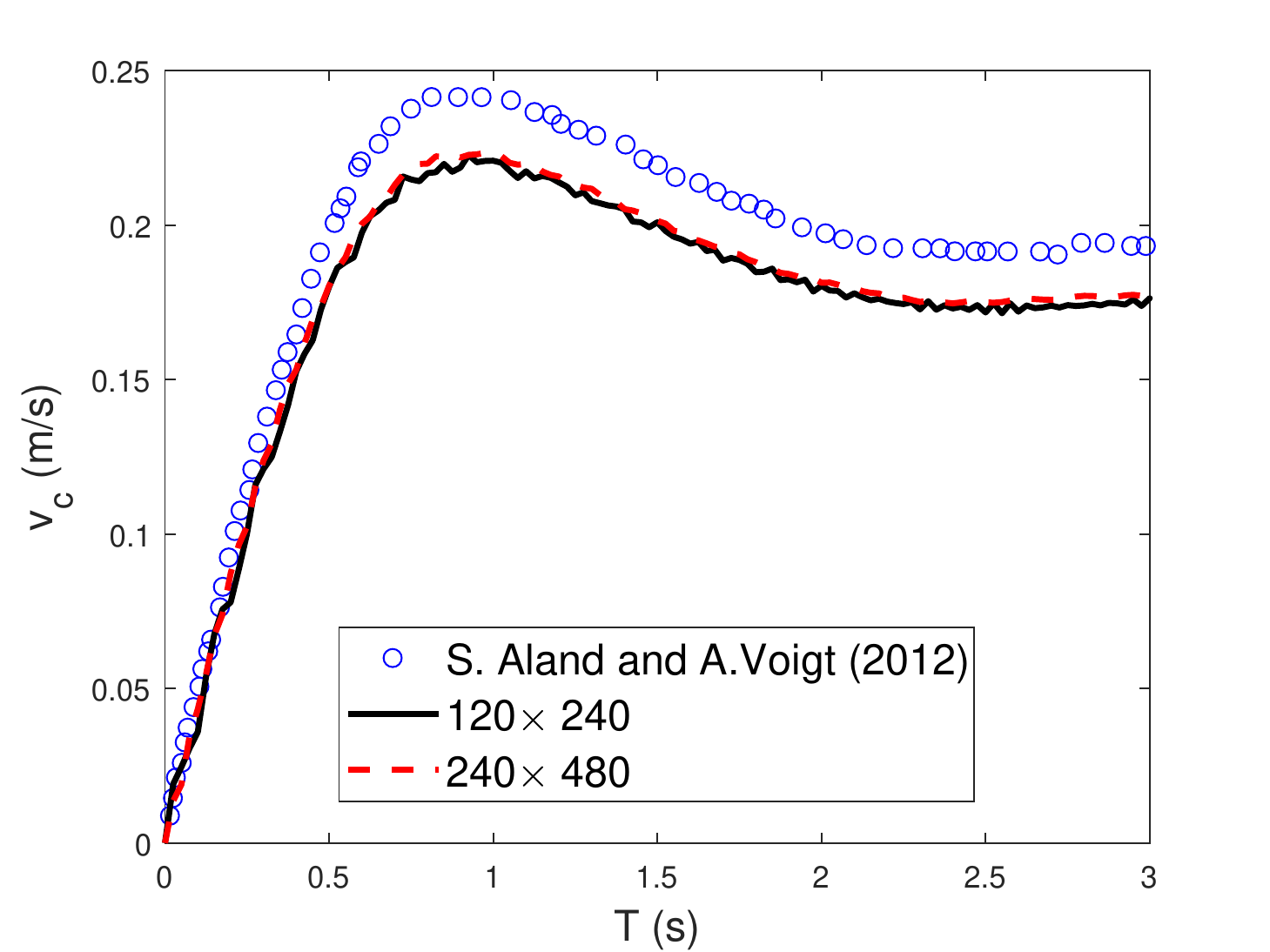}}\\
\subfloat[]{\includegraphics[width=0.25\textwidth]{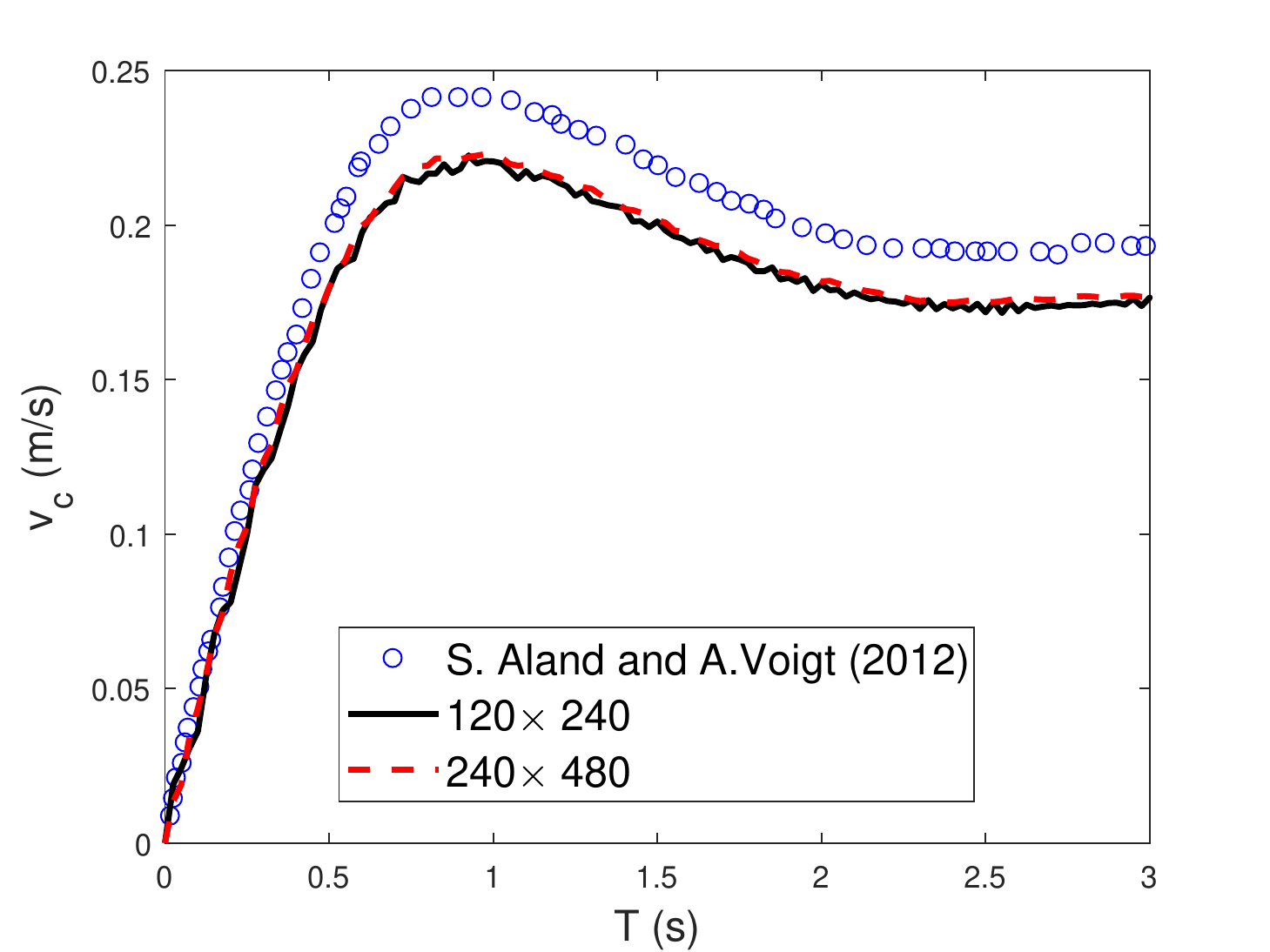}}~
\subfloat[]{\includegraphics[width=0.25\textwidth]{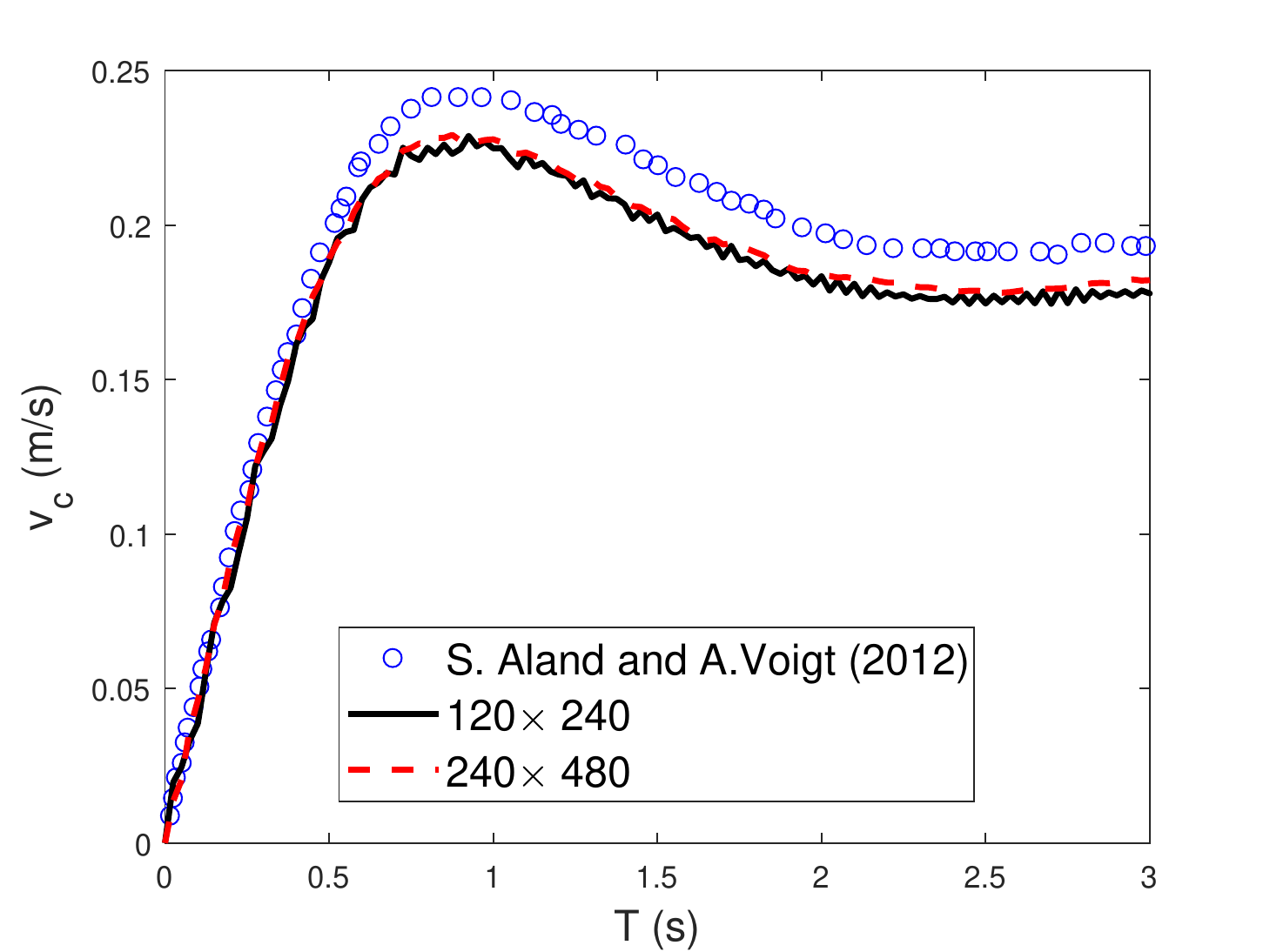}}~
\subfloat[]{\includegraphics[width=0.25\textwidth]{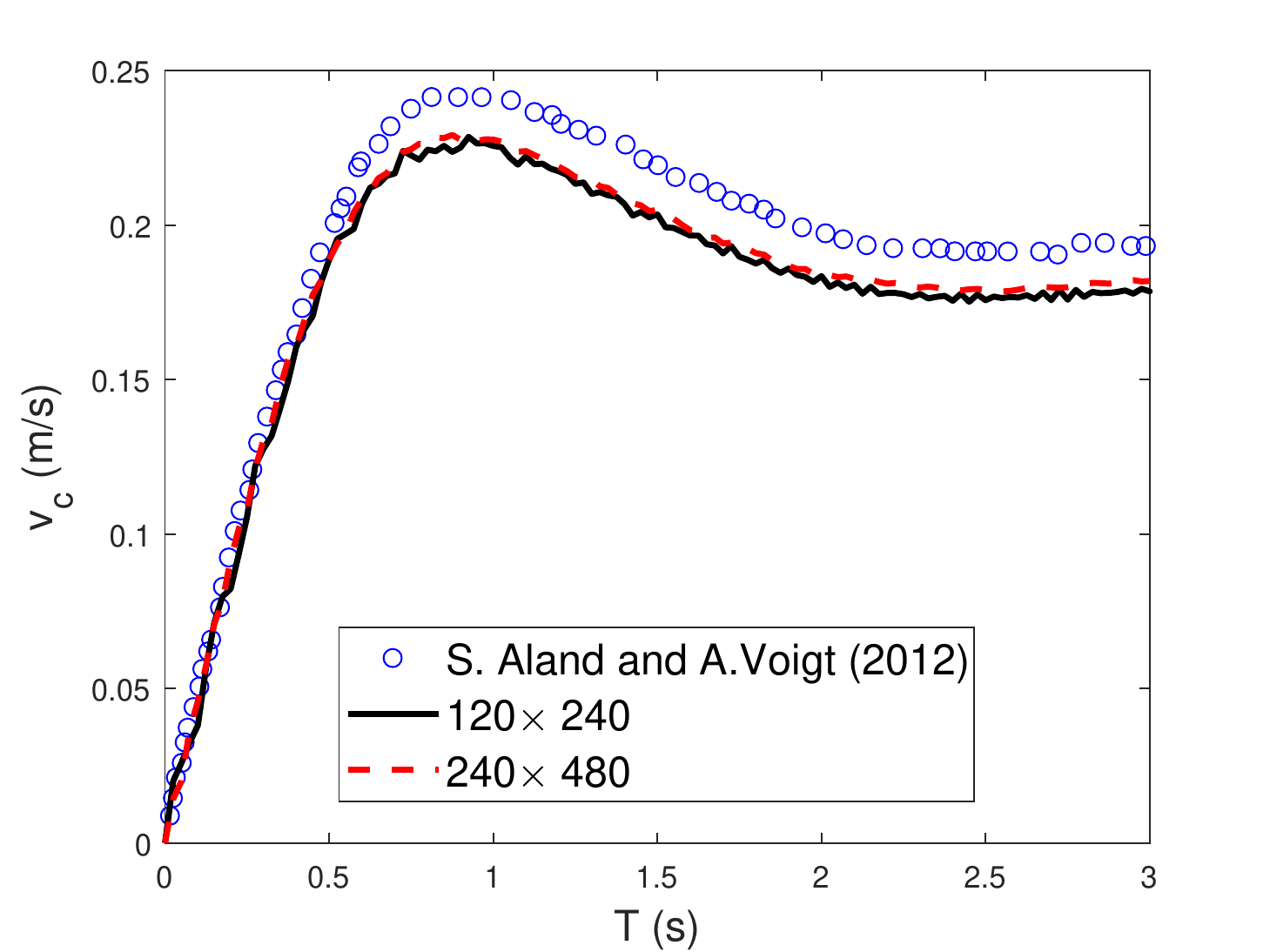}}~
\subfloat[]{\includegraphics[width=0.25\textwidth]{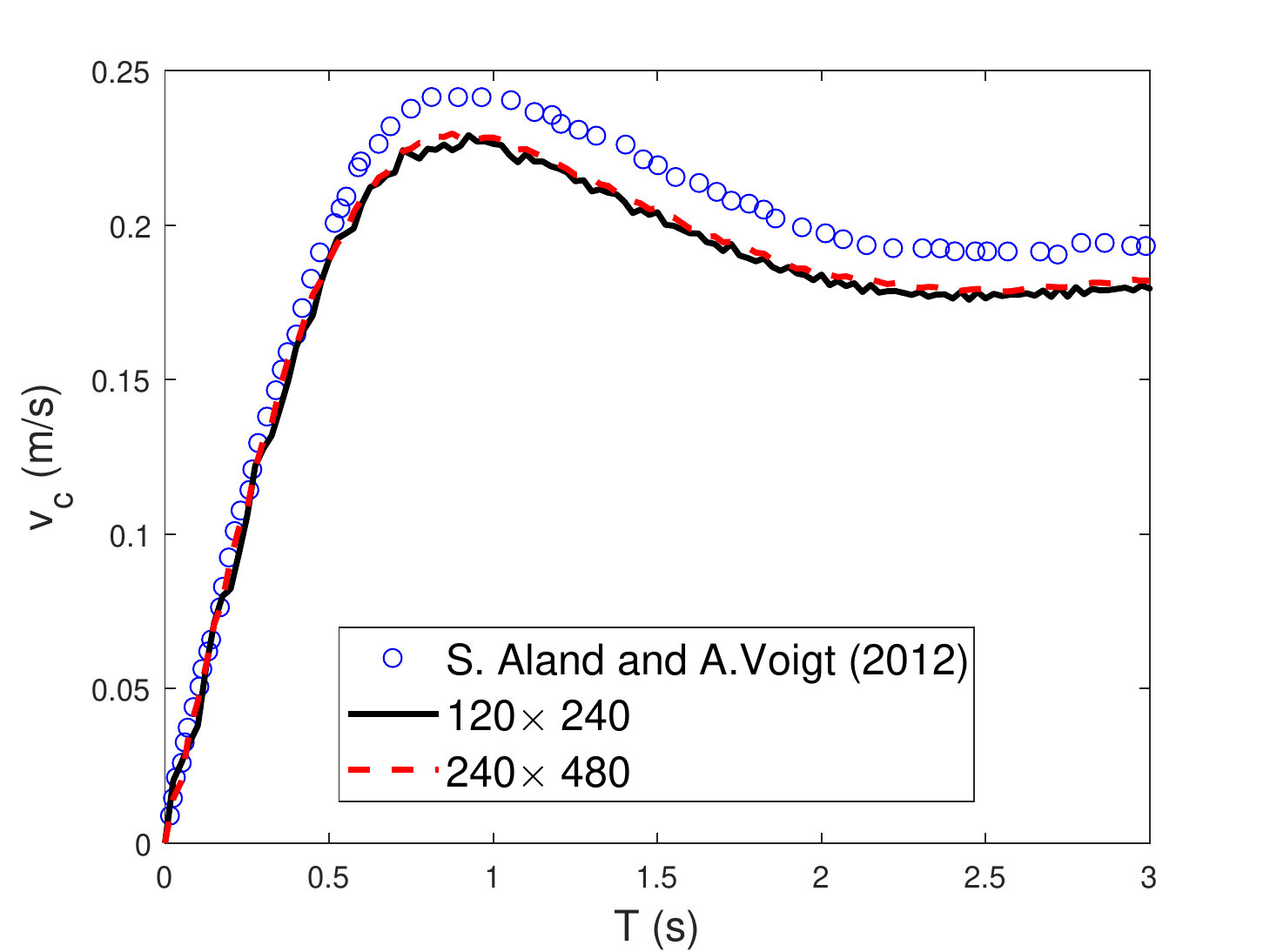}}~
\caption{The evolution of the rising velocity  for (a) $\bm F_{stf-1}$,(b) $\bm F_{stf-2}$,(c) $\bm F_{cpf-1}$,(d) $\bm F_{cpf-2}$,(e) $\bm F_{pf-1}$,(f) $\bm F_{pf-2}$,(g) $\bm F_{csf-1}$ and (h) $\bm F_{csf-2}$. }
\label{fig:bubblevelocity}
\end{figure}

\begin{figure}[!htb]
\centering
\subfloat[]{\includegraphics[width=0.25\textwidth]{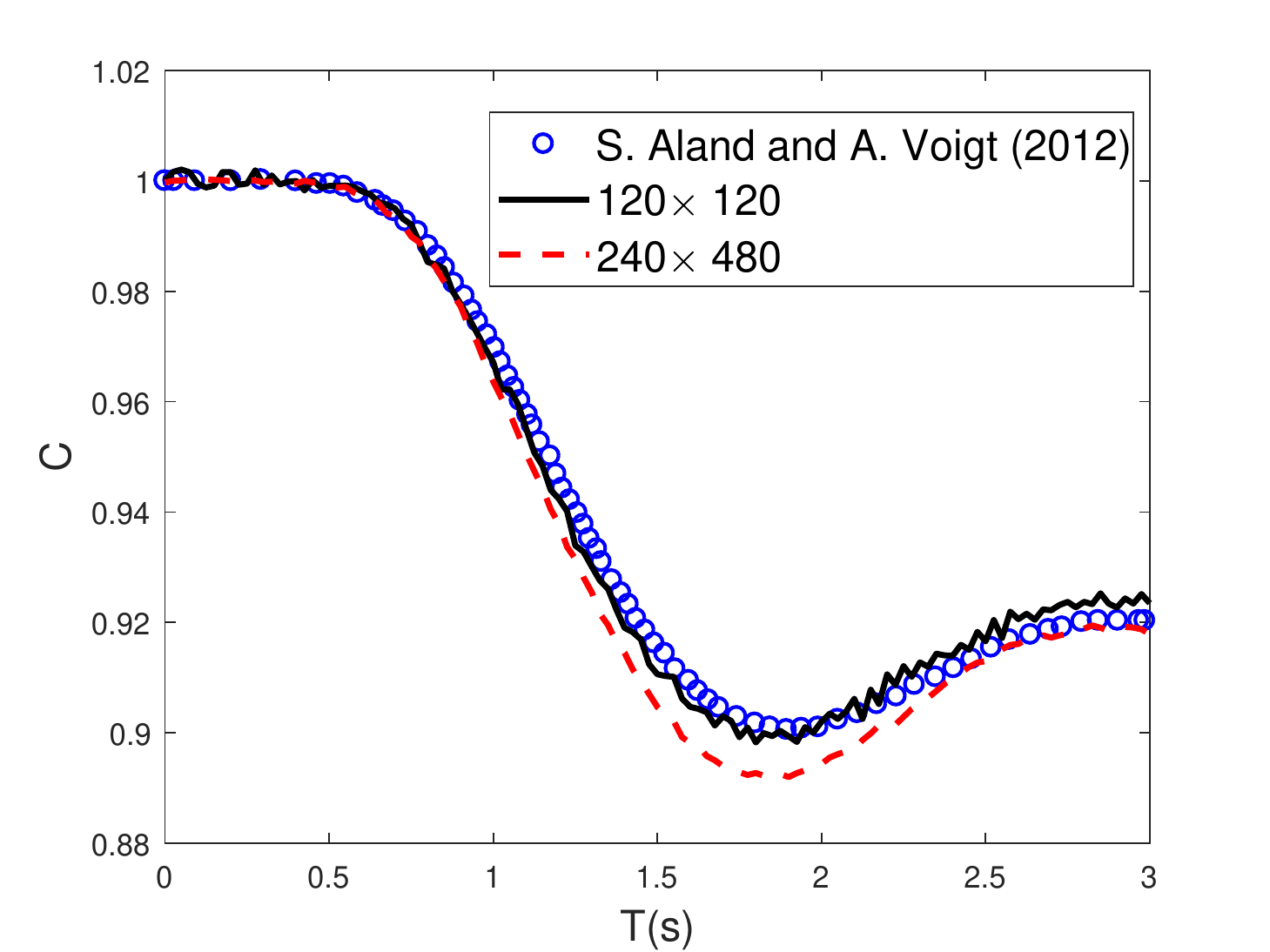}}~
\subfloat[]{\includegraphics[width=0.25\textwidth]{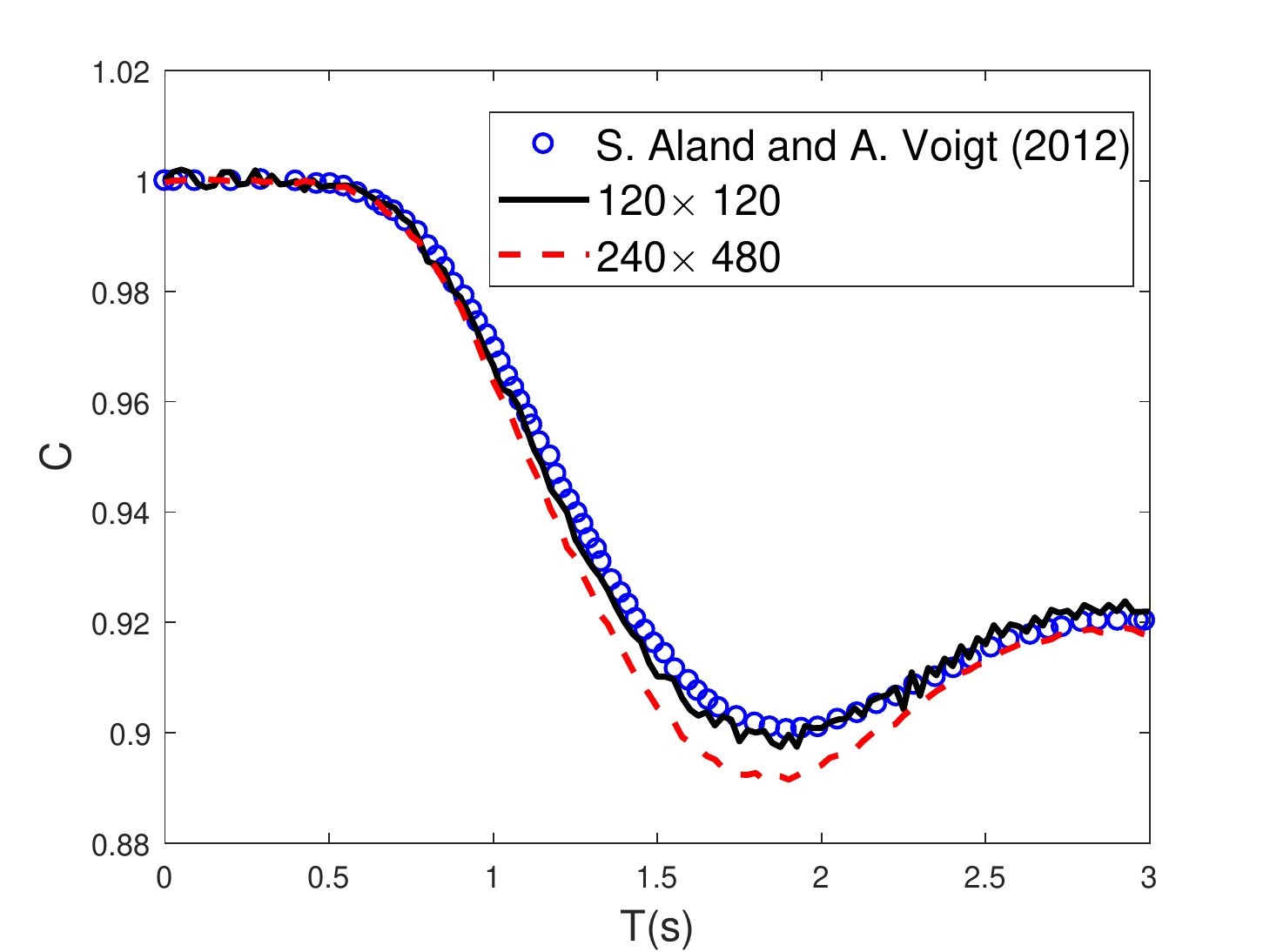}}~
\subfloat[]{\includegraphics[width=0.25\textwidth]{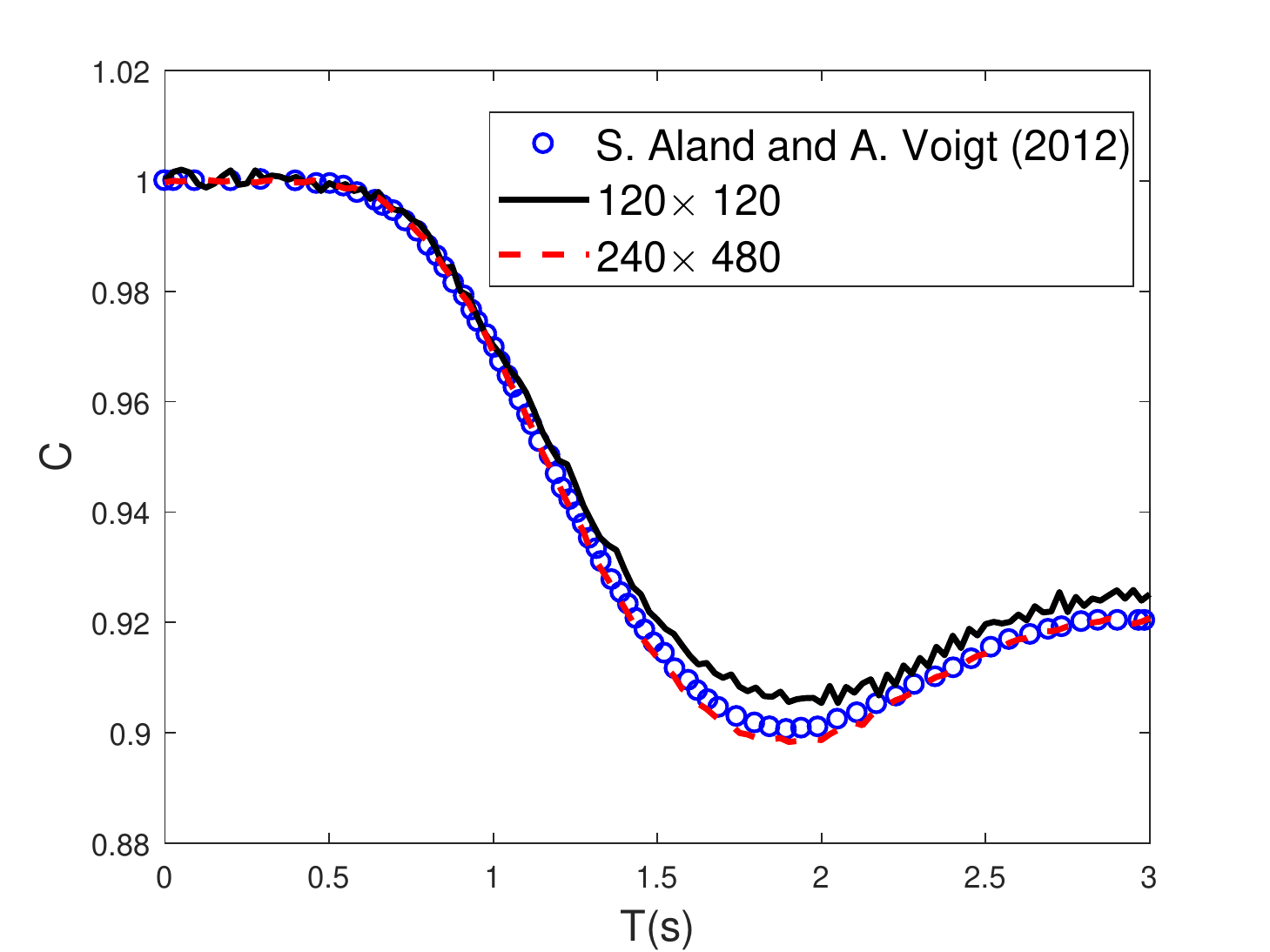}}~
\subfloat[]{\includegraphics[width=0.25\textwidth]{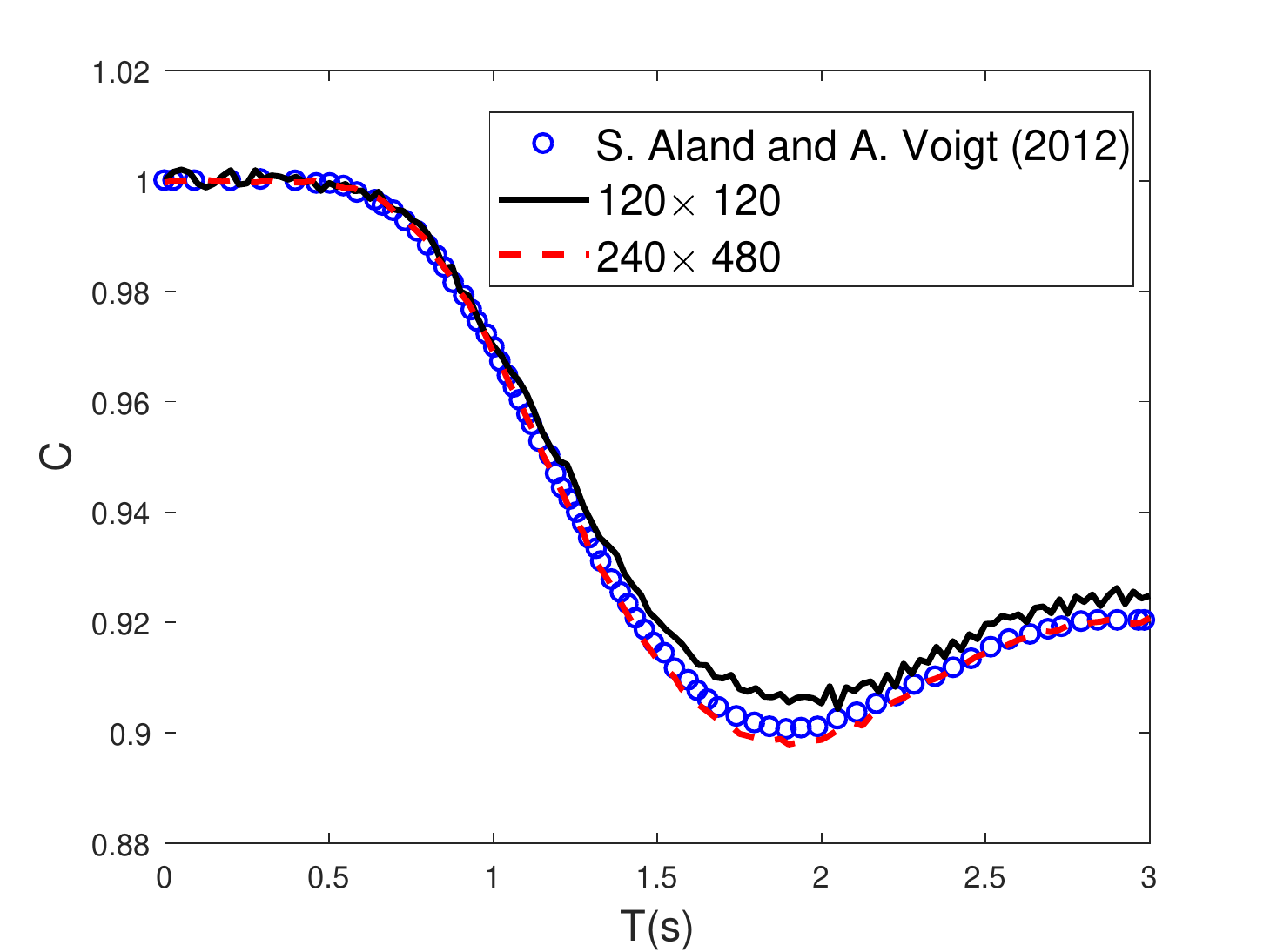}}\\
\subfloat[]{\includegraphics[width=0.25\textwidth]{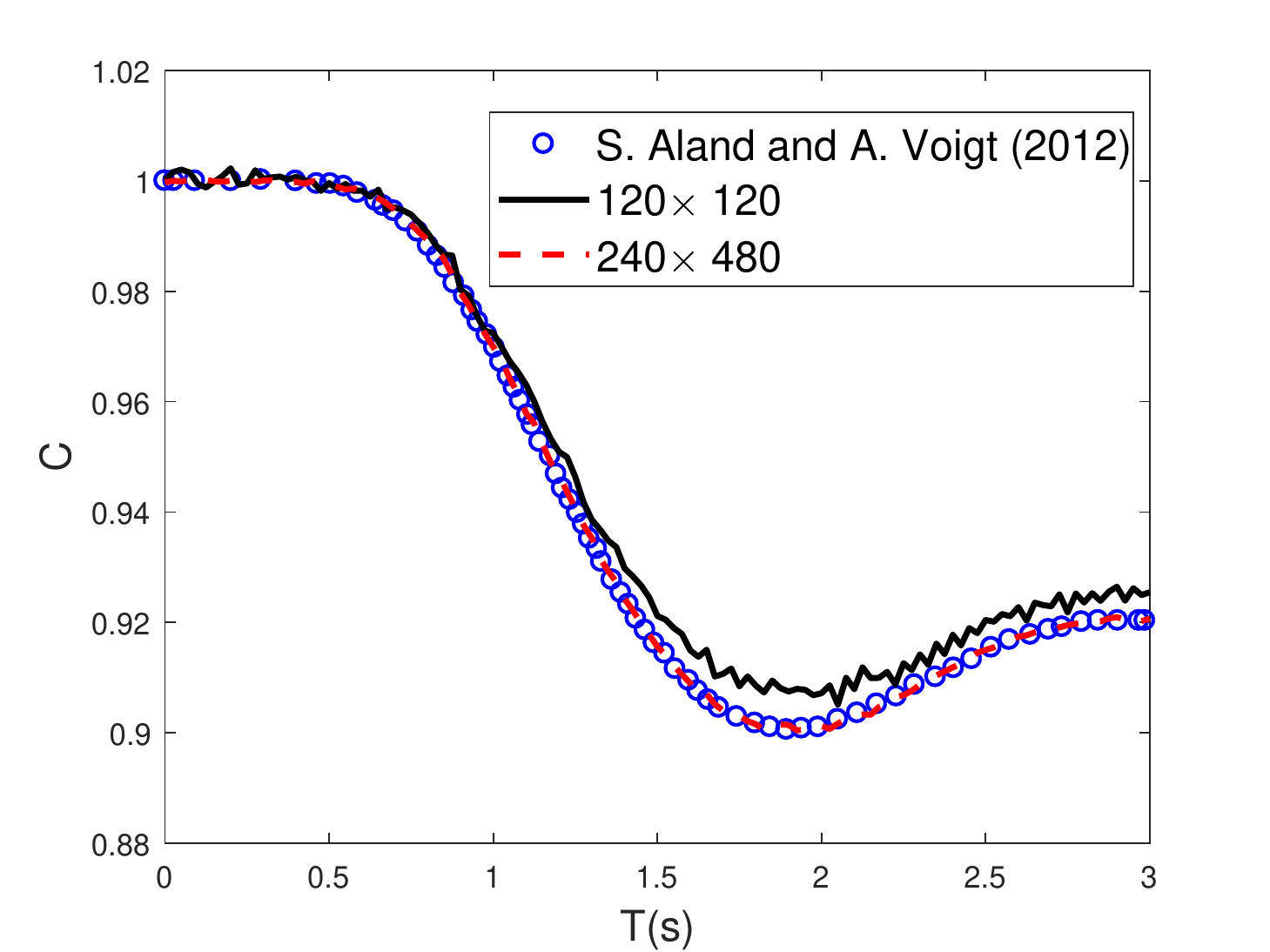}}~
\subfloat[]{\includegraphics[width=0.25\textwidth]{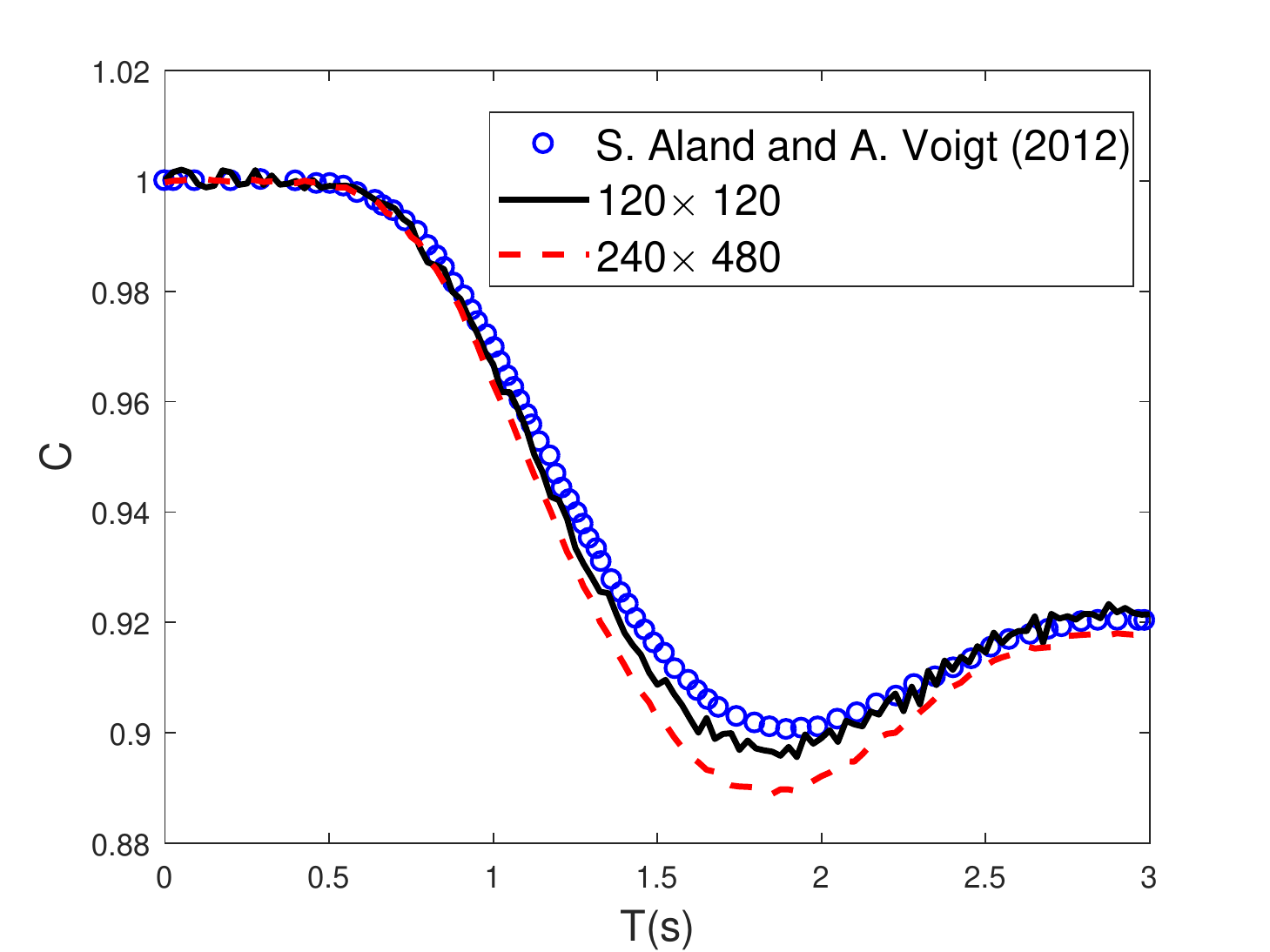}}~
\subfloat[]{\includegraphics[width=0.25\textwidth]{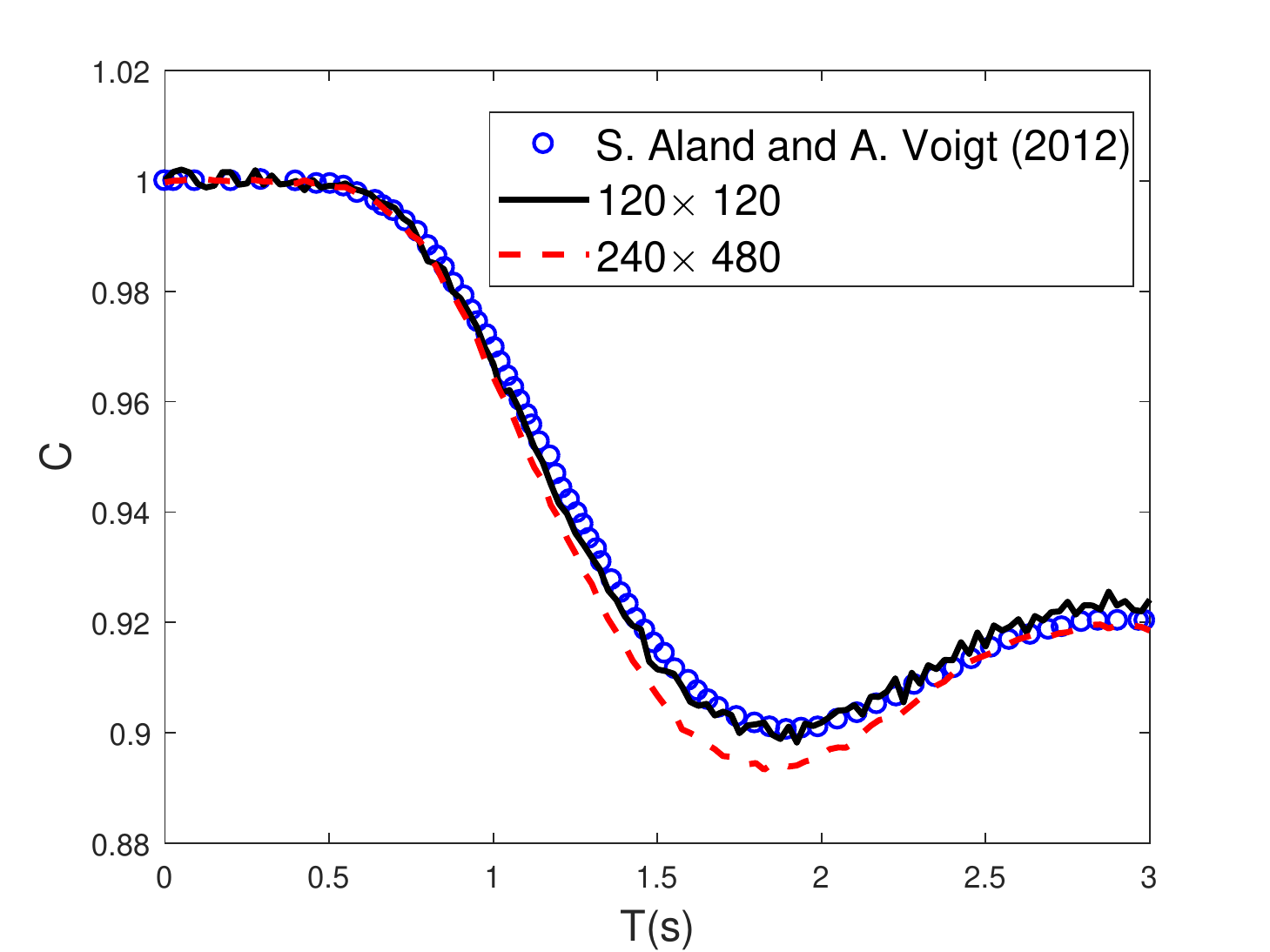}}~
\subfloat[]{\includegraphics[width=0.25\textwidth]{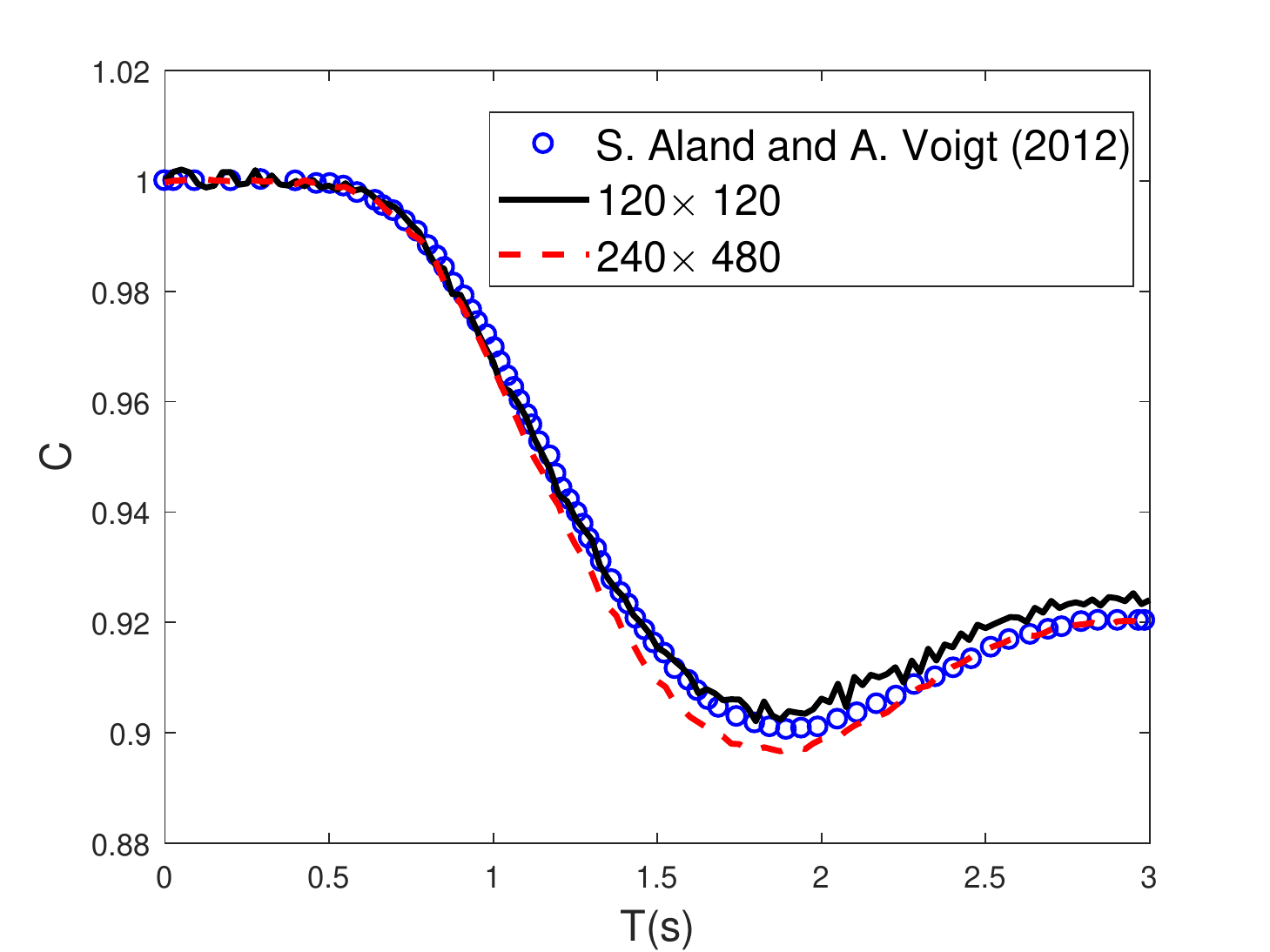}}~
\caption{The evolution of circularity for (a) $\bm F_{stf-1}$,(b) $\bm F_{stf-2}$,(c) $\bm F_{cpf-1}$,(d) $\bm F_{cpf-2}$,(e) $\bm F_{pf-1}$,(f) $\bm F_{pf-2}$,(g) $\bm F_{csf-1}$ and (h) $\bm F_{csf-2}$. }
\label{fig:bubblecircularity}
\end{figure}

Figure~\ref{fig:bubbleshape} shows the bubble shapes predicted by various interfacial force formulations at $t=3\text{s}$. It can be seen that all the results agree well with the benchmark solutions. However, the shapes of the bubble obtained by $\bm F_{cpf-1}, \bm F_{cpf-2}$ and $\bm F_{pf-1}$ are clearly lower than the reference solutions for both grids.
Figure~\ref{fig:bubbleYc} shows the time histories of the center of mass.
At the initial stage, all the results are in good agreement with those reported in \cite{aland2012benchmark}. However,  the discrepancy between the results with $\bm F_{cpf-1}, \bm F_{cpf-2}$ and $ \bm F_{pf-1}$ and the reference solutions becomes larger after $t=1.5\text{s}$.
Figure~\ref{fig:bubblevelocity} compares the rising velocity of the bubble.
 All the results are similar and lower than the reference solutions when the bubble velocity approaches its maximum value.  This may be caused by the interfacial compressibility effect of the  LBM~\cite{zu2013phase,kim2015lattice,zhang2019fractional}. In addition, the viscous effect caused by side walls may slow down the bubble~\cite{lee2010bubble}.
Figure \ref{fig:bubblecircularity} shows the circularity over time for all surface tension formulations, which clearly show that the data with all interfacial force formulations on both meshes agree well with the reference values. However,  the results with $\bm F_{pf-1}$ and $\bm F_{pf-2}$ on the coarse mesh deviate slightly from the reference solutions.
The minimum circularity on the finer mesh  is significantly lower than that of the reference solution except for $\bm F_{pf-1}$.

For quantitative comparison, the maximum mass center position, the maximum rising velocity and minimum circularity with each  force formulation are calculated and compared with the reference results. The results are presented in Table~\ref{tab:rising}. Overall, the values obtained by $\bm F_{stf-1}, \bm F_{stf-2},\bm F_{csf-1}$ and $ \bm F_{csf-2}$ are similar and  in better agreement with the reference data.
\begin{table}
\centering
\caption{Physical parameters and dimensionless numbers}
\setlength{\tabcolsep}{3mm}{
\label{tab:test_Paras}
\begin{tabular}{cccccccc}%
\hline
$\rho_1 (\text{kg}/\text{m}^3)$ &  $\rho_2(\text{kg}/\text{m}^3)$ & $\mu_1 (\text{Pa} \cdot \text{s})$ & $\mu_2(\text{Pa}\cdot \text{s})$
 & $g (\text{m/$s^2$})$ &$\sigma (\text{N/m})$ &\text{Eo} & \text{Re}  \\
\hline
 1000& 100& 10 &   1  & 0.98  & 24.5  & 10  & 35    \\
\hline
\end{tabular} }
\end{table}


\begin{table}[!htb]
  \centering
  \caption{ Benchmark quantities for rising bubble on $240\times 480$ meshes. }\label{tab:rising}
\setlength{\tabcolsep}{2mm}{%
\begin{tabular}{cccccccccc}
\hline
Parameter & Ref.\cite{hysing2009quantitative} & $\bm F_{stf-1}$ &$\bm F_{stf-2}$& $\bm F_{cpf-1}$& $\bm F_{cpf-2}$& $\bm F_{pf-1}$&$\bm F_{pf-2}$ &$\bm F_{csf-1}$&$\bm F_{csf-2}$ \\
\hline
$y_{max}$& 1.0817  &1.0672     &1.0674   &1.0539    &1.0539
                    &1.0536     &1.0669   &1.0675    &1.0681   \\
$v_{max}$& 0.2417  &0.2278    &0.2279   &0.2229    &0.2228
                   &0.2223     &0.2276   &0.2279    &0.2284 \\
$C_{min}$& 0.9013  &0.8917    &0.8914   &0.8983    &0.8979
                   &0.9004     &0.8889   &0.8933    &0.8966 \\
\hline
\end{tabular}}
\end{table}

\subsection{Droplet deformation in shear flow}
\begin{figure}[htp]
\centering
\includegraphics[width=0.5\textwidth]{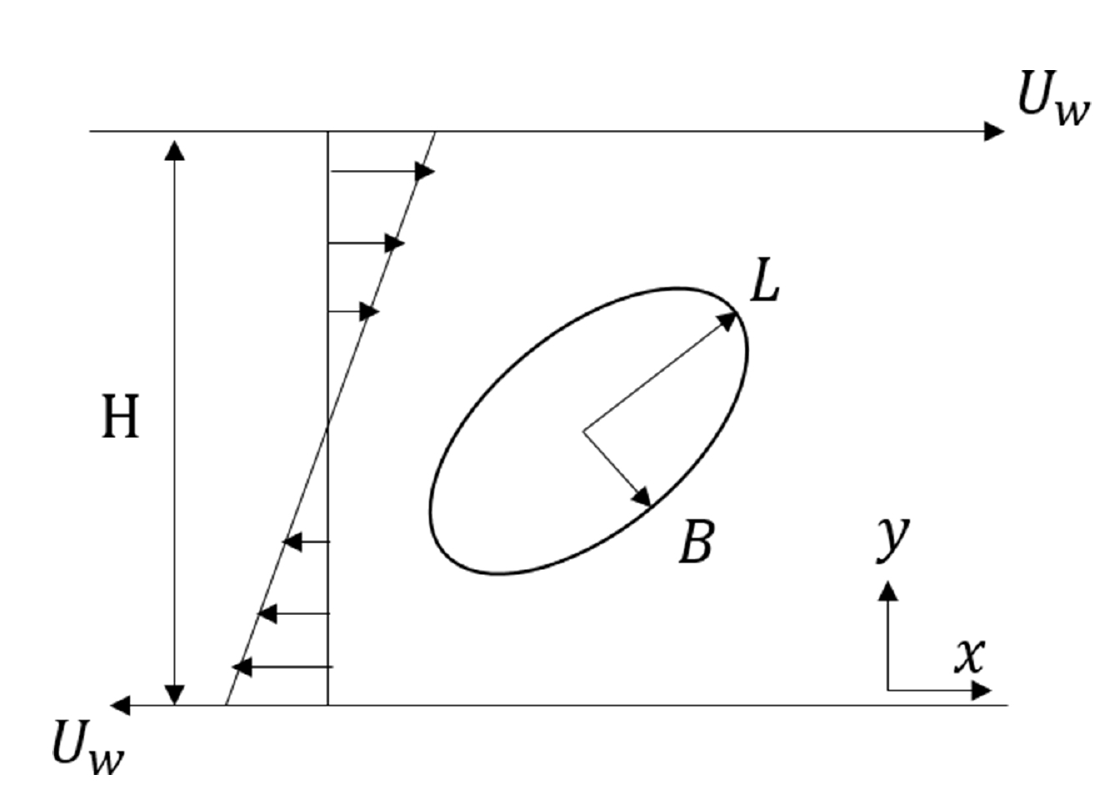}
\caption{ Drop deformation in a shear flow. L is the major axis and B is the minor axis.}
\label{fig:shearflow}
\end{figure}

\begin{figure}[htp]
\centering
\subfloat[]{\includegraphics[width=0.25\textwidth]{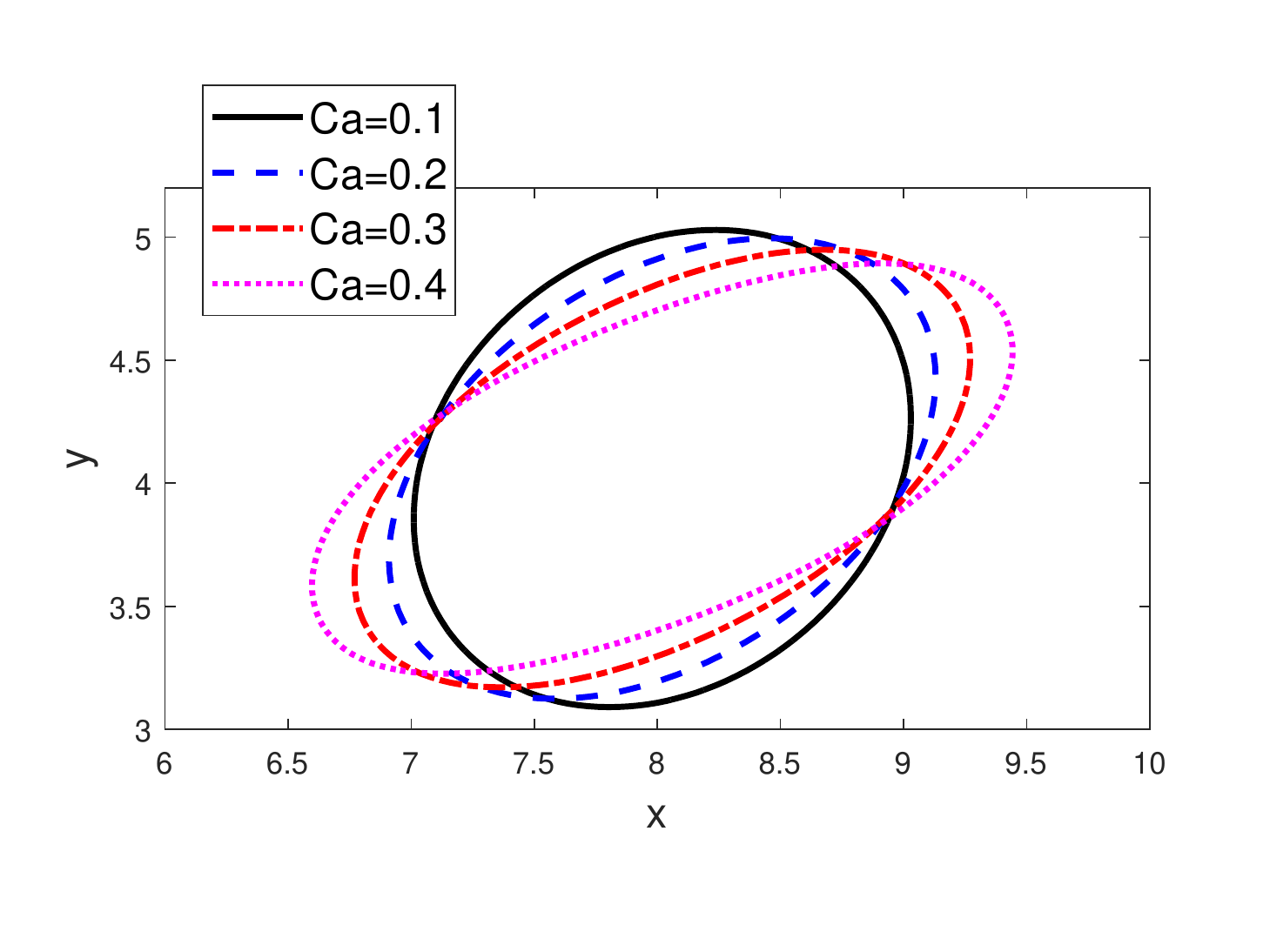}}~
\subfloat[]{\includegraphics[width=0.25\textwidth]{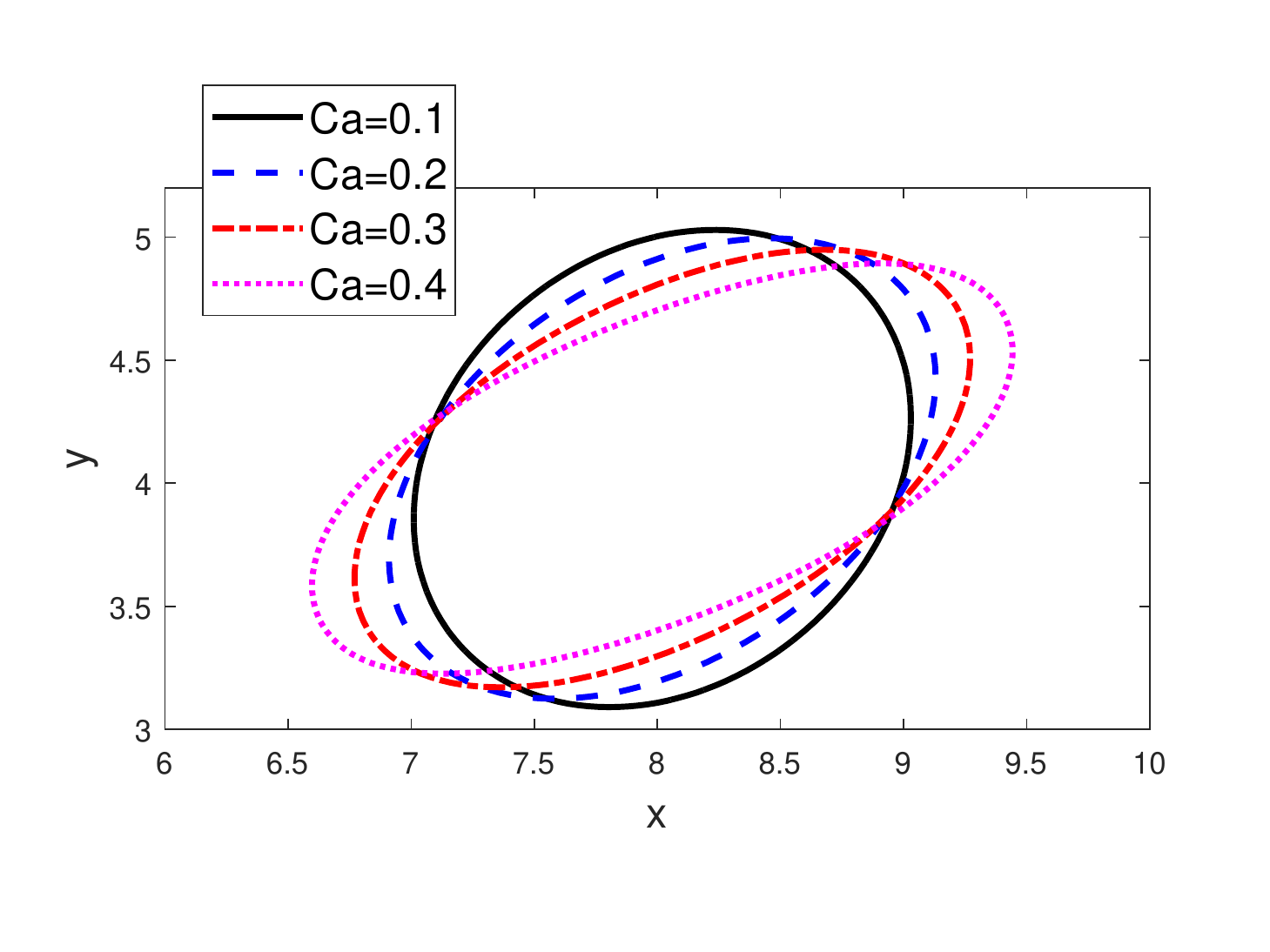}}~
\subfloat[]{\includegraphics[width=0.25\textwidth]{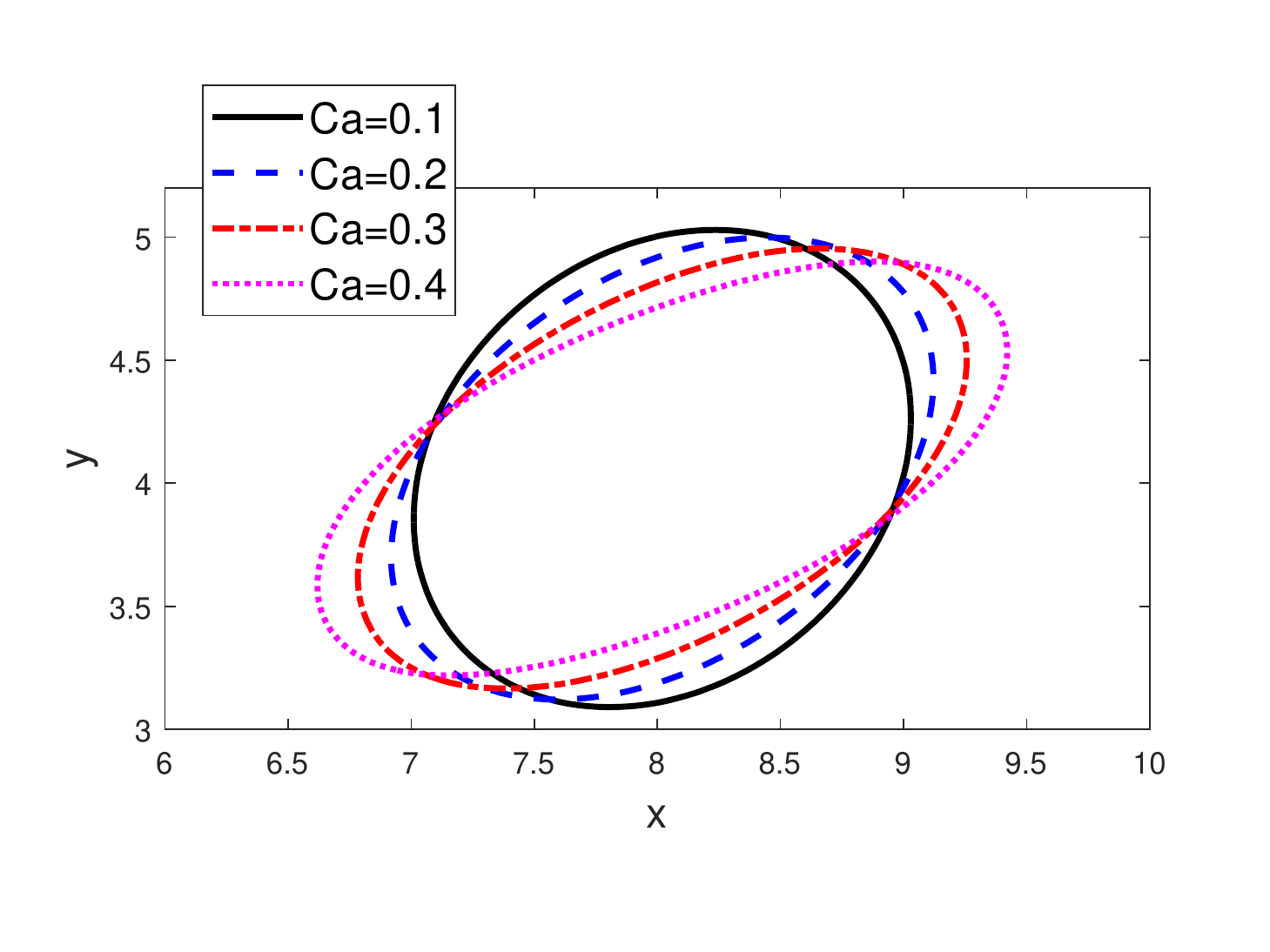}}~
\subfloat[]{\includegraphics[width=0.25\textwidth]{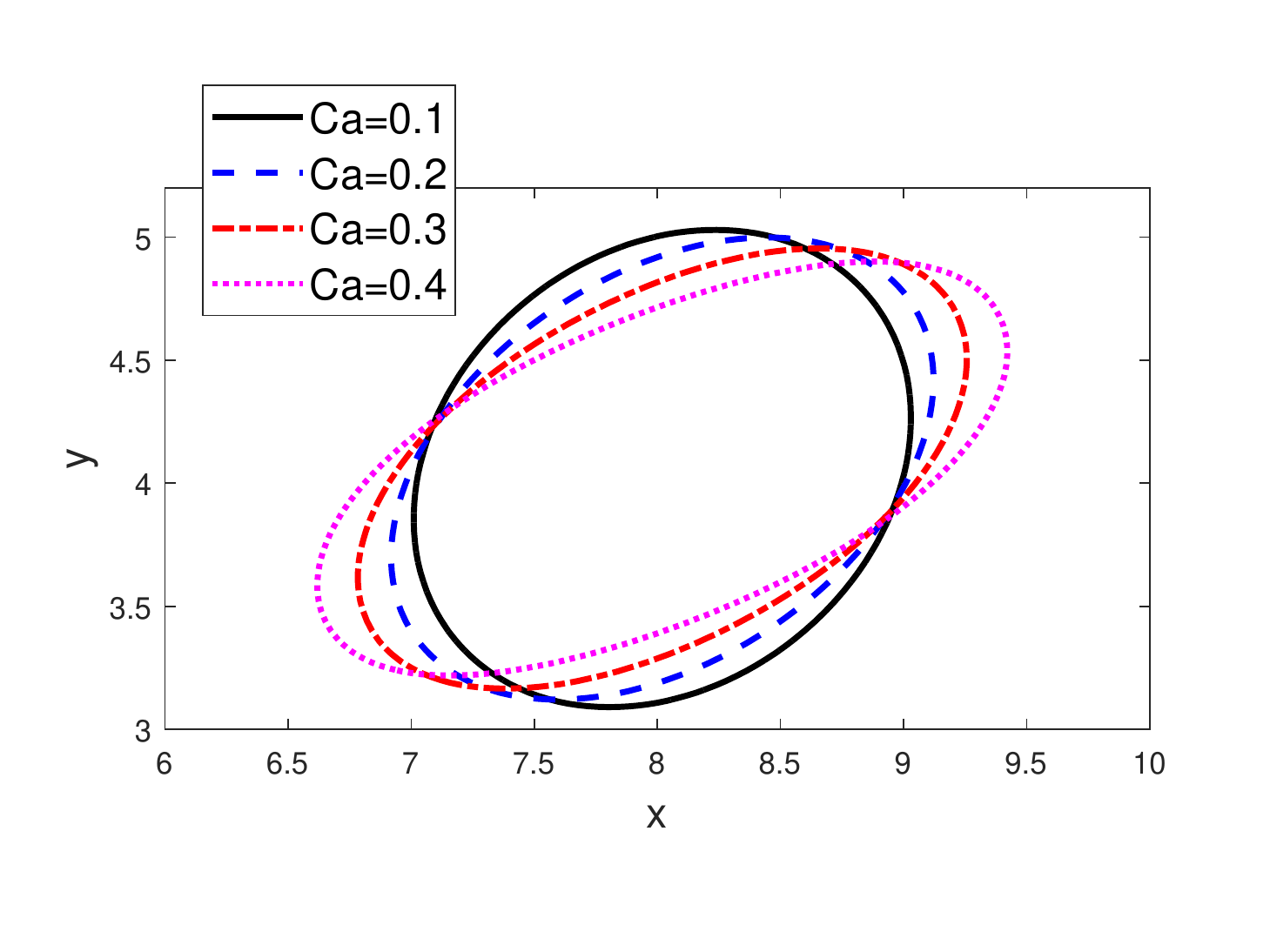}}\\
\subfloat[]{\includegraphics[width=0.25\textwidth]{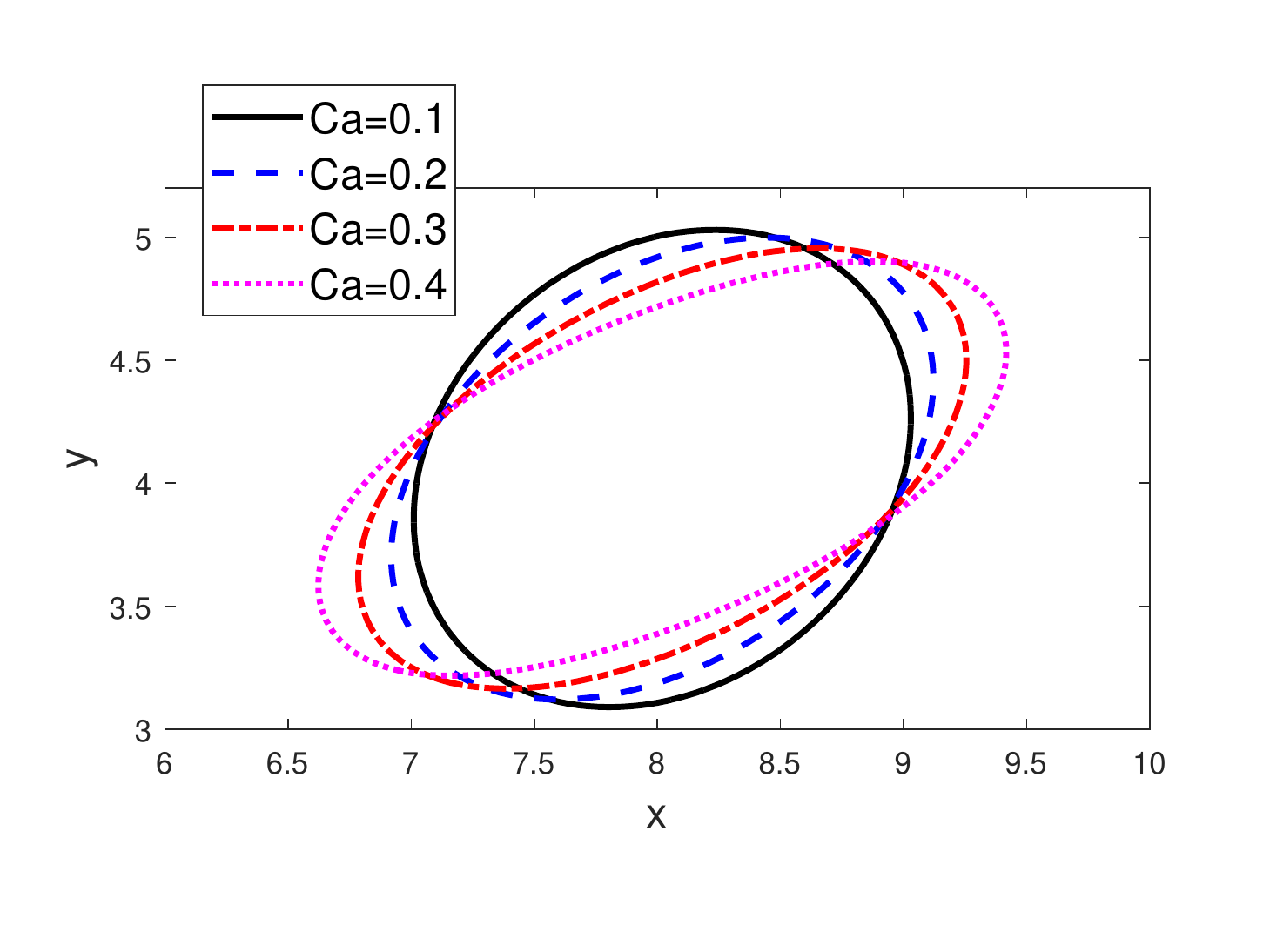}}~
\subfloat[]{\includegraphics[width=0.25\textwidth]{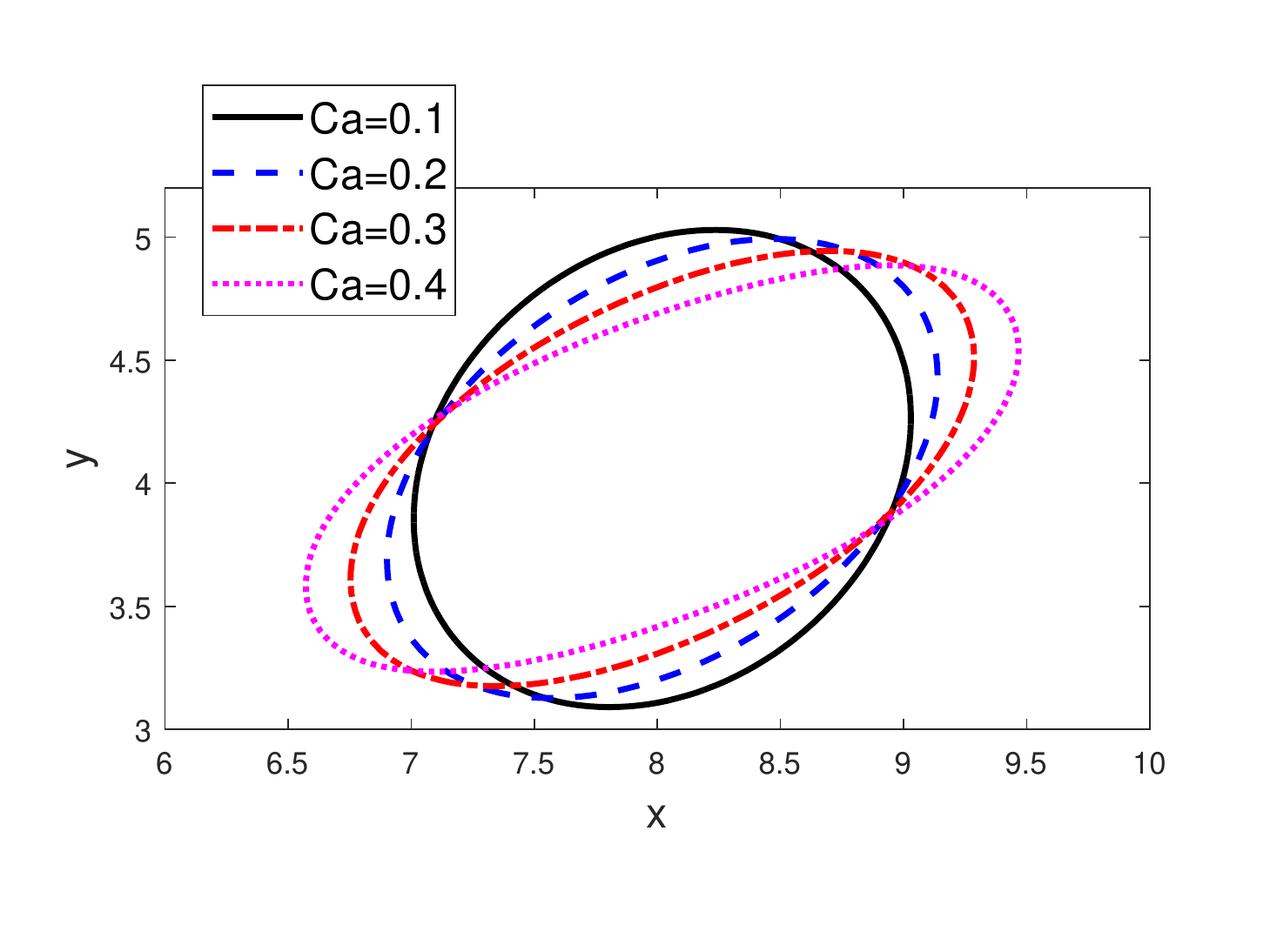}}~
\subfloat[]{\includegraphics[width=0.25\textwidth]{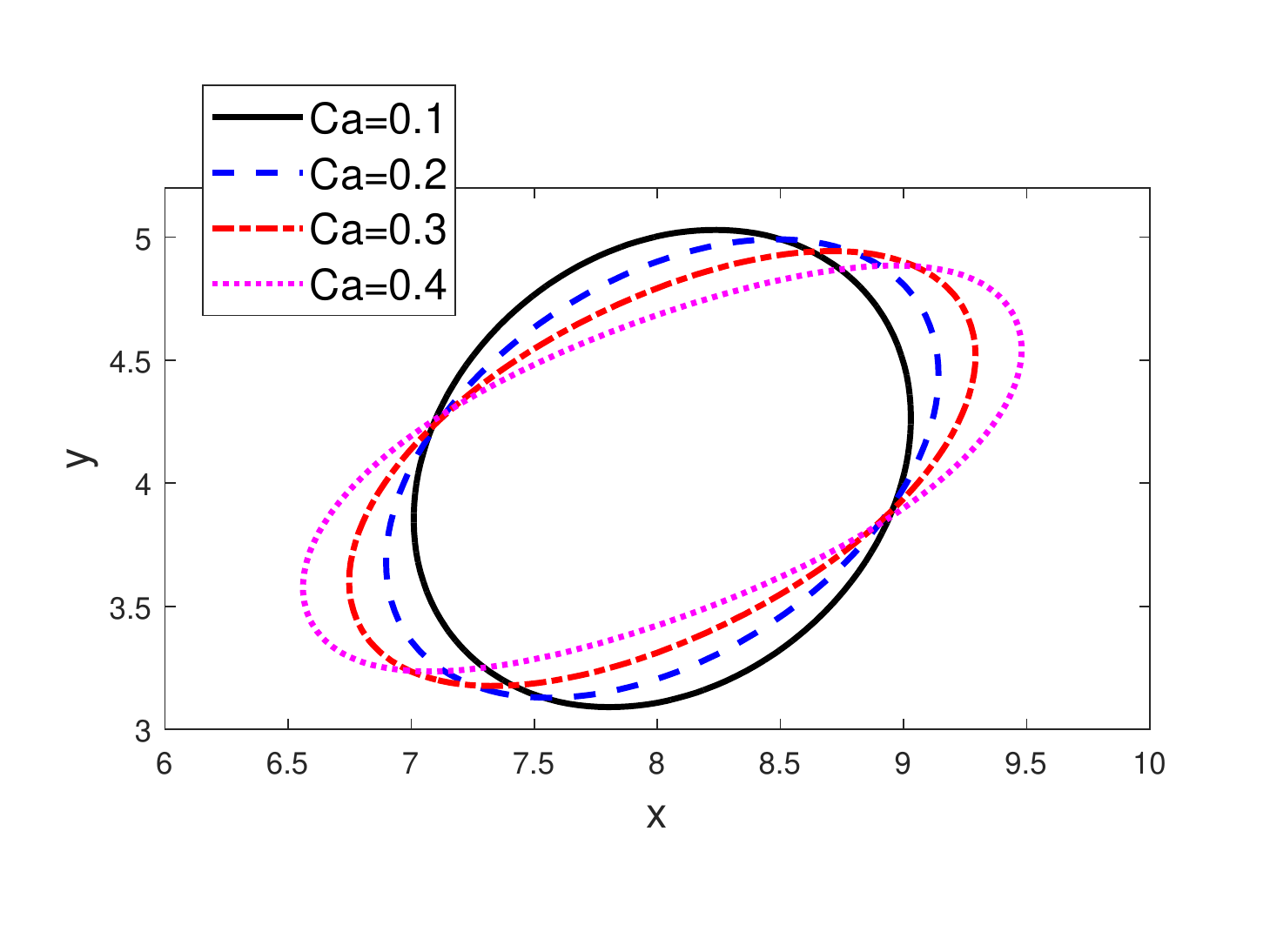}}~
\subfloat[]{\includegraphics[width=0.25\textwidth]{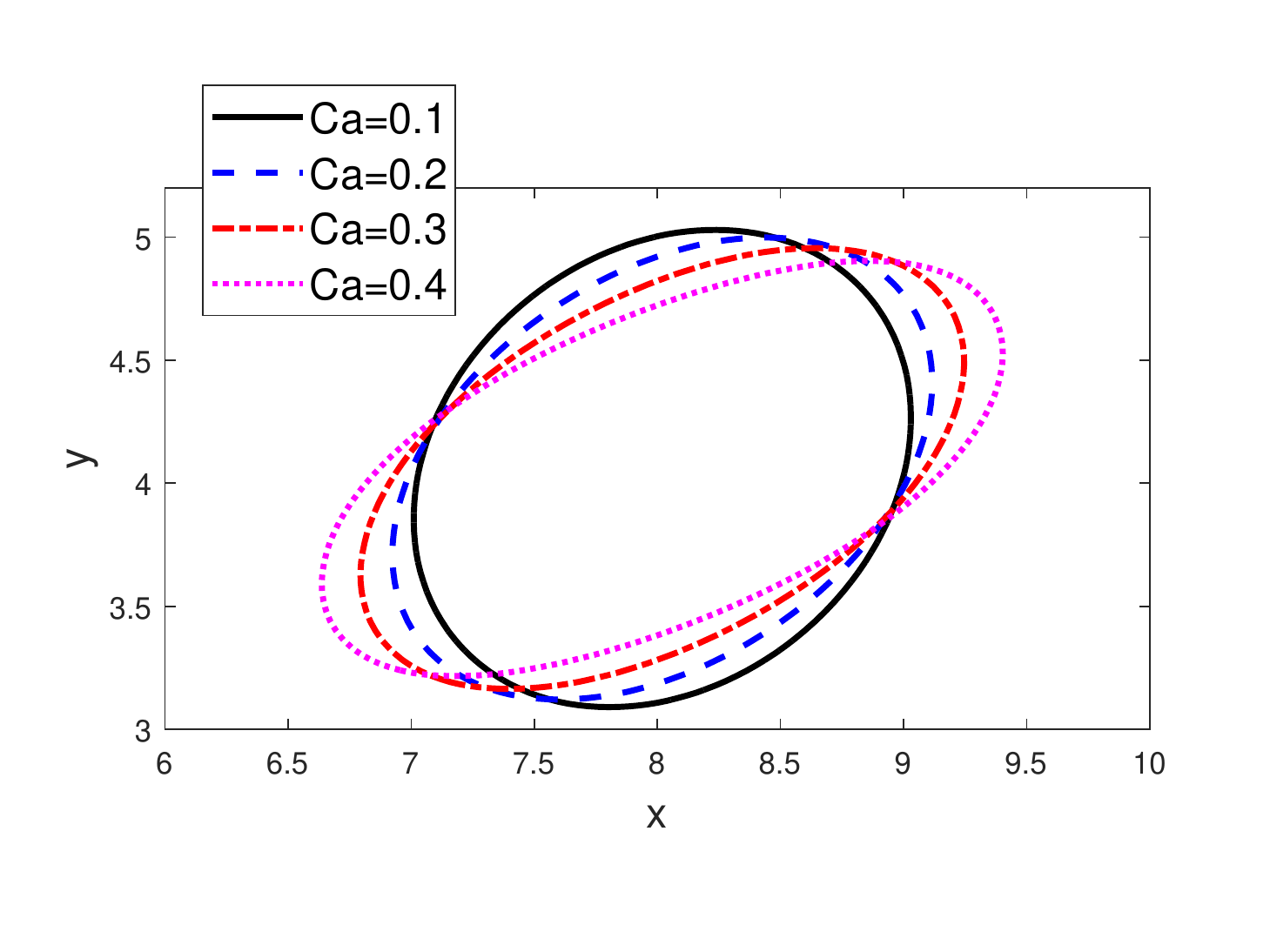}}~
\caption{The $\phi=0$ contours of the order parameter at $\text{Ca}=0.1, 0.2, 0.3$ on  $200\times200$ mesh  with (a) $\bm F_{stf-1}$,(b) $\bm F_{stf-2}$,(c) $\bm F_{cpf-1}$,(d) $\bm F_{cpf-2}$,(e) $\bm F_{pf-1}$,(f) $\bm F_{pf-2}$,(g) $\bm F_{csf-1}$ and (h) $\bm F_{csf-2}$.}
\label{fig:shear_shape}
\end{figure}

Finally, we consider a circle drop deformation in a shear flow. The schematic of the flow field is shown in Fig~\ref{fig:shearflow}.
Initially,  a circle drop is located at the center of a rectangle domain of $2H\times H$. The effect of gravity is ignored. The top and bottom walls maintain velocities $U$ and $-U$, respectively, leading to a shear rate $\text{E}=2U/H$. The periodic boundary conditions are applied to  the left and right boundaries. The same density and viscosity are specified for both the drop and surrounding fluid.  In the simulation, we set $H=8\text{m},R=1\text{m}, U_w=4\text{m/s},\rho_d=\rho_s=1\text{kg/$\text{m}^3$}$. The Reynolds number $\text{Re}=E \rho_d R^2/\mu_d=0.1$.
The capillary number $\text{Ca}=\mu E R/\sigma$ is varied from $0.1$ to $0.4$ by varying $\sigma$. The uniform grid size of $200\times 200$ is employed.
 The shapes of the deformed drop at steady state are illustrated in Fig.~\ref{fig:shear_shape}. It can be seen that the shapes of the drop given by all interfacial force formulations deform into an ellipsoidal one and are elongated as $\text{Ca}$ increases. In particular, the shapes of the drop obtained by $\bm F_{csf-1}$  are overstretched compared to other results.
The shape of the drop can be characterized by a Taylor  deformation parameter defined as $D=(L-B)/(L+B)$, where $L$ and $B$ are the lengths along the major axis and the minor axis of the droplet, respectively.
A theoretical solution derived on the assumptions of the Stokes flow and small deformation shows that  the Taylor  deformation parameter is related to the capillary number and the viscosity ratio
\cite{taylor1934formation,taylor1932viscosity}
\begin{equation}\label{eq:shear}
D=\frac{L-B}{L+B}=\text{Ca}\frac{19\lambda+16}{16\lambda+16},
\end{equation}
where $\lambda=\mu_d/\mu_f$ is the viscosity ratio between the drop fluid and the surrounding fluid.
Table~\ref{tab:linear_theory}  shows the Taylor deformation parameters  with different force formulations. It can be seen that the values predicted by $\bm F_{csf-1}$ are significantly higher than the theoretical values. Overall, the values with $\bm F_{stf-1}$ and $\bm F_{stf-2}$ are close to the theoretical ones.
\begin{table}[!htb]
\centering
\caption{Comparison of Taylor deformation number $D$ with linear theory}\label{tab:linear_theory}
\setlength{\tabcolsep}{4mm}{%
\begin{tabular}{ccccccccc}
\hline
$\bm F_{sf}$ & \multicolumn{4}{c}{$\rho_d/\rho_s=1$} & & \multicolumn{3}{c}{$\rho_d/\rho_s=0.1$} \\
\cline{2-5}  \cline{7-9}
    &$\text{Ca=0.1}$ &$\text{Ca=0.2}$ & $\text{Ca=0.3}$& $\text{Ca=0.4}$   &   &$\text{Ca=0.1}$ &$\text{Ca=0.2}$ & $\text{Ca=0.3}$\\
\hline
$\bm F_{stf-1}$ & 0.1094 & 0.2227 & 0.3280 &0.4302               &  & 0.1039 & 0.2028   &0.2989     \\
$\bm F_{stf-2}$ & 0.1094 & 0.2227 & 0.3280 &0.4302               &  & 0.1038 & 0.2027   &0.2988     \\
$\bm F_{cpf-1}$ & 0.1094 & 0.2161 & 0.3188 &0.4187               &  & 0.0994 & 0.1957   &0.2896     \\
$\bm F_{cpf-2}$ & 0.1094 & 0.2161 & 0.3189 &0.4190               &  & 0.0996 & 0.1961   &0.2901    \\
$\bm F_{pf-1}$  & 0.1094 & 0.2155 & 0.3174 &0.4164               &  & 0.1006 & 0.1963   &0.2894     \\
$\bm F_{pf-2}$  & 0.1094 & 0.2299 & 0.3381 &0.4440               &  & 0.1067 & 0.2087   &0.3084     \\
$\bm F_{csf-1}$ & 0.1094 & 0.2349 & 0.3430 &0.4516               &  & 0.1168 & 0.2163   &0.3144     \\
$\bm F_{csf-2}$ & 0.1094 & 0.2113 & 0.3114 &0.4085               &  & 0.0961 & 0.1891   &0.2818     \\
$\text{Eq.}(\ref{eq:shear})$ & 0.1094  & 0.2188 &  0.3281 &0.4375 &  & 0.1017 & 0.2034   &0.3051     \\
\hline
\end{tabular}}
\end{table}

\section{Conclusions}
In this paper, we successfully implemented  the phase-field-based lattice Bolzmann method  with different interfacial force formulations for two phases flow. The performance of each surface tension formulation has been validated and compared.  For a stationary drop,  $\bm F_{csf-2}$ provides the most accurate prediction in terms of the surface tension coefficient.
The potential form  can generate  small spurious currents.  $\bm F_{stf-2}, \bm F_{csf-1}$ and $\bm F_{csf-2}$ produce a smooth pressure field across the interface and symmetric distribution of the interfacial force.  The distributions of  $\bm F_{stf-1}$, $\bm F_{cpf-1}$  $\bm F_{pf-1}$ and $\bm F_{pf-2}$ become symmetric with respect to the phase interface.
For the droplets merging problems, there are obvious differences for the interface shapes of the droplets during coalescence. The droplets is more prone to merge due to the surface tension effects when $\bm F_{cpf-1}, \bm F_{pf-2}$ and $\bm F_{csf-2}$ are used. In particular, the unexpected movement of droplets with unequal sizes occurs when $\bm F_{cpf-1}$, $\bm F_{cpf-2}$, $\bm F_{csf-1}$ and $\bm F_{csf-2}$ are used.
It is also found that  $\bm F_{stf}$ and $\bm F_{csf}$ show  better numerical stability than $\bm F_{cpf}$ and $\bm F_{csf}$.
For the test of capillary wave, the evolution processes of the interface amplitude from $\bm F_{pf-1}$ and $\bm F_{csf-2}$ are closer to the analytical solutions in all formulations. It is worth noting that $\bm F_{cpf-1}$ can yield  good results but the Peclet number should be carefully chosen. For the simulation of a rising bubble, both the stress form and CSF form give a good results in terms of the mass center. $\bm F_{cpf-1},\bm F_{cpf-2}$ and $\bm F_{pf-1}$ clearly underestimate the center  at the late stage. For the  rising velocity, all formulations underestimate the maximum rise velocity.
In terms of the circularity, only $\bm F_{cpf-2}$ and $\bm F_{csf-2}$ give the predictions closer to the reference solutions. For shear flow, all formulations  give accurate predictions in comparison with the linear theory at $\text{Ca}=0.1$. With the increase of capillary number, $\bm F_{csf-1}$ produces a larger  deformation than the theoretical predictions. For all the considered capillary number, $\bm F_{stf-1}$ and $\bm F_{stf-2}$ can give a satisfactory prediction.

In summary, it seems that no surface tension force formulation can give satisfactory results in all tests. Different forms may be considered for different problems. Overall,  $\bm F_{cpf}$ is good for calculating multiphase flows with   small interface deformation. Both $\bm F_{stf}$ and $\bm F_{csf}$ are good for dynamical situations.  We hope the present comparison can provide insights into the advantages and limitations of each formulation.

\section*{DATA AVAILABILITY}
The data that support the findings of this study are available from the corresponding author
upon reasonable request.
\section*{ACKNOWLEDGEMENTS}
This study was supported by the National Science Foundation of China(51836003).

\begin{appendix}
\section{Relations among different interfacial force formulations}
\label{eq:derivationNS}
This appendix presents the relations among different interfacial force formulations. In Eq.(\ref{eq:SF7}), the curvature term can be written as
\begin{equation}\label{ap1:curvature}
\nabla\cdot \bm n=\nabla\cdot  \left( \frac{\nabla\phi}{|\nabla\phi|}\right)=\frac{1}{|\nabla\phi|}
\left(\nabla^2\phi-\frac{\nabla\phi\cdot \nabla|\nabla\phi|}{|\nabla\phi|} \right).
\end{equation}
Substituting the above equation into Eq.(\ref{eq:SF7}) yields
\begin{equation}\label{eq}
\begin{aligned}
\bm F_{csf-1}&=-\kappa\nabla\phi|\nabla\phi|\nabla\cdot\bm n \\
&=-\kappa\nabla\phi \left(\nabla^2\phi-\frac{\nabla\phi\cdot \nabla|\nabla\phi|}{|\nabla\phi|} \right), \\
&=\bm F_{pf-1}+ \kappa \frac{\nabla\phi(\nabla\phi\cdot\nabla|\nabla\phi|)}{|\nabla\phi|},
\end{aligned}
\end{equation}
By using the equality  $\nabla\phi\nabla^2\phi=\nabla(\phi\nabla^2\phi)-\phi\nabla\nabla^2\phi$, one can obtain the following relationship,
\begin{equation}
\bm F_{csf-1}=\bm F_{pf-2}- \kappa\nabla (\phi\nabla^2\phi) + \kappa \frac{\nabla\phi(\nabla\phi\cdot\nabla|\nabla \phi|)}{|\nabla\phi|},
\end{equation}
Based on Eq.(\ref{eq:chemical_potential}),  $\bm F_{csf-1}$ can be rewritten as
\begin{equation}
\begin{aligned}
\bm F_{csf-1} =&\nabla\phi \left(\mu_{\phi}- \frac{\partial f_0}{\partial \phi} \right)+  \kappa \frac{\nabla\phi(\nabla\phi\cdot\nabla|\nabla\phi|)}{|\nabla\phi|},\\
  =& \bm F_{cpf-2}- \nabla f_0 + \kappa \frac{\nabla\phi(\nabla\phi\cdot\nabla|\nabla\phi|)}{|\nabla\phi|}, \\
  =& \bm F_{cpf-1}+ \nabla(\phi\mu_{\phi}) - \nabla f_0 + \kappa \frac{\nabla\phi(\nabla\phi\cdot\nabla|\nabla\phi|)}{|\nabla\phi|},
\end{aligned}
\end{equation}
where we have used the equality $\nabla(\phi\mu_\phi)=\mu_\phi \nabla\phi + \phi \nabla\mu_{\phi} $.
By using the following equality,
\begin{equation}
-\kappa \nabla\phi \Delta\phi=\frac{\kappa}{2}\nabla|\nabla\phi|^2 -\nabla\cdot \kappa (\nabla\phi\otimes\nabla\phi),
\end{equation}
one can obtain
\begin{equation}\label{ap:FcsfFstf}
\begin{aligned}
\bm F_{csf-1} &=-\kappa\nabla\phi\nabla^2\phi+ \kappa \frac{\nabla\phi(\nabla\phi\cdot\nabla|\nabla\phi|)}{|\nabla\phi|},\\
 &=\bm F_{stf-1}+\frac{\kappa}{2}\nabla|\nabla\phi|^2+ \kappa \frac{\nabla\phi(\nabla\phi\cdot\nabla|\nabla\phi|)}{|\nabla\phi|},\\
\end{aligned}
\end{equation}
By virtue of Eq.(\ref{eq:equilibrium_profile}), we have
\begin{equation}
\begin{aligned}
|\nabla\phi|=\frac{\partial\phi}{\partial r}=\frac{2}{W}(1-\phi^2), \\
\frac{\nabla\phi (\nabla\phi \cdot \nabla|\nabla\phi| )}{|\nabla\phi|}=\frac{1}{2}\nabla|\nabla\phi|^2=\frac{8}{W^2}\phi(\phi^2-1)\nabla\phi,
\end{aligned}
\end{equation}
Eq.(\ref{ap:FcsfFstf}) is then derived as
\begin{equation}
\bm F_{csf-1}=-\nabla\cdot\kappa (\nabla\phi\otimes\nabla\phi) +\kappa\nabla|\nabla\phi|^2=\bm F_{stf-2}.
\end{equation}

\section{The Dirac function in $\bm F_{csf-1}$ and $\bm F_{csf-2}$}\label{ap3:dirac}
In $\bm F_{csf-1}$, the surface Dirac function is chosen as $\alpha |\nabla \phi|^2$ with $\alpha $ being an undetermined parameter.
Based on Eq.(\ref{eq:equilibrium_profile}),
\begin{equation}\label{eq}
|\nabla\phi|=\frac{\partial \phi}{\partial r}=\frac{2}{W}(1-\phi^2),
\end{equation}
Inserting the above equation into Eq.(\ref{eq:diracfunction}) yields,
\begin{equation}\label{eq}
\begin{aligned}
\int_{-\infty}^{\infty} \alpha|\nabla \phi|^2 dr
&=  \int_{-\infty}^{\infty} \alpha \frac{2}{W}(1-\phi^2) \frac{\partial \phi}{\partial r} dr  \\
&=\int_{-\infty}^{\infty} \alpha \frac{2}{W}(1-\phi^2)  d\phi \\
&= \alpha\frac{2}{W}  \int_{-\infty}^{\infty}  d\left(\phi-\frac{\phi^3}{3} \right) \\
&=\frac{8\alpha}{3W}=1,
\end{aligned}
\end{equation}
where $\phi|_{r=\infty}=1$ and $\phi|_{r=-\infty}=-1$ are used. As a result, $\alpha=\frac{3W}{8}$.

In $\bm F_{csf-2}$, the surface Dirac function is chosen as $\alpha |\nabla \phi| $. Analogously, one can have
\begin{equation}
\begin{aligned}
\int_{-\infty}^{\infty} \alpha|\nabla \phi| dr
=  \int_{-\infty}^{\infty} \frac{\partial \phi}{\partial r} dr =\int_{-\infty}^{\infty} \alpha   d\phi
= 2\alpha=1.
\end{aligned}
\end{equation}
This leads to $\alpha=\frac{1}{2}$.

\end{appendix}
\section*{References}
\bibliography{comparisonSFzchclean}
\end{document}